\documentclass[aps, pra, twocolumn, notitlepage,superscriptaddress,nofootinbib]{revtex4-2}

\usepackage{amsmath}
\usepackage{amsfonts}
\usepackage{amsthm}
\usepackage{amssymb}

\usepackage{bm}
\usepackage{physics}

\usepackage{graphicx}
\usepackage[caption=false]{subfig}

\usepackage[breaklinks=true,colorlinks,citecolor=blue,linkcolor=red,urlcolor=blue]{hyperref}

\newcommand{\ii}{\mathrm{i}}
\newcommand{\ee}{\mathrm{e}}
\newcommand{\ce}{\varepsilon}
\newcommand{\gt}{\Tilde{g}}
\newcommand{\jp}{J_{\perp}}

\newcommand{\one}{\openone}
\newcommand{\ash}{\mathrm{arcsinh}}
\newcommand{\ach}{\mathrm{arccosh}}
\newcommand{\qs}{{q\sigma}}
\newcommand{\qns}{{q\neq0,\sigma}}
\newcommand{\qsprime}{{q^\prime\sigma^\prime}}
\newcommand{\fig}{Fig.}
\newcommand{\figs}{Figs.}
\newcommand{\eq}{Eq.}
\newcommand{\eqs}{Eqs.}

\newcommand{\sect}{section}

\newcommand{\app}{appendix}
\newcommand{\apps}{appendices}

\begin{document}
\title{A full view on the dynamics of an impurity coupled to two one-dimensional fermionic baths}
\author{Martino Stefanini}
\affiliation{International School for Advanced Studies (SISSA), Via Bonomea 265, 34136 Trieste (Italy) }
\author{Massimo Capone}
\affiliation{International School for Advanced Studies (SISSA), Via Bonomea 265, 34136 Trieste (Italy) }
\affiliation{CNR-IOM Democritos, Via Bonomea 265, 34136 Trieste (Italy) }
\author{Alessandro Silva}
\affiliation{International School for Advanced Studies (SISSA), Via Bonomea 265, 34136 Trieste (Italy) }

\date{\today}

\begin{abstract}
We consider a model for the motion of an impurity interacting with two parallel, one-dimensional (bosonized) fermionic baths. The impurity is able to move along the baths, and to jump from one to the other. We provide a perturbative expression for the 
evolution of the system when the impurity is injected in one of the baths, with a given wave packet. 
We obtain an approximation formally of infinite-order in the impurity-bath coupling, which allows us to reproduce the orthogonality catastrophe. We monitor and discuss 
the dynamics of the impurity and its effect on the baths, in particular for a Gaussian wave packet. 
Besides the motion of the impurity, we also analyze the dynamics of the bath density and momentum density (i.e. the particle current), and show that it fits an intuitive semi-classical interpretation. We also quantify the correlation that is established between the baths by calculating the inter-bath, equal-time spatial correlation functions of both bath density and momentum, finding a complex pattern. We show that this pattern contains information on both the impurity motion and on the baths, and that these can be unveiled by taking appropriate ''slices'' of the time evolution.
\end{abstract}

\maketitle

\section{Introduction}
Impurity problems have been a source of fruitful ideas in condensed matter physics since decades. The most famous one is probably the Kondo effect \cite{Kondo, Mahan, Hewson}, in which a fixed spin interacts with a bath of noninteracting electrons, causing an anomalous resistive behavior at low temperature.
While in the Kondo problem the impurity is immobile, a large literature has been devoted to see what happens if the foreign particle can move. The idea traces back at least to Landau \cite{Landau33, LandauPekar48} who introduced the concept of  "polaron" in a solid-state framework to describe the quasiparticle arising from a strong coupling of an electron with lattice phonons \cite{Mahan, DevreesePolarons}.
\par In the last three decades, the progress of ultra-cold atom experiments \cite{ColdAtoms-LewensteinSanperaAhufinger,RevModPhys.80.885-ColdAtoms,PethickSmith} offered new possibilities in the study of impurity problems. The experimental flexibility and control over the various parameters has allowed for a precise investigation of both immobile and mobile impurities, such as foreign atoms in Bose-Einstein condensates \cite{grusdt2015new} and in Fermi gases \cite{Massignan_2014}. In particular, a one-dimensional (1D) setting, in which impurities are restricted to move along elongated baths, has revealed some remarkable phenomena, such as Bloch oscillations even in the absence of a lattice \cite{Meinert945}. Indeed, the specificity of one spatial dimension has stimulated a large array of theoretical and experimental studies \cite{PhysRevA.85.023623, PhysRevLett.103.150601}. 
\par In a recent paper \cite{PhysRevB.103.094310}, the authors considered a model of a mobile impurity that is able to jump between two 1D fermionic baths. This model was meant to provide a simplified perspective of the dynamics of an excitation (the impurity) in a heterostructure, with the baths playing the role of the ''layers''. This investigation was inspired by the budding field of oxide heterostructure engineering \cite{Zubko2011,Hwang2012}, and in particular by the question of the conditions under which quantum-mechanical coherence can improve transport of the excitation through the heterostructure \cite{Kropf2019}. In Ref. \cite{PhysRevB.103.094310} the topic was addressed by computing the Green's function of the impurity.
\par In the present paper, we propose an improved perturbative treatment that is able to reproduce the Green's function calculated in \cite{PhysRevB.103.094310}, while providing the time evolution for the whole system formed by the impurity and the baths.
Thus, we can access the time evolution of any observable (while correlation functions at different times are not directly accessible). Moreover, the numerical effort required is modest enough to allow the study of a large subset of the possible initial momentum distributions of the impurity. We will present the time evolution of some observables for a Gaussian wave packet. We will discuss the dynamics of the probability distribution for the impurity, observing how the motion and the spreading of the wave packet is influenced by the baths. We will also take a different perspective on the polaron dynamics, focusing on the effects of the impurity on the baths. Hence, we will show how the density and current in the baths are modified, and how the exchange of the impurity between the baths is reflected in the shape of their density correlations.
\par The paper is organized as follows: in \sect~\ref{sec:definition of the model} we introduce the model, and prepare the ground for the perturbative treatment by performing a suitable unitary transformation on the Hamiltonian, obtaining a simpler one. We summarize the main results in \sect~\ref{sec:overview of the results}. The following two sections are technical in nature. In \sect~\ref{sec:the perturbative technique} we explain the improved perturbative technique that we used to obtain the evolution of the full impurity-bath state. In \sect~\ref{sec: observables} we illustrate the observables that we will focus on in the rest of the paper, and we provide the expressions of their expectation values. Finally, in \sect~\ref{sec:numerical results} we present the results of our numerical computations of the various observables. In \sect~\ref{sec:conclusions} we sum up our findings and provide some outlook. We have confined some more technical points to the \apps.
\section{Definition of the model}\label{sec:definition of the model}
In this section we define our model and the main assumptions and approximations behind it. We also employ a widely known unitary transformation to simplify the Hamiltonian in a form that is more suitable for the perturbative calculation of the dynamics.
\par We consider a fermionic impurity which moves along a ladder, which hosts two one-dimensional (1D) interacting fermionic baths on its two legs. These baths are independent of each other, and the impurity interacts with each of them. We take the length of the system to be $L=Na$ ($a$ being the lattice spacing), and we work in periodic boundary conditions (pbc). The Hamiltonian is
\begin{equation}
    \mathcal{H}=\mathcal{H}_{\textup{imp}}+\mathcal{H}_{\textup{bath}}+\mathcal{H}_{\textup{c}}~,
\end{equation}
where the noninteracting impurity part is
\begin{align}
    \mathcal{H}_{\textup{imp}}&=2J_\parallel\sum_{j\sigma}{d_{j\sigma}^\dag d_{j\sigma}}-J_\parallel\sum_{j\sigma}{(d_{j+1,\sigma}^\dag d_{j,\sigma}+\textup{h.c.})}+\notag\\
    &-\jp\sum_{j\sigma}{d^\dag_{j\bar{\sigma}}d_{j\sigma}}~.
\end{align}
The index $j$ enumerates the lattice sites along the chains, while the pseudo-spin $\sigma=\uparrow,\downarrow$ (or equivalently $\sigma=\pm1$ used in the equations) specifies the chain. We will assume that the inter-bath hopping $\jp$ is (much) smaller than the intra-bath one $J_\parallel$.
\par We are interested in the low-energy, long-wavelength behavior, therefore we use bosonization (see \cite{Giamarchi, GogolinNersesyanTsvelik}) to write the bath Hamiltonian as a pair of Tomonaga-Luttinger Liquids (TLLs) with sound speeds $v_\sigma$ and Luttinger parameters $K_\sigma$ (encoding the intra-bath interactions):
\begin{equation}
    \mathcal{H}_{\textup{bath}}=\sum_{\sigma}{v_{\sigma}\int{\frac{\dd{x}}{2\pi}}\bigg[K_{\sigma}\bigg(\dv{\theta_{\sigma}}{x}(x)\bigg)^{2}+\frac{1}{K_\sigma}\bigg(\dv{\phi_{\sigma}}{x}(x)\bigg)^{2}\bigg]}~,
\end{equation}
with $\comm{\phi_\sigma(x)}{\theta_{\sigma^\prime}(y)}=-\ii\delta_{\sigma,\sigma^\prime}\arg(\alpha+\ii(x-y))$, $\alpha$ being a small length scale acting as a UV cutoff. Finally, in the spirit of a low-energy approximation, we use a simple density-density contact interaction,
\begin{equation}\label{eq: Hint}
    \mathcal{H}_{\textup{c}}=\sum_{j\sigma}{g_{\sigma}d_{j\sigma}^{\dag}d_{j\sigma}\rho_{\sigma}(ja)}~,
\end{equation}
where we take the long-wavelength approximation $\rho_\sigma(x)=\bar{\rho}_\sigma-1/\pi\dv*{\phi_\sigma(x)}{x}$ \cite{Giamarchi}. The first term is the average density. If the baths have identical properties, this term can be discarded because it is only an overall energy shift. This will be the case for most numerical calculations of this work. 
\par Expressing the bath fields $\phi_\sigma(x)$, $\theta_\sigma(x)$ in terms of the boson modes $b_\qs$ that diagonalize $\mathcal{H}_{\textup{bath}}$ (we follow the conventions of \cite{Giamarchi}), we find
\begin{align}
    \mathcal{H}_{\textup{bath}}&=\sum_{q\neq0,\sigma}{v_\sigma \abs{q}b_\qs^\dag b_\qs}~,\\
    \mathcal{H}_c&=\sum_{j\sigma}{\sum_{q\neq0}{\tfrac{W_\qs}{L^{1/2}}d_{j\sigma}^\dag d_{j\sigma}(\ee^{-\ii q a j}b_\qs^\dag+\ee^{\ii q a j}b_\qs)}}~,\label{eq: H coupling}
\end{align}
where we defined 
\begin{equation}
    W_\qs\equiv g_\sigma K_\sigma^{1/2}\frac{\abs{q}^{1/2}}{(2\pi)^{1/2}}~.
\end{equation}
The description of the baths as TLLs is an effective field theory, valid up to a momentum (energy) cutoff $\alpha^{-1}$ ($\Lambda$). Therefore, we will endow $W_\qs \to W_\qs \ee^{-\alpha\abs{q}/2}$ with a momentum cutoff whenever we will find divergent expressions. 
\par We will often use a momentum space representation for the impurity,
\begin{equation}
    d_{j\sigma}=\frac{1}{\sqrt{N}}\sum_{p}{\ee^{\ii paj}d_{p\sigma}}~,
\end{equation}
so that
\begin{equation}
    \mathcal{H}_{\textup{imp}}=\sum_{p\sigma}{\qty(E(p)d^{\dag}_{p\sigma}d_{p\sigma}-\jp d_{p\bar{\sigma}}^{\dag}d_{p\sigma})}~,
\end{equation}
where $E(p)=2J_\parallel(1-\cos(pa))$. In the low-energy limit we will approximate
\begin{equation}
    E(p)\simeq\frac{p^2}{2M}~,
\end{equation}
where $M^{-1}=2J_\parallel a^2$. We diagonalize $\mathcal{H}_{\textup{imp}}$ in terms of even ($e$) and odd ($o$) modes 
\begin{align}
    d_{p,\mu=e/o}&=\tfrac{1}{\sqrt{2}}(d_{p\,,1}\pm d_{p\,,-1})\!\Rightarrow\\
    \Rightarrow\; \mathcal{H}_{\textup{imp}}&=\sum_{p\mu}{\lambda_{p\mu} d_{p\mu}^\dag d_{p\mu}}~,
\end{align}
where $\lambda_{p, e/o}=E(p)\mp \jp$.

Notice that we neglect the term $\propto \cos(2\bar{\rho}_\sigma x-2\phi_\sigma(x))$ \cite{Giamarchi} in the long-wavelength expansion of the bosonized bath density in \eq~\eqref{eq: Hint}. This term would imply back-scattering of the impurity \cite{Meinert945, Giamarchi, GogolinNersesyanTsvelik} with momentum exchange $2\pi\bar{\rho}_\sigma=2k_{F\sigma}$. Our approximation is justified by the fact that we assume that the impurity momentum is much smaller than $2\pi\bar{\rho}_\sigma$. 
In addition, while the calculations described below in \sect~\ref{sec:the perturbative technique} are non-perturbative in $\jp$, since the deexcitation of the odd mode is accompanied by the emission of phonons of energy of about $2\jp$, this energy (or, equivalently, the wave number $q\sim2\jp/v_\sigma$) must be small enough so that the bosonized description of the baths in terms of sound modes applies. At the same time
 $2\jp\ll2k_{F\sigma} v_\sigma$ in order to ensure that we can neglect the above-mentioned cosine term in the bosonized density.
\subsection{The Lee-Low-Pines transformation}\label{subsec:LLP transformation}
The Hamiltonian presented in the previous paragraphs can be cast in a simpler form that we will use in the perturbative calculation. In polaron problems, it has long been known \cite{LLP} that it is possible to take advantage of the conservation of the total polaron momentum $P_{tot}=\sum_{p\sigma}{p\, d_{p\sigma}^\dag d_{p\sigma}}+\sum_\qns{q\, b_\qs^\dag b_\qs}$. This is achieved by performing a unitary transformation, first introduced by Lee, Low and Pines (LLP):
\begin{equation}
    \mathcal{H}_{LLP}\equiv U_{LLP}^\dag \mathcal{H} U_{LLP},\quad U_{LLP}=\ee^{-\ii X P_b}~,
\end{equation}
with 
\begin{equation}
    X=a\sum_{j\sigma}{j\, d_{j\sigma}^\dag d_{j\sigma}}
\end{equation}
being the impurity position operator\footnote{In pbc the position operator is of course ill-defined. However, the LLP operator is perfectly defined because $P_b$ is quantized in multiples of $2\pi/L$.} and 
\begin{equation}
    P_b=\sum_\qns{q\, b_\qs^\dag b_\qs}
\end{equation}
the total momentum of the baths. This transformation acts as
\begin{subequations}\label{eq: LLP action}
\begin{gather}
    U_{LLP}^\dag d_{j\sigma} U_{LLP}=\ee^{-\ii aj P_b}d_{j\sigma}~,\\
    U_{LLP}^\dag b_{q\sigma} U_{LLP}=\ee^{-\ii q X}b_{q\sigma}~.
\end{gather}
\end{subequations}
Using the property that in a single-impurity subspace $d_{j\sigma}^\dag d_{j\sigma}\ee^{\ii qX}=d_{j\sigma}^\dag d_{j\sigma}\ee^{\ii qaj}$, we obtain
\begin{multline}\label{eq: Hllp}
    \mathcal{H}_{LLP}=\sum_{p\sigma}{E(p-P_b)d_{p\sigma}^\dag d_{p\sigma}}-\jp\sum_{p\sigma}{d_{p\bar{\sigma}}^\dag d_{p\sigma}}+\\
    +\sum_\qns{v_\sigma\abs{q} b_\qs^\dag b_\qs}+\sum_\qns{\tfrac{W_\qs}{L^{1/2}}\sum_j{d_{j\sigma}^\dag d_{j\sigma}}(b_{q\sigma}^\dag+b_{-q\sigma})}~.
\end{multline}
We can see that the impurity momentum $P=\sum_{p\sigma}{p\, d_{p\sigma}^\dag d_{p\sigma}}$ is now conserved. Indeed, in the LLP basis it coincides with the total polaron momentum. As the dynamics does not couple different momentum sectors, we can work in a given sector, and substitute $P$ with its eigenvalue $p$. Then, the only "active" impurity degree of freedom is the bath index, that we describe via pseudo-spin variables
\begin{equation}
    \sigma_i\equiv\sum_{p,\sigma,\sigma^\prime}{(\tau_i)_{\sigma\sigma^\prime}d_{p\sigma}^\dag d_{p\sigma^\prime}}=\sum_{j,\sigma,\sigma^\prime}{(\tau_i)_{\sigma\sigma^\prime}d_{j\sigma}^\dag d_{j\sigma^\prime}}
\end{equation}
In the above equation, $(\tau_i)_{\sigma\sigma^\prime}$ is the i-th Pauli matrix. Taking into account that $\sum_{p\sigma}{d_{p\sigma}^\dag d_{p\sigma}}=\one$ (i.e. there is only one impurity in the system), we can write
\begin{multline}\label{eq: LLP Hamiltonian}
    \mathcal{H}_{LLP}(p)=\tfrac{(p-P_b)^2}{2M}-\jp\sigma_1+\sum_\qns{v_\sigma\abs{q} b_\qs^\dag b_\qs}+\\
    +\sum_\qns{\tfrac{W_\qs}{L^{1/2}}\tfrac{\one+\sigma \sigma_3}{2}(b_{q\sigma}^\dag+b_{-q\sigma})}~.
\end{multline}
This final form of the Hamiltonian, that we will study in the following, is somewhat reminiscent of a spin-boson model \cite{RevModPhys.59.1}. 
\section{Overview of the results}\label{sec:overview of the results}
\begin{figure*}
    \centering
    \subfloat[\label{fig: density comparison 0.1 up}]{\includegraphics[width=0.4\linewidth]{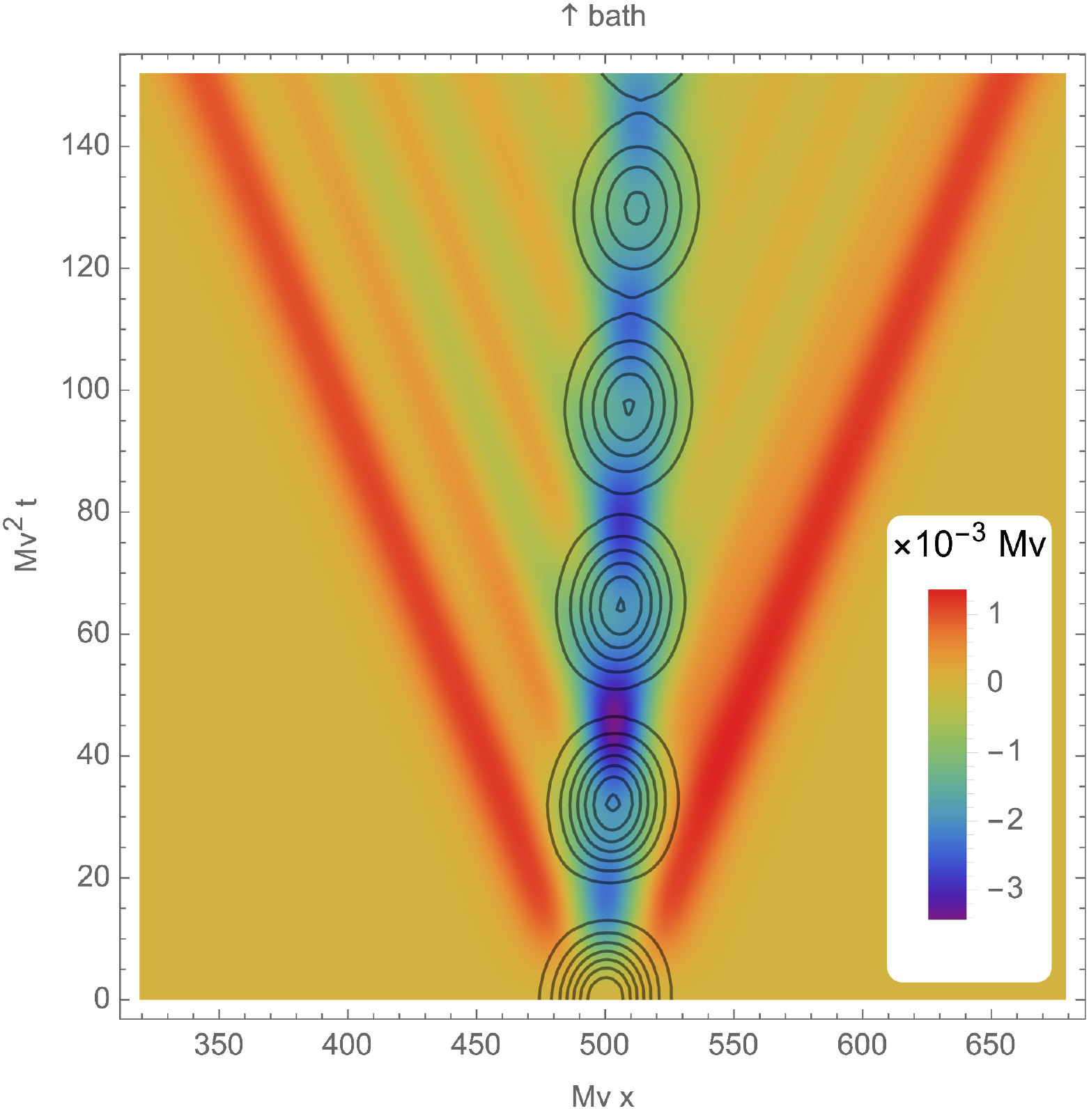}}
    \subfloat[\label{fig: density comparison 0.1 down}]{\includegraphics[width=0.4\linewidth]{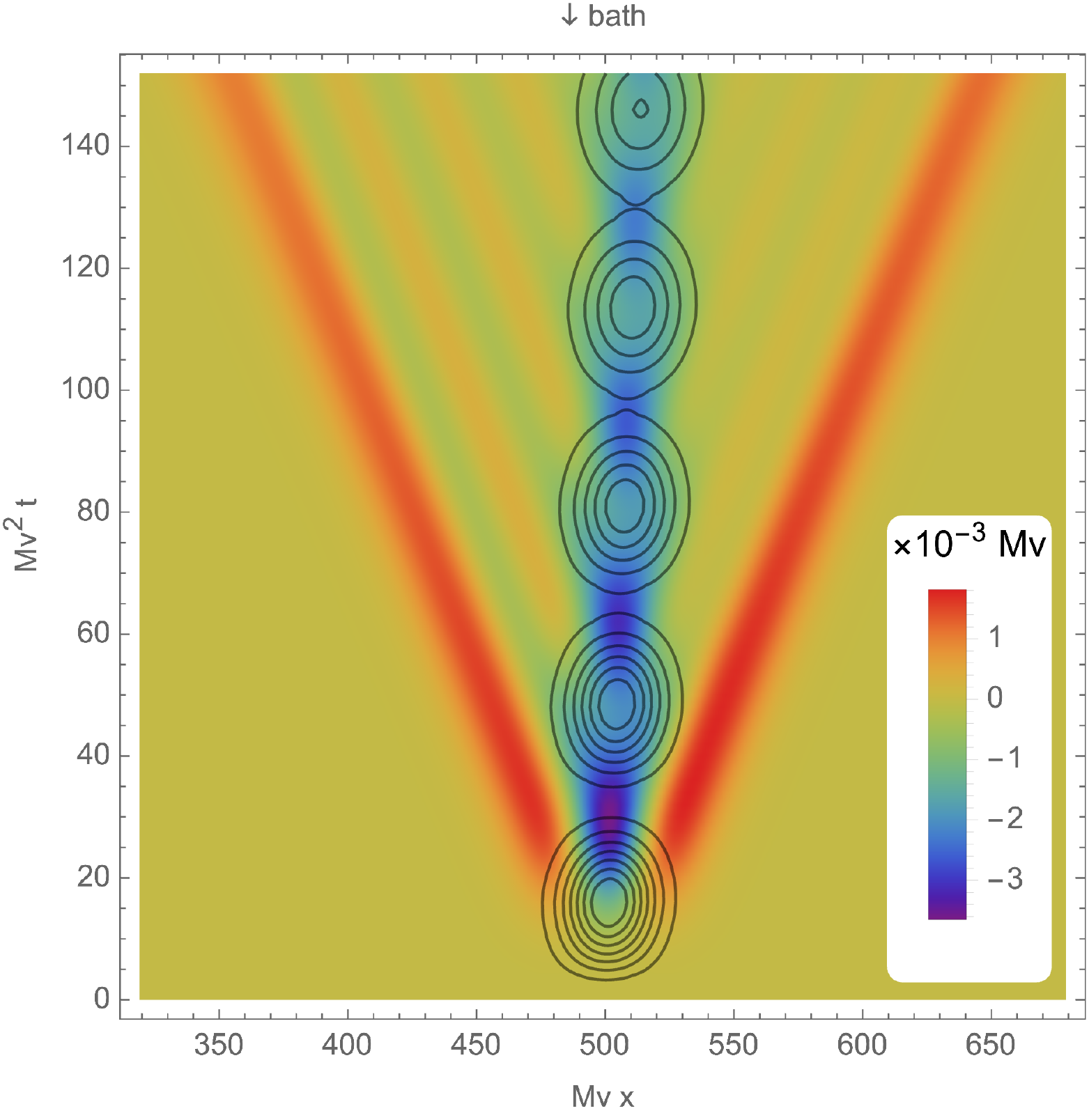}}\\
    \subfloat[\label{fig: density comparison 0.03 up}]{\includegraphics[width=0.4\linewidth]{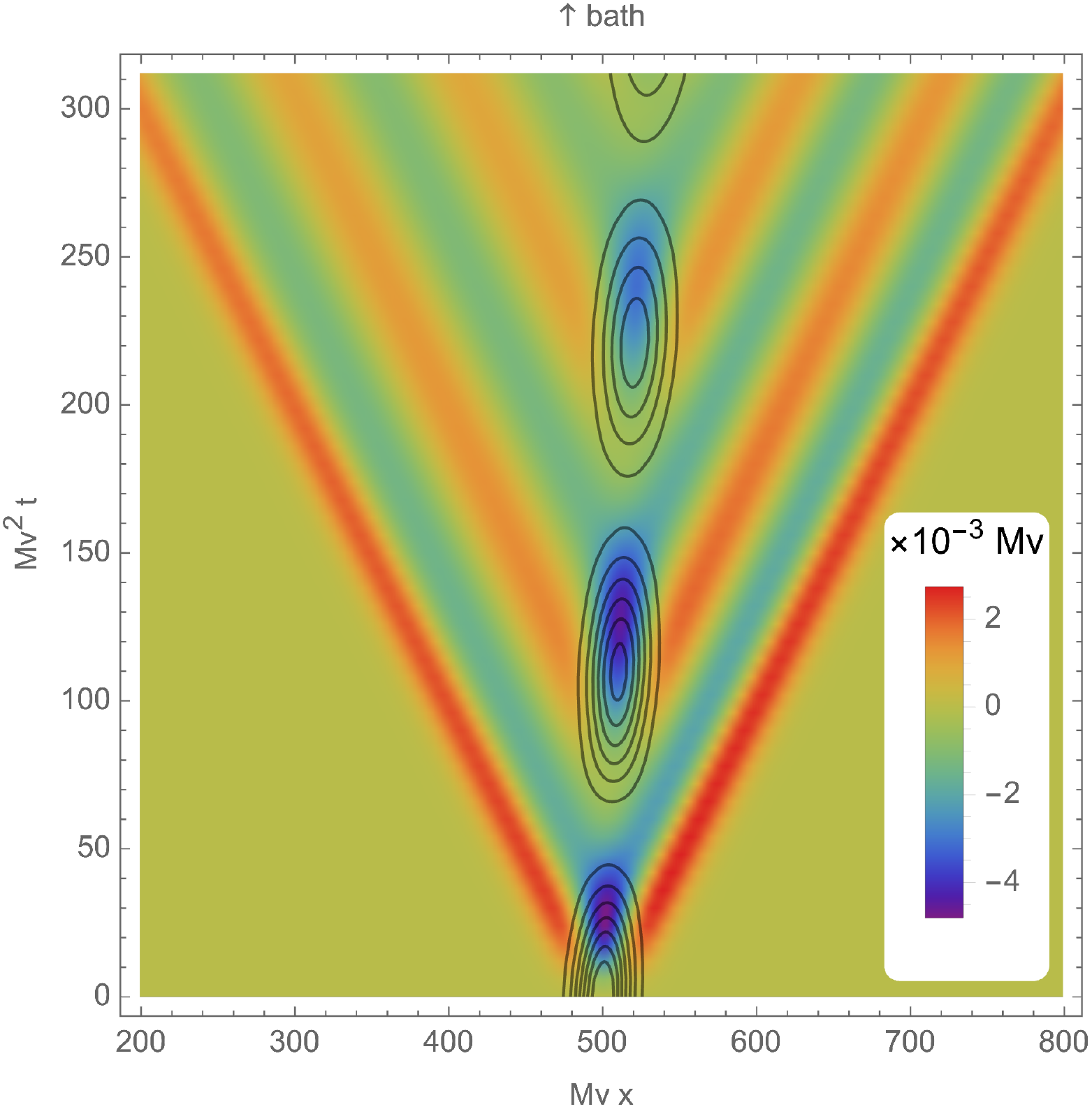}}
    \subfloat[\label{fig: density comparison 0.03 down}]{\includegraphics[width=0.4\linewidth]{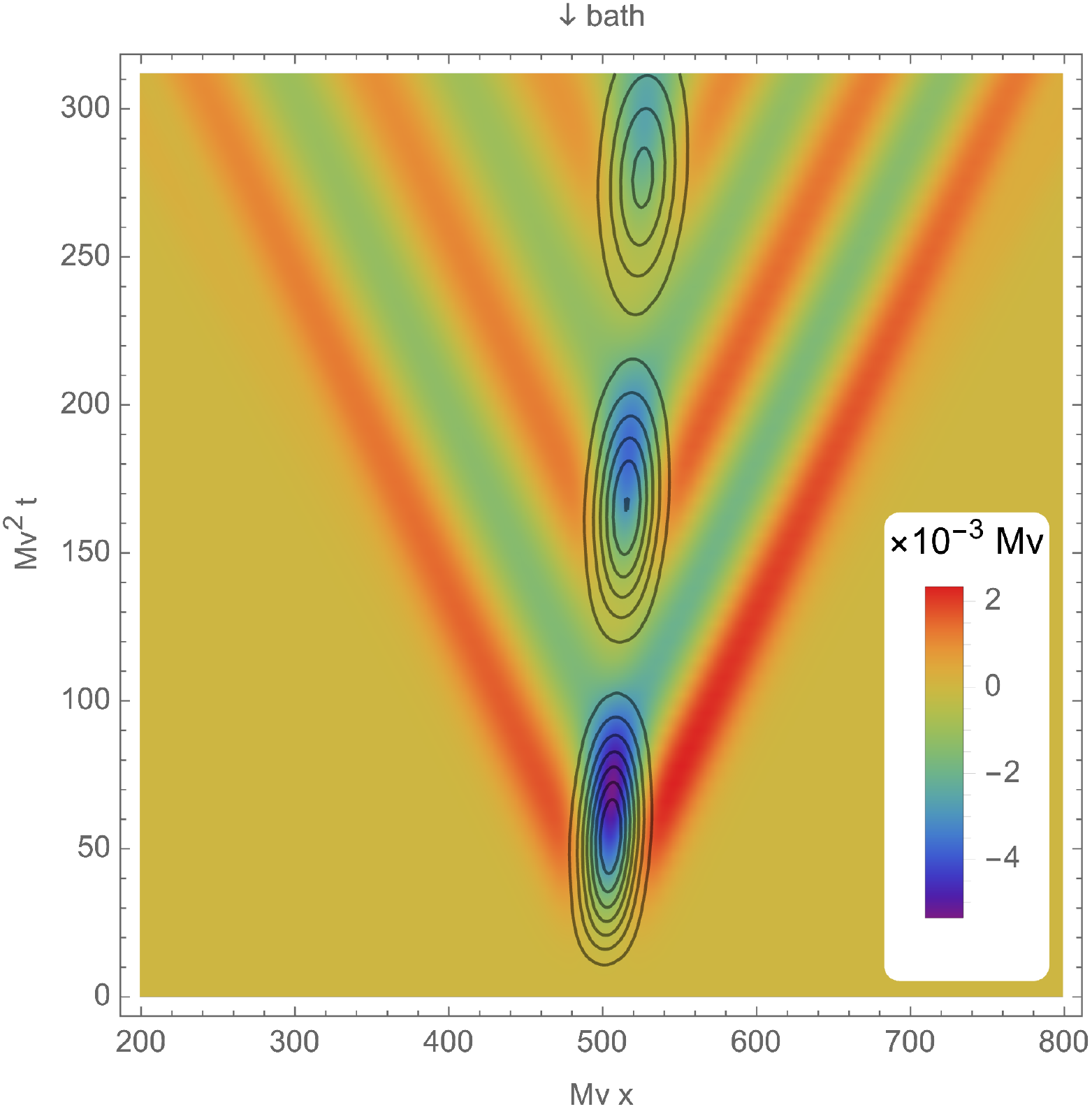}}
    \caption{Comparison of the dynamics of the density fluctuation in the baths (color scale) and of the impurity density (black contours) for $g^2K=0.5v^2$ and (a, b) $\jp=0.1Mv^2$ or (c, d) $\jp=0.03Mv^2$. The dynamics of impurity and baths and are more synchronized for low $\jp$. The wave packet is Gaussian, composed of $N_p=64$ momenta distributed around $p_0=0.1Mv$ with a standard deviation of $\delta p=0.04Mv$. The color scale of (a) [(c)] is the same of (b) [(d)].}
    \label{figs: density comparison}
\end{figure*}
In the next section, we will study the dynamics of the model \eqref{eq: LLP Hamiltonian} using an improved, infinite-order perturbative expansion in the impurity-bath coupling $g_\sigma K^{1/2}_\sigma/v_\sigma$. Since the formalism is quite involved, we will start presenting the physical picture that emerges from the results we obtain. We assume that initially the baths are in their ground state $\ket{\omega}$, while the impurity is introduced in the system in an arbitrary wave packet
\begin{equation}
    \ket{\Psi(0)}=\sum_{p,\mu\in\{e,o\}}{c_{p\mu}\ket{p ,\mu}_d}\ket{\omega}_b~.
\end{equation}
When the baths are in the ground state the LLP transformation acts as the identity, so the state above can be considered as the representation of the wave function both in the laboratory and in the LLP frames.
\par The main achievement of this paper will be a perturbative expression of the state of the impurity and baths, which allows us to obtain the dynamics of the baths in response to the introduction and motion of the impurity. The advantage of having an approximate representation for the wave function is that we will be able to understand the dynamics of impurity and baths on the same footing. The case of a Gaussian impurity wave packet is illustrated in \fig~\ref{figs: density comparison}. In these plots, the contour lines depict the probability density of finding the impurity at a given position, while the color scale represents the fluctuation of the density of the baths. The plots on the left depict the $\sigma=1$ (or $\uparrow$) bath, while the ones on the right represent the situation in the $\sigma=-1$ ($\downarrow$) bath. The impurity is initialized in a Gaussian wave packet with a standard deviation of position of about $12.5 (Mv)^{-1}$ around the center of the $\uparrow$ bath, with an average momentum $p_0=0.1Mv$. 
\par
\begin{figure}
    \centering
    \includegraphics[width=.9\linewidth]{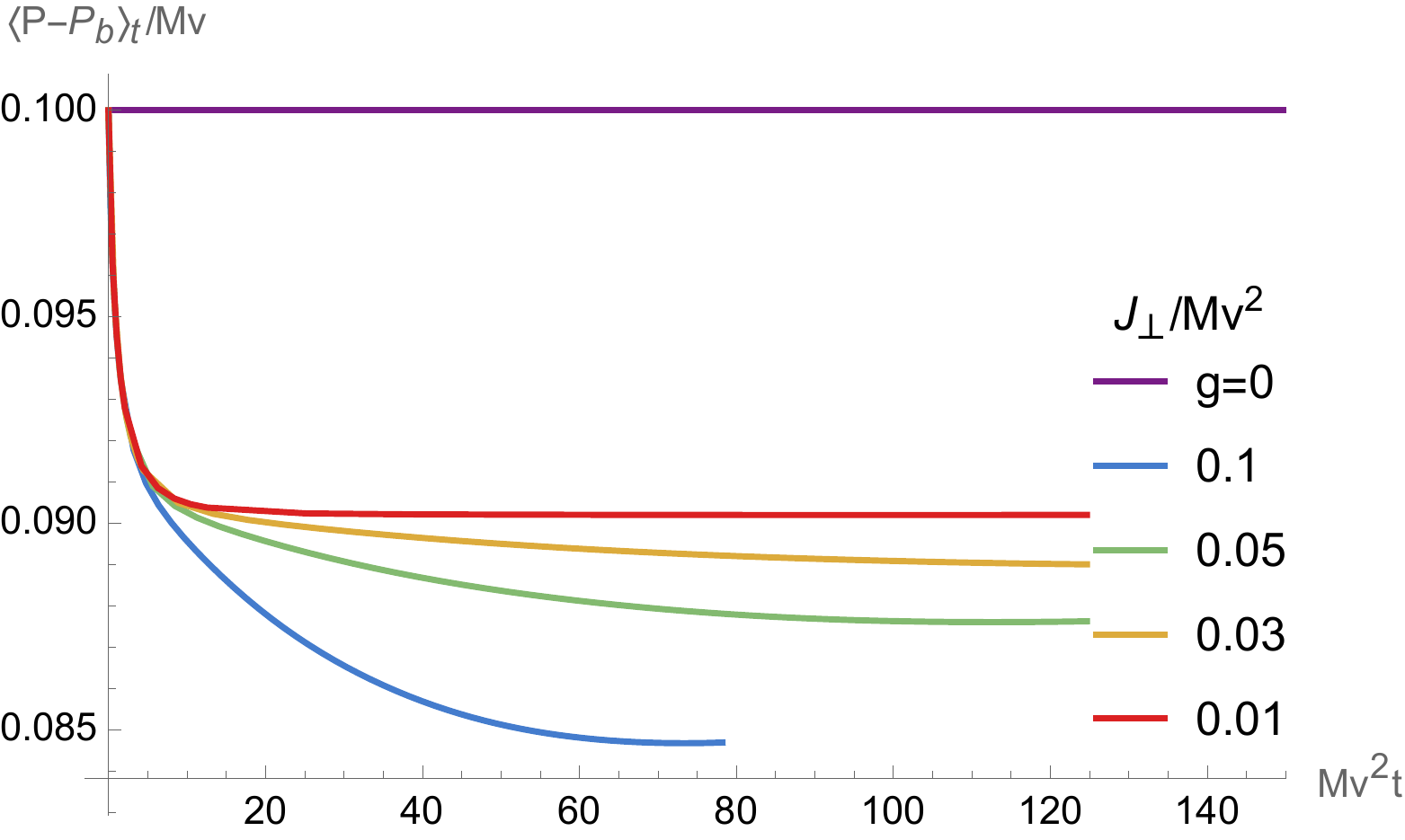}
   \caption{Time evolution of the impurity momentum for the initial state $\ket{p_0\uparrow}_d$, and symmetric baths. The plots are for $p_0=0.1Mv$, $g^2K=0.5v^2$ and show the effect of increasing $\jp$. The $\jp=0.1Mv^2$ curve is shown only within the time interval when the perturbative solution is valid.}
    \label{fig: impurity momentum J}
\end{figure}

 The motion of the impurity is qualitatively similar to what we would expect in the absence of the interaction: the whole wave packet oscillates from one bath to the other, while drifting because of the finite momentum. There is a little distortion in the shape of the wave packet, with a tendency of the peaks to spread in the time direction (namely, the impurity never completely leaves a bath for the other). At the same time, there is a net momentum transfer between impurity and baths, as detailed in \fig~\ref{fig: impurity momentum J}. This momentum transfer is clearly seen also in the dynamics of the baths. The latter evolve in an intuitive, semiclassical fashion: when the impurity appears for the first time in one of the baths, the latter responds by forming a density depletion in the position of the impurity, while at the same time two wave fronts form on its sides and propagate away at the speed of sound. Then, as the impurity oscillates from one bath to the other, the depth of the density depletion oscillates as well, and these oscillations propagate in the baths in the form of two trains of ripples. The wavelength of the latter corresponds to that of the phonons emitted from the decay of the odd mode of the impurity, which is different for backward and forward emission because of the finite impurity momentum. 
Notice that the dynamics of the baths lags behind the time evolution of the impurity. This effect is particularly visible for larger inter-bath hopping (\figs~\ref{fig: density comparison 0.1 up} and \ref{fig: density comparison 0.1 down}), while for smaller $\jp$ the baths and the impurity are almost synchronized (\figs~\ref{fig: density comparison 0.03 up} and \ref{fig: density comparison 0.03 down}). 
\par The relation of the densities of the baths with the impurity density, presented in the figures above, can be understood by noticing that the equation of motion linking them at operator level is
\begin{equation}\label{eq: eq of motion rho}
    \left(\tfrac{1}{v_\sigma^2}\partial_t^2-\partial_x^2\right)\rho_\sigma(x,t)=\tfrac{g_\sigma K_\sigma}{\pi v_\sigma}\partial_x^2 N_\sigma(x,t)~,
\end{equation}
where $N_\sigma(x)\equiv\sum_{j}\delta(x-ja)d_{j\sigma}^\dag d_{j\sigma}$. The above equation is valid within the approximation that the bosonized density retains only the longest-wavelength contribution $-1/\pi\partial_x\phi_\sigma(x)$, which guarantees that the equations of motion for the densities are linear. Solving \eq~\eqref{eq: eq of motion rho}
one easily obtains 
\begin{multline}\label{eq: linear relation densities}
    \rho_\sigma(x,t)= \rho_\sigma(x,0)+\\
    +g_\sigma \int_\mathbb{R}\dd{x^\prime}\int_0^t\dd{t^\prime}\chi_\sigma(x-x^\prime,t-t^\prime)N_\sigma(x^\prime,t^\prime)~,
\end{multline}
where $\rho_\sigma(x,0)$ is the noninteracting density. The kernel $\chi_\sigma(x,t)$ is the retarded density-density linear response function for the baths,
\begin{equation}
 \begin{aligned}
    \chi_\sigma(x,t)&\equiv-\ii\theta(t)\ev{\comm{\rho_\sigma(x,t)}{\rho_\sigma(0,0)}}{\omega}=\notag\\
    &=\theta(t)\tfrac{K_\sigma}{2\pi}\left[\delta_\alpha^\prime(x+v_\sigma t)-\delta_\alpha^\prime(x-v_\sigma t)\right]~,
\end{aligned}   
\end{equation}

where the prime indicates a derivative with respect to the argument of the function
\begin{equation}
    \delta_\alpha(x)\equiv\frac{1}{\pi}\frac{\alpha}{x^2+\alpha^2}~.
\end{equation}
This function is a smeared representation of the Dirac delta function, whose smearing parameter $\alpha$ is the length that serves as a UV cutoff for the TLL.
\par If we take the average of \eq\eqref{eq: linear relation densities} over the initial state $\ket{\Psi(0)}=\ket{\chi}_d\ket{\omega}_b$, we obtain the relation
\begin{equation}
    \expval{\rho_\sigma(x)}_t=g_\sigma \int_\mathbb{R}\dd{x^\prime}\int_0^t\dd{t^\prime}\chi_\sigma(x-x^\prime,t-t^\prime)\expval{N_\sigma(x^\prime)}_{t^\prime}
\end{equation}
between the bath and impurity density. We have introduced the notation $\expval{\mathcal{O}(x,t)}\equiv\expval{\mathcal{O}(x)}_t$ to indicate the expectation value at time $t$ of the operator $\mathcal{O}(x)$ in the Schr\"odinger picture. Substituting the expression for the retarded response function with $\alpha\to0$ in the previous \eq~we get 
\begin{multline}\label{eq: continuum density linear response}
    \expval{\rho_\sigma(x)}_t=\tfrac{g_\sigma K_\sigma}{2\pi}\int_0^t\dd{t^\prime}\bigg[\partial_{x^\prime}\expval{N_\sigma(x^\prime)}_{t^\prime}\eval_{x^\prime=x+v_\sigma(t-t^\prime)}+\\-\partial_{x^\prime}\expval{N_\sigma(x^\prime)}_{t^\prime}\eval_{x^\prime=x-v_\sigma(t-t^\prime)}\bigg]~,
\end{multline}
where the spatial arguments $x\pm v_\sigma(t-t^\prime)$ have to be interpreted modulo translation by the length of the system, $L$, because of the pbc. Although the above equation looks like a linear response formula, we stress that it is actually non-perturbative, because it comes from the Heisenberg equation of motion, \eqref{eq: eq of motion rho}. It establishes an exact relation between the \emph{interacting} impurity and bath density, valid within our long-wavelength description of the system. It shows that the bath density a coordinates $(x,t)$ is a \emph{superposition} of the all the values of the gradient of the impurity density on the light-cone of the given space-time point. Besides indicating that the baths have rather long ''memory'', the superposition of the various images of the non-positive-definite gradient of the impurity density paves the way to interference effects. Equation \eqref{eq: continuum density linear response} also provides a hint that the time evolution of the bath density will have a semi-classical character, in the sense that it follows the motion of the impurity.
\par In the following paragraphs, we will build a perturbative approximation of the time evolution of the system state. This solution will give us access to the dynamics of both the density of the impurity and of the baths, and we will observe the realization of the semi-classical behavior entailed by \eq~\eqref{eq: continuum density linear response} [albeit \eq~\eqref{eq: continuum density linear response} itself will be satisfied only approximately].
\section{The perturbative technique}\label{sec:the perturbative technique}

We will now discuss the technique with which the results exposed in the previous section were obtained, and construct a perturbative expansion for the impurity-bath wave function.
The first useful step is to rewrite the LPP in a given momentum sector, \eq~(\ref{eq: LLP Hamiltonian}) separating the \it noninteracting \rm part ${\cal H}_0(p)$ and the perturbation $V$
\begin{equation}\label{eq:separation of HLLP}
\mathcal{H}_{LLP}(p)=\mathcal{H}_{0}(p)+\sigma_{3}V+:\tfrac{P_b^{2}}{2M}:~,
\end{equation}
where the colons $:\cdot :$ stand for normal-ordering with respect to the phononic vacuum $\ket{\omega}$. The noninteracting Hamiltonian is  
\begin{subequations}
\begin{gather}
\mathcal{H}_{0}(p)=
h_0(p)
+E(p)-\jp\sigma_{1}~,
\label{eq: H0}
\end{gather}
\end{subequations}
with $h_0(p)$ being the bare bath Hamiltonian
\begin{equation}
    h_0(p)\equiv\sum_{\qns}{\big[\Omega_{\qs}(p)b^{\dag}_{\qs}b_{\qs}+\tfrac{W_{\qs}}{2\sqrt{L}}(b^{\dag}_{\qs}+b_{\qs})\big]}~,
\end{equation}
with 
\begin{equation}
\Omega_\qs(p)\equiv v_\sigma \abs{q} -\tfrac{qp}{M} +\tfrac{q^2}{2M}~.    
\end{equation}
The perturbation  $V$ is\footnote{We are using $W_{\qs}=W_{-\qs}$ to change the sign of the momentum label of the annihilation operator.} 
\begin{equation}\label{eq:V}
V=\sum_{q\neq 0, \sigma}\sigma \frac{W_{q\sigma}}{2\sqrt{L}}(b^{\dagger}_{q\sigma}+b_{q\sigma}).    
\end{equation}
\par The idea is to obtain a perturbative expansion for the time evolved wave function in powers of $g_\sigma K_\sigma^{1/2}/v_\sigma$, treating $\mathcal{H}_0$ as unperturbed Hamiltonian. We notice that $\mathcal{H}_0$ contains $g_\sigma K_\sigma^{1/2}/v_\sigma$, hence the resulting perturbative approximation will be actually of infinite order. A physical picture behind this choice is described in \app~\ref{app: symmetric baths}. In intuitive terms, the unperturbed Hamiltonian contains the bath-induced transitions in which the band index $(e,o)$ of the impurity does not change (intra-band transitions), whereas the $\sigma_3 V$ term describes inter-band transitions. We will apply the approach described in Ref. \cite{RevModPhys.44.602}, that is designed to avoid the appearance of secular terms, i.e. terms that grow indefinitely in time, whose presence would invalidate the perturbative treatment \cite{BenderOrszag}.
\par Let us begin from the case in which the initial condition is factorized as $\ket{p\mu}_d \ket{\omega}_b\equiv\ket{p\mu,\omega}\equiv\ket{\Psi_{p\mu}(0)}$. Then, we go to a modified interaction picture
\begin{equation}
\ket{\Psi_{p\mu}(t)}=\ee^{-\ii\mathcal{H}_{0} t}\ket{\psi^I_{p\mu}(t)}=\ee^{-\ii\mathcal{H}_{0} t}a_{p\mu}(t)\ket{\phi_{p\mu}(t)}~,
\end{equation}
in which the vector $\ket{\psi^I_{p\mu}(t)}$ is split into a complex function $a_{\mu}(t)$ and a state $\ket{\phi_{p\mu}}$. This splitting is specified imposing   
\begin{equation}\label{eq: unit condition}
    \braket{p\mu,\omega}{\phi_{p\mu}(t)}\equiv 1
\end{equation}
at all times, implying that 
\begin{equation}
\ket{\phi_{p\mu}(t)}=\ket{p\mu,\omega}+\ket{\delta \phi_{p\mu}},
\end{equation}
with $\ket{\delta\phi_{p\mu}}$ orthogonal to $\ket{p\mu,\omega}$. This condition ensures that secular terms will be resummed to all orders into $a_{p\mu}(t)$. 

Substituting the \eqs~above into the Schr\"odinger equation we obtain
\begin{subequations}\label{eq: equation set}
\begin{align}
\ii\dv{}{t}\ket{\phi_{p\mu}(t)}&=(\Delta\mathcal{H}(t)-\Delta E_{p\mu}(t))\ket{\phi_{p\mu}(t)}~,\\
\ii\dv{a_{p\mu}}{t}&=\Delta E_{p\mu}(t) a_{p\mu},\, a_{p\mu}(0)=1~,\\
\Delta E_{p\mu}(t)&\equiv\mel{p\mu,\omega}{\Delta\mathcal{H}(t)}{\phi_{p\mu}(t)}~,
\end{align}
\end{subequations}
where the equation for $a_{p\mu}$ is obtained by projecting onto $\ket{p\mu,\omega}$, and $\Delta\mathcal{H}(t)$ is the interaction-picture perturbation 
\begin{equation}
\Delta\mathcal{H}(t)= \ee^{\ii \mathcal{H}_{0} t}\big(\sigma_{3}V+:\tfrac{P_b^{2}}{2M}:\big)\ee^{-\ii \mathcal{H}_{0} t}.
\end{equation}
Notice that we chose to treat $:P_b^2:/2M$ as a perturbation, despite its formal independence from the coupling constant. This choice is justified when the initial bath state is the vacuum, as in this case the phonons modes will start to be populated only because of the interaction. This approach is akin to a spin-wave expansion in magnetic systems \cite{Auerbach}.

The equation for $a_{p\mu}$ is integrated straightforwardly:
\begin{equation}\label{eq: full solution a}
    a_{p\mu}(t)=\ee^{-\ii\int_0^t\dd{t^\prime}\Delta E_{p\mu}(t^\prime)}~.
\end{equation}
The equation for $\ket{\phi_{p\mu}(t)}$ will in turn be  solved perturbatively, assuming the expansion
\begin{equation}
    \ket{\delta\phi_{p\mu}}=\sum_{n=1}^\infty\ket{\phi_{p\mu}^{(n)}}.
\end{equation}
where $\ket{\small\phi_{p\mu}^{(n)}}=\order{[g_\sigma K_\sigma^{1/2}/v_\sigma]^{n}}$. As usual in perturbative treatments, this Ansatz generates a hierarchy of equations for $\ket{\phi^{(n)}_{p\mu}(t)}$, which have to be solved by matching terms of the same order. Consequently, the function $\Delta E_{p\mu}(t)$ becomes a series in powers of the coupling, which is then substituted into \eq~\eqref{eq: full solution a}.
In our case, the matching of powers of $g K^{1/2}/v$ is non-trivial, because $\mathcal{H}_0$ already contains the coupling constant.

The details of the expansion can be now worked out order by order. The interaction-picture perturbation perturbation is
\begin{align}
\Delta\mathcal{H}(t)=\sigma_{3}(t)(\hat{V}(t)+\expval{V(t)})+:\tfrac{P_b^{2}}{2M}:(t).
\end{align}
Here\footnote{Notice that $\expval{V(t)}$ is non-vanishing only if the baths are asymmetric, in which case it is logarithmically divergent in the TLL cutoff.}
\begin{subequations}
\begin{align}
\sigma_{3}(t)&=\sum_{\mu=e/o}{\ee^{-2\ii\mu\jp t}\dyad{p\mu}{p\bar{\mu}}}~,\\
\hat{V}(t)&=\sum_{\qns}{\sigma\tfrac{W_{\qs}}{2L^{1/2}}(b_{\qs}\ee^{-\ii\Omega_{\qs}t}+b_{\qs}^{\dag}\ee^{\ii\Omega_{\qs}t})}~,\\
\expval{V(t)}&=-2\sum_{\qns}{\sigma\tfrac{W_{\qs}^{2}}{4L}\frac{1-\cos{\Omega_{\qs}t}}{\Omega_{\qs}}}~.
\end{align}
\end{subequations}
The last term originates from the inclusion of a term $\propto g_\sigma K_\sigma^{1/2}/v_\sigma$ in $h_0(p)$. 
Notice that
\begin{equation}
\sigma_{3}(t)\ket{p\mu}=\ee^{2\ii\mu\jp t}\ket{p\bar{\mu}}
\end{equation}
where for $\mu=e/o$ we have $\bar{\mu}=o/e$.
\par Now we plug the expansion of $\ket{\phi_{p\mu}}$ in \eq~(\ref{eq: equation set}a). At first order, we have
\begin{equation}
i\frac{d}{dt}\ket{\phi^{(1)}_{p\mu}(t)}=
(\Delta{\cal H}(t)-\Delta E^{(1)}_{p\mu}(t))\ket{p\mu,\omega},
\end{equation}
with 
\begin{equation}
 \Delta E^{(1)}_{p\mu}(t)=\bra{p\mu,\omega}   \Delta{\cal H}(t) \ket{p\mu,\omega}.
\end{equation}
The action of $\Delta\mathcal{H}(t)$ on the initial state reads 
\begin{multline}\label{eq: deltaH phi0}
    \Delta\mathcal{H}(t)\ket{p\mu,\omega}=\sum_\qns{\sigma\tfrac{W_{\qs}}{2L^{1/2}}\ee^{\ii\Omega_\qs^{+\mu}(p)t}b^{\dag}_\qs}\ket{p\bar{\mu},\omega}+\\
    +\ee^{2\ii\mu\jp t}\expval{V(t)}\ket{p\bar{\mu},\omega}+\tfrac{:P_b^2:}{2M}(t)\ket{p\mu,\omega}~.
\end{multline}
We have introduced the convenient shorthands
\begin{equation}
    \begin{aligned}
        \Omega_\qs^{\pm e}(p)&\equiv&  \Omega_\qs(p)\pm 2\jp~,\\
    \Omega_\qs^{\pm o}(p)&\equiv&  \Omega_\qs(p)\mp 2\jp~.
    \end{aligned}
\end{equation}
The lowest-order term generated by $\tfrac{:P_b^2:}{2M}(t)\ket{p\mu,\omega}$ is of order 2, so it is easy to verify that $\Delta E^{(1)}_{p\mu}(t)=0$. Moreover, we assume that $\tfrac{:P_b^2:}{2M}(t)\ket{\phi^{(1)}_{p\mu}(t)}$ is of second order, as well. We will verify the consistency of this assumption \textit{a posteriori}. Since $\langle V(t)\rangle$ is a second-order contribution we have that the first-order equation is simply
\begin{gather}
\ii\dv{}{t}\ket{\phi_{p\mu}^{(1)}(t)}=\sigma_{3}(t)\hat{V}(t)\ket{p\mu,\omega}~,
\end{gather}
which is straightforwardly integrated:
\begin{equation}\label{eq: first order phi}
    \ket{\phi_{\mu}^{(1)}(t)}=\sum_{\qns}{\sigma\tfrac{W_{\qs}}{2L^{1/2}}\tfrac{1-\ee^{\ii\Omega_{\qs}^{+\mu}t}}{\Omega_{\qs}^{+\mu}}b_{\qs}^{\dag}\ket{p\bar{\mu}\omega}}~.
\end{equation}
From the above equation, it can be verified that $\tfrac{:P_b^2:}{2M}(t)\ket{\phi^{(1)}_{p\mu}(t)}$ is of second order, as claimed before. The correct normalization of the state is guaranteed by computing $\Delta E_{p\mu}(t)$ to the second order: 
\begin{gather*}
\Delta E_{p\mu}(t)=\Delta E_{p\mu}^{(2)}(t)=\mel{\mu\omega}{\sigma_{3}(t)\hat{V}(t)}{\phi_{\mu}^{(1)}(t)}=\\
=-\sum_{\qns}{\tfrac{W_{\qs}^{2}}{4L}\tfrac{1-\ee^{-\ii\Omega^{+\mu}_{\qs}t}}{\Omega^{+\mu}_{\qs}}}~.
\end{gather*}
The normalization factor is:
\begin{align}\label{eq: apmu}
a_{p\mu}(t)&=\ee^{-\ii\int_{0}^{t}\dd{t_1}\Delta E^{(2)}_{p\mu}(t_{1})}=\notag\\
&=\exp(-\sum_{\qns}{\tfrac{W_{\qs}^{2}}{4L}\tfrac{1-\ii\Omega^{+\mu}_\qs t-\ee^{-\ii\Omega^{+\mu}_{\qs}t}}{(\Omega^{+\mu}_{\qs})^2}})~.
\end{align}

\par The second-order correction is quite more involved. There is one contribution coming from the second term of \eq~\eqref{eq: deltaH phi0}:
\begin{multline}\label{eq: delta1 phi2}
\ket{\delta_1\phi_{p\mu}^{(2)}(t)}=-\ii\int_0^t \dd{t^\prime}\ee^{2\ii\mu\jp t}\expval{V(t^\prime)}\ket{p\bar{\mu},\omega}=\\
=\tfrac{1}{2\mu\jp}\sum_\qns{}\sigma\tfrac{W_\qs^2}{4L}\big[\ee^{2\ii\mu\jp t}2\Re\chi_t(\Omega_\qs(p))+\\
-\chi_t(\Omega_\qs^{-\mu}(p))-\chi_t^\ast(\Omega_\qs^{+\mu}(p))\big]\ket{p\bar{\mu},\omega}\equiv\\
\equiv B_{p\mu}(t)\ket{p\bar{\mu},\omega}~.
\end{multline}
The other contribution involves bosonic creation operators and, in analogy with the first-order correction \eq~\eqref{eq: first order phi}, it has the form
\begin{equation}\label{eq: second order term}
    \ket{\delta_2\phi_{p\mu}^{(2)}(t)}=\sum_{\nu}{\sum_{\qs,\qsprime}{A^{\mu\nu}_{\qs,\qsprime}(p,t)b_\qs^\dag b_\qsprime^\dag \ket{p\nu,\omega}}}~,
\end{equation}
where $A^{\mu\nu}_{\qs,\qsprime}(p,t)$ is taken to be of order two, and symmetric upon exchange of $\qs$ with $\qsprime$. Then, it is easy to see that $:P_b^2/2M:(t)\ket{\delta_2\phi_{p\mu}^{(2)}(t)}$ generates a term of second order. Thus, the usual hierarchical structure of perturbative expansion (in which terms of order $n$ are the sources for the next order) is lost, but the equation for $A_{\qs,\qsprime}^{\mu\nu}(t)$ is still solvable. The details can be found in \app~\ref{app: second order terms}, here we only quote the results:
\begin{multline}\label{eq: Amumu}
    A_{\qs,\qsprime}^{\mu\mu}(p,t)=\tfrac{W_\qs W_\qsprime}{4L}\big\{\tfrac{q q^\prime}{2M} \tfrac{1}{\Omega_\qs   \Omega_\qsprime  }\times \\
    \times\big[\ee^{\ii\Omega_\qs  t}\chi_t(\Omega_\qs  +\tfrac{q q^\prime}{M})+\ee^{\ii\Omega_\qsprime  t}\chi_t(\Omega_\qsprime  +\tfrac{q q^\prime}{M})+ \\
    -\chi_t(\tfrac{q q^\prime}{M})-\ee^{\ii(\Omega_\qs  +\Omega_\qsprime  )t}\chi_t(\Omega_\qs  +\Omega_\qsprime  +\tfrac{q q^\prime}{M})\big]+ \\
    \tfrac{\sigma\sigma^\prime}{2}\big[\tfrac{1}{\Omega_\qsprime^{+\mu}  }\big(\ee^{\ii(\Omega_\qs  +\Omega_\qsprime  )t}\chi_t(\Omega_\qs  +\Omega_\qsprime  +\tfrac{q q^\prime}{M})+ \\
    -\ee^{\ii\Omega_\qs^{-\mu}  t}\chi_t(\Omega_\qs^{-\mu}  +\tfrac{q q^\prime}{M})
    \big)+ \\
    +\tfrac{1}{\Omega_\qs^{+\mu}  }\big(\ee^{\ii(\Omega_\qs  +\Omega_\qsprime  )t}\chi_t(\Omega_\qs  +\Omega_\qsprime  +\tfrac{q q^\prime}{M})+ \\
    -\ee^{\ii\Omega_\qsprime^{-\mu}  t}\chi_t(\Omega_\qsprime^{-\mu}  +\tfrac{q q^\prime}{M})
    \big)\big] \big\}
\end{multline}
\begin{multline}\label{eq: Amubarmu}
    A_{\qs,\qsprime}^{\mu\bar{\mu}}(p,t)=\tfrac{1}{2}\tfrac{W_\qs W_\qsprime}{4L}\big\{\tfrac{\sigma}{\Omega_\qs^{+\mu}  \Omega_\qsprime  }\times \\
    \big[\ee^{\ii\Omega_\qs^{+\mu}  t}\chi_t(\Omega_\qs  ^{+\mu}+\tfrac{q q^\prime}{M})+\ee^{\ii\Omega_\qsprime  t}\chi_t(\Omega_\qsprime  +\tfrac{q q^\prime}{M})+ \\
    -\chi_t(\tfrac{q q^\prime}{M})-\ee^{\ii(\Omega_\qs^{+\mu}  +\Omega_\qsprime  )t}\chi_t(\Omega_\qs  ^{+\mu}+\Omega_\qsprime  +\tfrac{q q^\prime}{M})\big]+ \\
    +\tfrac{\sigma^\prime}{\Omega_\qsprime^{+\mu}  \Omega_\qs  }\times \\
    \big[\ee^{\ii\Omega_\qsprime^{+\mu}  t}\chi_t(\Omega_\qsprime  ^{+\mu}+\tfrac{q q^\prime}{M})+\ee^{\ii\Omega_\qs  t}\chi_t(\Omega_\qs  +\tfrac{q q^\prime}{M})+ \\
    -\chi_t(\tfrac{q q^\prime}{M})-\ee^{\ii(\Omega_\qs  +\Omega_\qsprime^{+\mu}  )t}\chi_t(\Omega_\qs  +\Omega_\qsprime^{+\mu}  +\tfrac{q q^\prime}{M})\big]
    \big\}~.
\end{multline}
In the above equations, all $\Omega_\qs$s and $\Omega_\qs^{\pm\mu}$ are evaluated at momentum $p$,
and we defined the function
\begin{equation}
    \chi_t(\ce)\equiv\frac{1-\ee^{-\ii\ce t}}{\ce}~.
\end{equation}
\par To sum up, when $\ket{\Psi_{p\mu}(0)}=\ket{p\mu,\omega}$, the state evolution is approximated by
\begin{multline}\label{eq: sol pmu1}
    \ket{\Psi_{p\mu}(t)}=a_{p\mu}(t)\ee^{-\ii h_0(p)t}\big[\ee^{-\ii\lambda_{p\mu}t}\ket{p\mu,\omega}+\\
    +\ee^{-\ii\lambda_{p\bar{\mu}}t}\sum_{\qns}{\sigma\tfrac{W_{\qs}}{2L^{1/2}}\tfrac{1-\ee^{\ii\Omega_{\qs}^{+\mu}(p)t}}{\Omega_{\qs}^{+\mu}(p)}b_{\qs}^{\dag}\ket{p\bar{\mu}\omega}}+\\
    +\ee^{-\ii\lambda_{p\bar{\mu}}t}B_{p\mu}(t)\ket{p\bar{\mu},\omega}+\\
    +\sum_{\nu}{\sum_{\qs,\qsprime}{\ee^{-\ii\lambda_{p\nu}t}A^{\mu\nu}_{\qs,\qsprime}(p,t)b_\qs^\dag b_\qsprime^\dag \ket{p\nu,\omega}}}\big]~.
\end{multline}
Letting $\ee^{-\ii h_0(p) t}$ act on the bath states, we can also write
\begin{multline}\label{eq: sol pmu2}
     \ket{\Psi_{p\mu}(t)}=a_{p\mu}(t)\big[\ee^{-\ii\lambda_{p\mu}t}\ket{p\mu}\ket{\omega_p(t)}+\\
    -\ee^{-\ii(\lambda_{p\bar{\mu}}-2\mu\jp)t}\sum_{\qns}{\sigma\tfrac{W_{\qs}}{2L^{1/2}}\tfrac{1-\ee^{-\ii\Omega_{\qs}^{+\mu}(p)t}}{\Omega_{\qs}^{+\mu}(p)}b_{\qs}^{\dag}\ket{p\bar{\mu}}\ket{\omega_p(t)}}+\\
    +\ee^{-\ii\lambda_{p\bar{\mu}}t}\tilde{B}_{p\mu}(t)\ket{p\bar{\mu}}\ket{\omega_p(t)}+\\
    +\sum_{\nu}{}\sum_{\qs,\qsprime}{\ee^{-\ii\lambda_{p\nu}t}A^{\mu\nu}_{\qs,\qsprime}(p)\ee^{-\ii(\Omega_\qs(p)+\Omega_\qsprime(p))t}}\times\\ \times b_\qs^\dag b_\qsprime^\dag \ket{p\nu}\ket{\omega_p(t)}
    \big]~,
\end{multline}
where
\begin{align}
    \ket{\omega_{p}(t)}&\equiv\ee^{-\ii h_{0} t}\ket{\omega}=\notag\\
    &=\ee^{\ii\alpha_p(t)}\ket{\textup{coh}\big[-\tfrac{W_{\qs}}{2L^{1/2}}\tfrac{1-\ee^{-\ii\Omega_{\qs}(p)t}}{\Omega_{\qs}(p)}\big]}~,
\end{align}
is the "zeroth order" evolution of the initial bath state. The notation $\ket{\mathrm{coh}[z_\qs]}$ means a coherent state
\begin{equation}
    \ket{\mathrm{coh}[z_\qs]}\equiv\ee^{\sum_\qns{(z_\qs b_\qs^\dag-z^\ast_\qs b_\qs)}}\ket{\omega}~,
\end{equation}
and the phase is
\begin{equation}
    \alpha_p(t)\equiv\sum_{\qns}{\tfrac{W_{\qs}^{2}}{4L}\tfrac{\Omega_\qs(p) t-\sin(\Omega_{\qs}(p)t)}{(\Omega_{\qs}(p))^2}}~.
\end{equation}
Notice that commuting $\ee^{-\ii h_0(p)t}$ with $b^{\dagger}_{p\sigma}$ introduces a shift in $B_{p\mu}(t)$,
hence in \eq~\ref{eq: sol pmu2} we have
\begin{equation}
    \tilde{B}_{p\mu}(t)=B_{p\mu}(t)-\sum_\qns{\sigma\tfrac{W^2_\qs}{4L}\chi_t^\ast(\Omega_\qs^{+\mu}(p))}\chi_t(\Omega_\qs(p))~.
\end{equation}
The analogous shifts in the last term of \eq~(\ref{eq: sol pmu2}) have been neglected for consistency, since they result in higher-order contributions in our small parameter. The shift above is logarithmically divergent in the TLL cutoff, and vanishes whenever the two baths are identical, that is, they have the same properties ($v_\sigma,\, K_\sigma,\, g_\sigma$). In the following, we will work mainly in this symmetric case, hence we will soon discard it.
\par The perturbative solution \eqref{eq: sol pmu2} has a clear physical interpretation: the state evolution is approximated by the emission of phonons from the deexcitation $\ket{o}\to\ket{e}$ above a coherent-state background, which will turn out to embody the orthogonality catastrophe. The solution is also reminiscent of a popular Ansatz used to describe Fermi polarons in higher dimensions, in which the impurity state is expanded in terms with an increasing number of particle-hole pairs excited from the Fermi sphere \cite{Massignan_2014, PhysRevA.74.063628}. As the TLL bosons are descendants of particle-hole pairs, we see that the expression in \eq~\eqref{eq: sol pmu2} is indeed the 1D analogue of the above-mentioned Ansatz, albeit it is perturbative and not variational in nature. The main difference between the two is that instead of adding excitations on the Fermi sphere (i.e. the bosonic ground state $\ket{\omega}$), in \eq~\eqref{eq: sol pmu2} the bosons are added on top of a coherent state which embodies the physics of the OC. 
\par From the above results, the time evolution of a wave packet $\ket{\Psi(0)}=\sum_{p\mu}{c_{p\mu}\ket{p\mu,\omega}}$ is simply given by the superposition of \eq~\eqref{eq: sol pmu1} or \eqref{eq: sol pmu2}:
\begin{multline}\label{eq: sol wave packet1}
    \ket{\Psi(t)}=\sum_{p\mu}{\ee^{-\ii\lambda_{p\mu}t}\ket{p\mu}\ee^{-\ii h_0(p)t}}\times\\
    \times\big[(c_{p\mu}a_{p\mu}+c_{p\bar{\mu}}a_{p\bar{\mu}}B_{p\bar{\mu}}(t))\ket{\omega}+\\
    +c_{p\bar{\mu}}a_{p\bar{\mu}}\sum_{\qns}{\sigma\tfrac{W_{\qs}}{2L^{1/2}}\tfrac{1-\ee^{\ii\Omega_{\qs}^{-\mu}t}}{\Omega_{\qs}^{-\mu}}b_{\qs}^{\dag}\ket{\omega}}+\\
    +\sum_{\qs,\qsprime}{}\big(c_{p\mu}a_{p\mu}A^{\mu\mu}_{\qs,\qsprime}(p,t)+\\
    +c_{p\bar{\mu}}a_{p\bar{\mu}}A^{\bar{\mu}\mu}_{\qs,\qsprime}(p,t)\big)b_\qs^\dag b_\qsprime^\dag\ket{\omega}\big]~,
\end{multline}
or
\begin{multline}\label{eq: sol wave packet2}
\ket{\Psi(t)}=\sum_{p\mu}{\ee^{-\ii\lambda_{p\mu}t}\ket{p\mu}}\times\\
\times\big[(c_{p\mu}a_{p\mu}+c_{p\bar{\mu}}a_{p\bar{\mu}}\tilde{B}_{p\bar{\mu}}(t))\ket{\omega_p(t)}\\
    +c_{p\bar{\mu}}a_{p\bar{\mu}}\ee^{-2\ii\mu\jp t}\sum_{\qns}{\sigma\tfrac{W_{\qs}}{2L^{1/2}}\tfrac{1-\ee^{-\ii\Omega_{\qs}^{-\mu}t}}{\Omega_{\qs}^{-\mu}}b_{\qs}^{\dag}\ket{\omega_p(t)}}+\\
    +\sum_{\qs,\qsprime}{}\big(c_{p\mu}a_{p\mu}A^{\mu\mu}_{\qs,\qsprime}(p,t)+c_{p\bar{\mu}}a_{p\bar{\mu}}A^{\bar{\mu}\mu}_{\qs,\qsprime}(p,t)\big)\times\\
    \times\ee^{-\ii(\Omega_\qs(p)+\Omega_\qsprime(p))t}b_\qs^\dag b_\qsprime^\dag\ket{\omega_p(t)}\big]~.
\end{multline}
\par The calculation and characterization of the normalization factors $a_{p\mu}(t)$ can be found in \app~\ref{app: normalization factors}. For now, we just point out that they have the long-time asymptotic behaviors
\begin{subequations}\label{eq: asymptotics apmu}
\begin{align}
    a_{pe}(t)&\sim\ee^{z_{pe}-\ii\Delta\lambda_{pe}t+\order{1/t}}\\
    a_{po}(t)&\sim\ee^{z_{po}-\ii\Delta\lambda_{po}t-2\gamma_p t+\order{1/t}}~,
\end{align}
\end{subequations}
where $z_{p\mu}$ are two complex constants. These behaviors are established rather quickly (for small momentum, already for $\jp t\sim0.1$). The expressions for the parameters appearing in the equations above can be found in \app~\ref{app: normalization factors}.
\par We remark that the validity of the perturbative solution is limited in time because the state normalization is guaranteed only up to the fourth order. As a consequence, the maximum time before the normalization is significantly lost is of the order of the odd mode decay time, $1/(2\gamma_p)$, which is usually much longer than the period of the hopping between baths. 
\par As a check of the reliability of this solution, we calculated the "Green's function" $G_\parallel(p,t)=-\ii\mel{p\sigma,\omega}{\ee^{-\ii\mathcal{H}t}}{p\sigma,\omega}$ and compared the result with the one from the Linked Cluster Expansion (previously computed by the present authors in \cite{PhysRevB.103.094310}), finding perfect agreement in the symmetric case. More details can be found in \app~\ref{app: comparison with LCE}
\section{Observables}\label{sec: observables}
In this \sect, we report the expressions of some interesting observables of the impurity-baths systems, obtained from our perturbative solution \eqs~\eqref{eq: sol wave packet1} or \eqref{eq: sol wave packet2}.
\subsection{Impurity observables}\label{subsec:impurity observables}
The simplest observable is the probability that the impurity is found in the bath $\sigma$, which is the expectation value of the operator
\begin{equation}
    n_\sigma\equiv\sum_j{d_{j\sigma}^\dag d_{j\sigma}}~.
\end{equation}
After the LLP transformation, it reads [compare with \eq~\eqref{eq: Hllp} and \eq~\eqref{eq: LLP Hamiltonian}]
\begin{equation}
    n_\sigma=\frac{1}{2}(1+\sigma\sigma_3)~.
\end{equation}
Using the notation
\begin{equation}
    \ket{\Psi(t)}=\sum_{p\mu}{\ket{p\mu}_d\ket{\psi_{p\mu}(t)}_b}
\end{equation}
for the evolved state (notice that the bath states are not normalized to one), we find\footnote{Notice that we are setting explicitly $\braket{\Psi_t}\equiv 1$. However, this is true only for short times with respect to $\min_p(2\gamma_p)^{-1}$, because of the perturbative normalization. If we were to take this into account, we would have $\sum_\sigma\expval{n_\sigma}_t$ decreasing from $1$ in time.}
\begin{equation}\label{eq: nsigma}
    \expval{n_\sigma}_t=\frac{1}{2}+\frac{\sigma}{2}\sum_{p\mu}{\braket{\psi_{p\mu}(t)}{\psi_{p\bar{\mu}}(t)}}~,
\end{equation}
where $\expval{\cdot}_t\equiv\ev{\cdot}{\Psi(t)}$. The overlaps are easily calculated to be
\begin{multline}\label{eq: mu bar mu overlap}
    \braket{\psi_{p\mu}(t)}{\psi_{p\bar{\mu}}(t)}=\ee^{-2\ii\mu\jp t}\big[(c_{p\mu}a_{p\mu})^\ast c_{p\bar{\mu}}a_{p\bar{\mu}}+\\
    +\abs{c_{p\bar{\mu}}a_{p\bar{\mu}}}^2 \tilde{B}_{p\bar{\mu}}^\ast(t)+\abs{c_{p\mu}a_{p\mu}}^2\tilde{B}_{p\mu}(t)\big]+\\
    +\abs{c_{p\mu}a_{p\mu}}^2\sum_\qns{\sigma\tfrac{W_\qs^2}{4L}\tfrac{1-\ee^{\ii\Omega_\qs(p)t}}{\Omega_\qs(p)}\tfrac{1-\ee^{-\ii\Omega_\qs^{+\mu}(p)t}}{\Omega_\qs^{+\mu}(p)}}+\\
    +\abs{c_{p\bar{\mu}}a_{p\bar{\mu}}}^2\sum_\qns{\sigma\tfrac{W_\qs^2}{4L}\tfrac{1-\ee^{\ii\Omega_\qs^{-\mu}(p)t}}{\Omega_\qs^{-\mu}(p)}\tfrac{1-\ee^{-\ii\Omega_\qs(p)t}}{\Omega_\qs(p)}}+\\
    +\ee^{2\ii\mu\jp t}(c_{p\bar{\mu}}a_{p\bar{\mu}})^\ast c_{p\mu}a_{p\mu}\sum_\qns{\tfrac{W_\qs^2}{4L}\tfrac{1-\ee^{\ii\Omega_\qs^{-\mu}(p)t}}{\Omega_\qs^{-\mu}(p)}\tfrac{1-\ee^{-\ii\Omega_\qs^{+\mu}(p)t}}{\Omega_\qs^{+\mu}(p)}}+\\
    +\order{(gK^{1/2}/v)^3}
\end{multline}
at second order.
\par In problems concerning mobile impurities, a natural observable to be considered is the impurity momentum. In the LLP basis, it reads $P_d=P-P_b$, so, since $P$ is a constant of motion, all that is required to compute is the momentum carried by the baths. The latter is given by
\begin{equation}\label{eq: bath momentum}
    \begin{aligned}
    \expval{P_b}_t&=\sum_\qns{q \expval{b_\qs^\dag b_\qs}_t}=\\
    &=2\sum_{p\mu}\abs{c_{p\mu} a_{p\mu}}^2\sum_\qns{q\tfrac{W^2_\qs}{4L}\tfrac{1-\cos{\Omega_\qs^{+\mu}(p)t}}{(\Omega_\qs^{+\mu}(p))^2}}+\\
    &+2\Re\left[\sum_{p\mu}(c_{p\mu}a_{p\mu})^\ast c_{p\bar{\mu}}a_{p\bar{\mu}}\right.\times\\ &\times\left.\sum_\qns{\sigma q\tfrac{W^2_\qs}{4L}\tfrac{1-\ee^{-\ii\Omega_\qs(p) t}}{\Omega_\qs(p)}\tfrac{1-\ee^{\ii\Omega_\qs^{-\mu}(p)t}}{\Omega_\qs^{-\mu}(p)}}\right]+\\
    &+2\sum_{p\mu}\abs{c_{p\mu}}^2\sum_\qns{q\tfrac{W^2_\qs}{4L}\tfrac{1-\cos{\Omega_\qs(p) t}}{(\Omega_\qs(p))^2}}
\end{aligned}
\end{equation}
The sums over the bath momenta can be converted into energy integrals by introducing the appropriate density of states
\begin{align*}
    &\tfrac{1}{L}\sum_{q\neq0} q W_\qs^2 f(\Omega_\qs(p))=\int\dd{\ce}R_1^\sigma(\ce;p)f(\ce)~,\\
    &R_1^\sigma(\ce;p)\equiv\tfrac{1}{L}\sum_{q\neq0} q W_\qs^2\delta(\ce-\Omega_\qs(p))~,
\end{align*}
which in the continuum limit is given by
\begin{multline}
    R_1^\sigma(\ce; p)=\tfrac{M^2\gt_\sigma^2}{(2\pi)^2}\theta(\ce)\ee^{-\ce/\Lambda}\sum_{s=\pm1}\big[-s \big(v_\sigma+s\tfrac{p}{M}\big)\big]\times\\
    \times\bigg[\left(1+\tfrac{\ce}{k_{s\sigma}(p)}\right)^{1/2}-2+\left(1+\tfrac{\ce}{k_{s\sigma}(p)}\right)^{-1/2}\bigg]~.
\end{multline}
\par The expression in \eq~\eqref{eq: bath momentum} is composed of three contributions. The first two are contributions coming from the deexcitation of the odd mode, as signaled by their explicit dependence on $\jp$. The second of these two accounts for the asymmetries in the bath index, as it is vanishes if the baths are symmetric or if the impurity is initialized in one of its noninteracting eigenstates. The last part of \eq~\eqref{eq: bath momentum} comes from the coherent background $\ket{\omega_p(t)}$ term of the baths state, and is independent of the initial bath index of the impurity. It quantifies the momentum adsorbed by the baths as they adjust to the injection of the impurity.
\par Finally, we can calculate the probability of finding the impurity at the site $j$ and in the bath $\sigma$, namely the expectation value of the number operator $d_{j\sigma}^\dag d_{j\sigma}$. The latter is invariant under the LLP transformation [see \eq~\eqref{eq: LLP action}]. Using $d_{j\sigma}^\dag d_{j\sigma}=N^{-1}\sum_p{\ee^{\ii paj}\sum_k{d^\dag_{k-p\sigma}d_{k\sigma}}}$ and $d_{p\sigma}=(d_{pe}+\sigma d_{po})/\sqrt{2}$ we find
\begin{multline}\label{eq: impurity density}
    \expval{d_{j\sigma}^\dag d_{j\sigma}}_t=\tfrac{1}{2N}\sum_p\ee^{i p ja}\sum_{k,\mu}{\big(\braket{\psi_{k-p\mu}}{\psi_{k\mu}}}+\\
    +\sigma\braket{\psi_{k-p\mu}}{\psi_{k\bar{\mu}}}\big)~.
\end{multline}
Notice that \eq~\eqref{eq: impurity density} correctly reproduces \eq~\eqref{eq: nsigma} when summed on all sites. The fundamental ingredients of the above \eq~are the overlaps of the bath states, which read
\begin{subequations}
\begin{gather}
    \braket{\psi_{p\mu}}{\psi_{k\mu}}=\ee^{\ii(E_p-E_k)t}\braket{\omega_p(t)}{\omega_k(t)}\bigg[(c_{p\mu}a_{p\mu})^\ast c_{k\mu}a_{k\mu}+\notag\\
    +(c_{p\bar{\mu}}a_{p\bar{\mu}})^\ast c_{k\mu}a_{k\mu} \tilde{B}_{p\bar{\mu}}^\ast+(c_{p\mu}a_{p\mu})^\ast c_{k\bar{\mu}}a_{k\bar{\mu}} \tilde{B}_{k\bar{\mu}}+\notag\\
    +\ee^{-i\mu2\jp t}(c_{p\mu}a_{p\mu})^\ast c_{k\bar{\mu}}a_{k\bar{\mu}}\times\nonumber \\
    \times\sum_\qns{\sigma\tfrac{W_\qs^2}{4L}\tfrac{1-\ee^{\ii\Omega_\qs(p)t}}{\Omega_\qs(p)}\tfrac{1-\ee^{-\ii\Omega_\qs^{-\mu}(k)t}}{\Omega_\qs^{-\mu}(k)}}+\notag\\
    +\ee^{i\mu2\jp t}(c_{p\bar{\mu}}a_{p\bar{\mu}})^\ast c_{k\mu}a_{k\mu}\times\nonumber \\
    \times \sum_\qns{\sigma\tfrac{W_\qs^2}{4L}\tfrac{1-\ee^{\ii\Omega_\qs^{-\mu}(p)t}}{\Omega_\qs^{-\mu}(p)}\tfrac{1-\ee^{-\ii\Omega_\qs(k)t}}{\Omega_\qs(k)}}+\notag\\
    +(c_{p\bar{\mu}}a_{p\bar{\mu}})^\ast c_{k\bar{\mu}}a_{k\bar{\mu}}\sum_\qns{\tfrac{W_\qs^2}{4L}\tfrac{1-\ee^{\ii\Omega_\qs^{-\mu}(p)t}}{\Omega_\qs^{-\mu}(p)}\tfrac{1-\ee^{-\ii\Omega_\qs^{-\mu}(k)t}}{\Omega_\qs^{-\mu}(k)}}+\notag\nonumber \\
    +\order{(g K^{1/2}/v)^3}\bigg],\\
    \braket{\psi_{p\mu}}{\psi_{k\bar{\mu}}}=\ee^{\ii(E_p-E_k-\mu2\jp)t}\braket{\omega_p(t)}{\omega_k(t)}\times\notag\\
    \times\bigg[(c_{p\mu}a_{p\mu})^\ast c_{k\bar{\mu}}a_{k\bar{\mu}}+\notag\nonumber \\
    +(c_{p\bar{\mu}}a_{p\bar{\mu}})^\ast c_{k\bar{\mu}}a_{k\bar{\mu}} \tilde{B}_{p\bar{\mu}}^\ast+(c_{p\mu}a_{p\mu})^\ast c_{k\mu}a_{k\mu} \tilde{B}_{k\mu}+\notag\\
    +\ee^{i\mu2\jp t}(c_{p\mu}a_{p\mu})^\ast c_{k\mu}a_{k\mu}\times\notag\\
    \times\sum_\qns{\sigma\tfrac{W_\qs^2}{4L}\tfrac{1-\ee^{\ii\Omega_\qs(p)t}}{\Omega_\qs(p)}\tfrac{1-\ee^{-\ii\Omega_\qs^{+\mu}(k)t}}{\Omega_\qs^{+\mu}(k)}}+\notag \\
    +\ee^{i\mu2\jp t}(c_{p\bar{\mu}}a_{p\bar{\mu}})^\ast c_{k\bar{\mu}}a_{k\bar{\mu}} \times\notag\\
    \times\sum_\qns{\sigma\tfrac{W_\qs^2}{4L}\tfrac{1-\ee^{\ii\Omega_\qs^{-\mu}(p)t}}{\Omega_\qs^{-\mu}(p)}\tfrac{1-\ee^{-\ii\Omega_\qs(k)t}}{\Omega_\qs(k)}}+\notag\\
    +\ee^{\ii 4\mu\jp t}(c_{p\bar{\mu}}a_{p\bar{\mu}})^\ast c_{k\mu}a_{k\mu}\times\notag\\
    \times\sum_\qns{\tfrac{W_\qs^2}{4L}\tfrac{1-\ee^{\ii\Omega_\qs^{-\mu}(p)t}}{\Omega_\qs^{-\mu}(p)}\tfrac{1-\ee^{-\ii\Omega_\qs^{+\mu}(k)t}}{\Omega_\qs^{+\mu}(k)}}+\notag\\
    +\order{(g K^{1/2}/v)^3}\bigg]~,
\end{gather}
\end{subequations}
where the coherent-states overlap is given by
\begin{multline}
    \ln\braket{\omega_p(t)}{\omega_k(t)}=-\tfrac{1}{4L}\sum_{\qns}{W^2_\qs\bigg[ \tfrac{1+\ii\Omega_\qs(p)t-\ee^{\ii\Omega_\qs(p)t}}{(\Omega_\qs(p))^2}}+\\
    +\tfrac{1-\ii\Omega_\qs(k)t-\ee^{-\ii\Omega_\qs(k)t}}{(\Omega_\qs(k))^2}-\tfrac{1-\ee^{\ii\Omega_\qs(p)t}}{\Omega_\qs(p)}\tfrac{1-\ee^{-\ii\Omega_\qs(k)t}}{\Omega_\qs(k)}
    \bigg]
\end{multline}
For any $p\neq k$ it has a slow, power-law decrease in time and it is a non-analytic function of the momenta in $p=k$. Of course, for $p=k$ it is identically equal to $1$ (as $\ket{\omega_{p}(t)}$ is normalized), which is also its maximum absolute value.
\subsection{Bath observables}\label{subsec:bath observables}
We will look at correlation functions of the bath densities $\rho_\sigma(x)$ and conjugate momenta \cite{Giamarchi} $\Pi_\sigma(x)\equiv 1/\pi \dv*{\theta_\sigma(x)}{x}$,
\begin{subequations}\label{eq: expv densities}
\begin{align}
    \rho_\sigma(x)&=-\frac{1}{\pi}\dv{}{x}\phi_\sigma(x)=\notag\\
    &=\tfrac{K_\sigma^{1/2}}{L^{1/2}}\sum_{q\neq0}{V_q(\ee^{\ii q x}b_\qs+\ee^{-\ii q x}b_\qs^\dag)}\\
    \Pi_\sigma(x)&=\frac{1}{\pi}\dv{}{x}\theta_\sigma(x)=\notag\\
    &=\tfrac{1}{K_\sigma^{1/2}L^{1/2}}\sum_{q\neq0}{\mathrm{sgn}(q) V_q(\ee^{\ii q x}b_\qs+\ee^{-\ii q x}b_\qs^\dag)}~,
\end{align}
\end{subequations}
where $\mathrm{sgn}(q)$ is the sign function. We remark that we understand $\rho_\sigma(x)$ as the fluctuation part of the density, that is, we already subtracted the average density $\bar{\rho}_\sigma$ from it\footnote{If the physical meaning of $\rho_\sigma(x)$ is obvious, it may be less so for the momentum density $\Pi_\sigma(x)$. The equation of motion for the density (i.e. the continuity equation) shows that, within our long-wavelength approximation, $\Pi_\sigma(x)$ is proportional to the particle current: $j_\sigma(x)=v_\sigma K_\sigma \Pi_\sigma(x)$.}. In order to use the perturbative solution we found, we must first perform a LLP transformation [\eq~\eqref{eq: LLP action}], which replaces $b_\qs\to b_\qs\ee^{-\ii q X}$. Therefore, we have
\begin{subequations}
\begin{align}
    \expval{\rho_\sigma(x)}_t&=2\Re\Big[\tfrac{K_\sigma^{1/2}}{L^{1/2}}\sum_{q\neq0}{V_q \ee^{\ii q x}\expval{\ee^{-\ii q X}b_\qs}_t}\Big]~,\\
    \expval{\Pi_\sigma(x)}_t&=2\Re\Big[\tfrac{1}{K_\sigma^{1/2}L^{1/2}}\sum_{q\neq0}{\mathrm{sgn}(q) V_q \ee^{\ii q x}}\times\notag\\
    &\times\expval{\ee^{-\ii q X}b_\qs}_t \Big]~.
\end{align}
\end{subequations}
Using expression \eqref{eq: sol wave packet1} [or \eqref{eq: sol wave packet2}] and the property that $\ee^{-\ii q X}\ket{p\mu}=\ket{p-q\mu}$, we find
\begin{multline}\label{eq: expv b}
    \expval{\ee^{-\ii q X}b_\qs}_t=-\tfrac{W_\qs}{2 L^{1/2}}\sum_{p\mu}{\ee^{\ii(E_{p-q}-E_p)t}\braket{\omega_{p-q}(t)}{\omega_p(t)}}\\
    \times (c_{p-q,\mu}a_{p-q,\mu}(t))^\ast\big[c_{p\mu}a_{p\mu}(t)\chi_t(\Omega_{\qs}(p))+\\
    +\sigma\ee^{-\ii\mu2\jp t}c_{p\bar{\mu}}a_{p\bar{\mu}}(t)\chi_t(\Omega_{\qs}^{-\mu}(p))\big]+\order{(g K^{1/2}/v)^3}~,
\end{multline}
at the lowest order. It can be seen immediately that the two densities $\expval{\rho_\sigma(x)}_t$ and $\expval{\Pi_\sigma(x)}_t$ vanish unless the initial impurity state contains more than one momentum. This situation is physically consistent with the intuition that the impurity in a well-defined momentum state is equally distributed along the bath(s).
\par As the impurity is exchanged between the baths, these will become correlated. We will measure the amount of inter-bath correlation by computing the equal-time connected correlation functions
\begin{align}
    \expval{\rho_\sigma(x)\rho_{\bar{\sigma}}(y)}_t^c&\equiv \expval{\rho_\sigma(x)\rho_{\bar{\sigma}}(y)}_t-\expval{\rho_\sigma(x)}_t\expval{\rho_{\bar{\sigma}}(y)}_t\notag\\
     \expval{\Pi_\sigma(x)\Pi_{\bar{\sigma}}(y)}_t^c&\equiv \expval{\Pi_\sigma(x)\Pi_{\bar{\sigma}}(y)}_t-\expval{\Pi_\sigma(x)}_t\expval{\Pi_{\bar{\sigma}}(y)}_t~,
\end{align}
whose expressions are
\begin{subequations}
\begin{gather}
    \expval{\rho_\sigma(x)\rho_{\bar{\sigma}}(y)}_t=(K_\sigma K_{\bar{\sigma}})^{1/2}2\Re\bigg\{\tfrac{1}{L}\sum_{q,\bar{q}\neq0}{V_q V_{\bar{q}}}\times\notag\\ 
    \times\Big[\expval{b_\qs b_{\bar{q}\bar{\sigma}}\ee^{-\ii(q+\bar{q})X}}_t\ee^{\ii(qx+\bar{q}y)}+\notag\\
    +\expval{b_\qs^\dag b_{\bar{q}\bar{\sigma}}\ee^{\ii(q-\bar{q})X}}_t\ee^{-\ii(qx-\bar{q}y)}\Big]\bigg\}~,\\
    \expval{\Pi_\sigma(x)\Pi_{\bar{\sigma}}(y)}_t=(K_\sigma K_{\bar{\sigma}})^{-1/2}2\Re\bigg\{\tfrac{1}{L}\sum_{q,\bar{q}\neq0}{}V_q V_{\bar{q}}\times\notag\\
    \times\mathrm{sgn}(q\bar{q})\Big[\expval{b_\qs b_{\bar{q}\bar{\sigma}}\ee^{-\ii(q+\bar{q})X}}_t\ee^{\ii(qx+\bar{q}y)}+\notag\\
    +\expval{b_\qs^\dag b_{\bar{q}\bar{\sigma}}\ee^{\ii(q-\bar{q})X}}_t\ee^{-\ii(qx-\bar{q}y)}\Big]\bigg\}~.
\end{gather}
\end{subequations}
The relevant averages are given by
\begin{subequations}\label{eqs: expv bb}
\begin{align}
    &\expval{b_\qs b_{\bar{q}\bar{\sigma}}\ee^{-\ii(q+\bar{q})X}}_t=\notag\\
    &=\sum_{p,\bar{p},\mu}{}\delta_{\bar{p}-p,q+\bar{q}}\ee^{\ii(E_p-E_{\bar{p}})t}\braket{\omega_p(t)}{\omega_{\bar{p}}(t)}\notag\\
    &\times\big\{\tfrac{W_\qs W_{\bar{q}\bar{\sigma}}}{4L}\big[(c_{p\mu}a_{p\mu})^\ast c_{\bar{p}\mu}a_{\bar{p}\mu}\chi_t(\Omega_\qs(\bar{p}))\chi_t(\Omega_{\bar{q}\bar{\sigma}}(\bar{p}))+\notag\\
    &+\sigma\ee^{-\ii \mu2\jp t}(c_{p\mu}a_{p\mu})^\ast c_{\bar{p}\bar{\mu}}a_{\bar{p}\bar{\mu}}\big(\chi_t(\Omega_\qs^{-\mu}(\bar{p}))\chi_t(\Omega_{\bar{p}\bar{\sigma}}(\bar{p})+\notag\\
    &-\chi_t(\Omega_\qs(\bar{p}))\chi_t(\Omega^{-\mu}_{\bar{p}\bar{\sigma}}(\bar{p}))\big)\big]+\notag\\
    &+2(c_{p\mu}a_{p\mu})^\ast\big[c_{\bar{p}\mu}a_{\bar{p}\mu}A_{\qs,\bar{q}\bar{\sigma}}^{\mu\mu}(\bar{p},t)+c_{\bar{p}\bar{\mu}}a_{\bar{p}\bar{\mu}}A_{\qs,\bar{q}\bar{\sigma}}^{\bar{\mu}\mu}(\bar{p},t)\big]\big\}+\notag\\
    &+\order{(g K^{1/2}/v)^3}
    \intertext{and}
    &\expval{b_\qs^\dag b_{\bar{q}\bar{\sigma}}\ee^{\ii(q-\bar{q})X}}_t=\notag\\
    &=\tfrac{W_\qs W_{\bar{q}\bar{\sigma}}}{4L}\sum_{p,\bar{p},\mu}{}\delta_{\bar{p}-p,-q+\bar{q}}\ee^{\ii(E_p-E_{\bar{p}})t}\braket{\omega_p(t)}{\omega_{\bar{p}}(t)}\notag\\
    &\times\big[(c_{p\mu}a_{p\mu})^\ast c_{\bar{p}\mu}a_{\bar{p}\mu}\big(\chi_t^\ast(\Omega_\qs(p))\chi_t(\Omega_{\bar{q}\bar{\sigma}}(\bar{p}))+\notag\\
    &-\chi_t^\ast(\Omega^{+\mu}_\qs(p))\chi_t(\Omega^{+\mu}_{\bar{q}\bar{\sigma}}(\bar{p}))\big)\notag\\
    &+\sigma\ee^{-\ii \mu2\jp t}(c_{p\mu}a_{p\mu})^\ast c_{\bar{p}\bar{\mu}}a_{\bar{p}\bar{\mu}}\big(\chi_t^\ast(\Omega_\qs^{+\mu}(p))\chi_t(\Omega_{\bar{p}\bar{\sigma}}(\bar{p})+\notag\\
    &-\chi_t^\ast(\Omega_\qs(p))\chi_t(\Omega^{-\mu}_{\bar{p}\bar{\sigma}}(\bar{p}))\big)\big]+\order{(g K^{1/2}/v)^3}~.
\end{align}
\end{subequations}
We point out that in the limit $\jp\to0$ the baths become decoupled, therefore $\eval{\expval{\rho_\sigma(x)\rho_{\bar{\sigma}}(y)}_t}_{\jp=0}=\eval{\expval{\rho_\sigma(x)}_t}_{\jp=0}\eval{\expval{\rho_{\bar{\sigma}}(y)}_t}_{\jp=0}$ (and analogously for momentum). However, this property is not satisfied by the perturbative expressions given above. This violation comes about because in our perturbative solution we broke up the interaction in two terms, treating one exactly (intra-band processes) while expanding in the second one (inter-band processes). When $\jp= 0$, this separation is not justified, and it actually gives rise to a spurious (yet small) inter-bath correlation. Therefore, we choose to subtract the $\jp=0$ values of the correlation functions.
\section{Numerical results}\label{sec:numerical results}
In this section, we show numerical results for the evolution of the impurity probability density, the bath density and bath momentum density when the impurity is initialized in a wave packet within the $\uparrow$ bath ($\sigma=1$). The method we employed allows for virtually arbitrary wave packets, compatibly with the low-momentum conditions for the validity of the long-wavelength model. We choose a Gaussian profile in momentum space:
\begin{equation}\label{eq: wave packet}
    c_{p\mu}=2^{-1/2}N(p_0,\delta p)\ee^{-\tfrac{(p-p_0)^2}{4\delta p^2}-\ii x_0 p}~,
\end{equation}
where $p_0$ is the average momentum, $\delta p$ is the width of the distribution, $x_0$ is the average initial position and $N(p_0,\delta p)$ is chosen to ensure that $\sum_{p\mu}{\abs{c_{p\mu}}^2}=1$. Notice that the above momentum profile corresponds to a wave function that is factorized between space and bath indices. This choice is not required by our perturbative method, but it simplifies the analysis of the results. We work in a finite-size system of length $L$, with periodic boundary conditions. Momenta are then quantized according to $p_n=2\pi/L\cdot n$, $n\in\mathbb{Z}$, and we take a wave packet composed of $N_p$ momenta, distributed symmetrically around $p_0$\footnote{More specifically, we define $n_{p_0}\in\mathbb{Z}$ such that $p_0=2\pi/L\,n_{p_0}$, then we take $p_n=2\pi/L(n_{p_0}+n)$ where $n=-N_p/2+1,\dots, N_p/2$ when $N_p$ is even, and $n=-(N_p-1)/2,\dots, (N_p-1)/2$ when $N_p$ is odd.}. We usually take $N_p=32$ or $N_p=64$ momenta, and $x_0=L/2$. 
\subsection{Impurity oscillations}
\begin{figure}
    \centering
    \subfloat[\label{fig:impurity oscillations J}]{\includegraphics[width=0.9\linewidth]{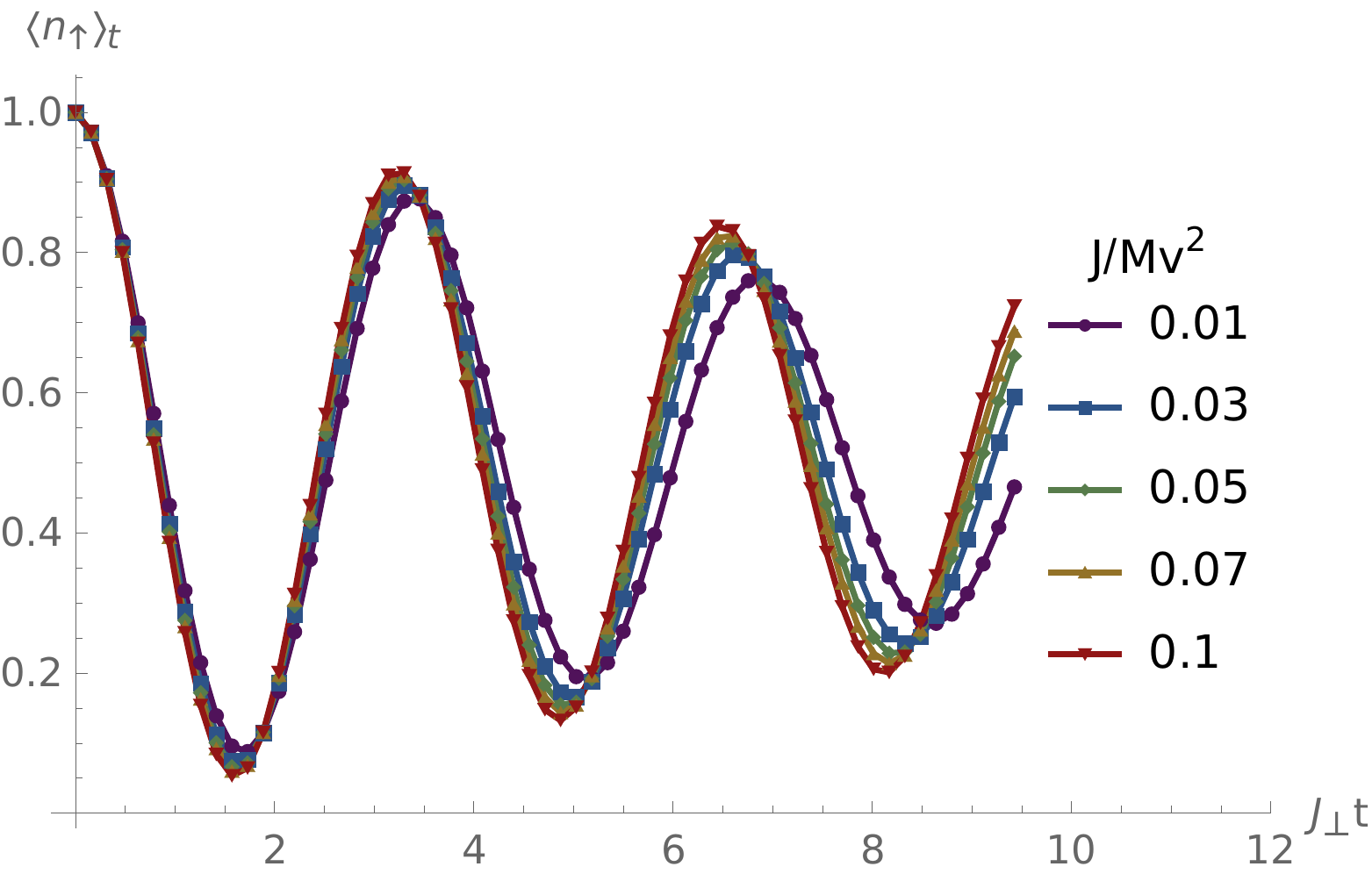}}\\
    \subfloat[\label{fig:impurity oscillations g}]{\includegraphics[width=0.9\linewidth]{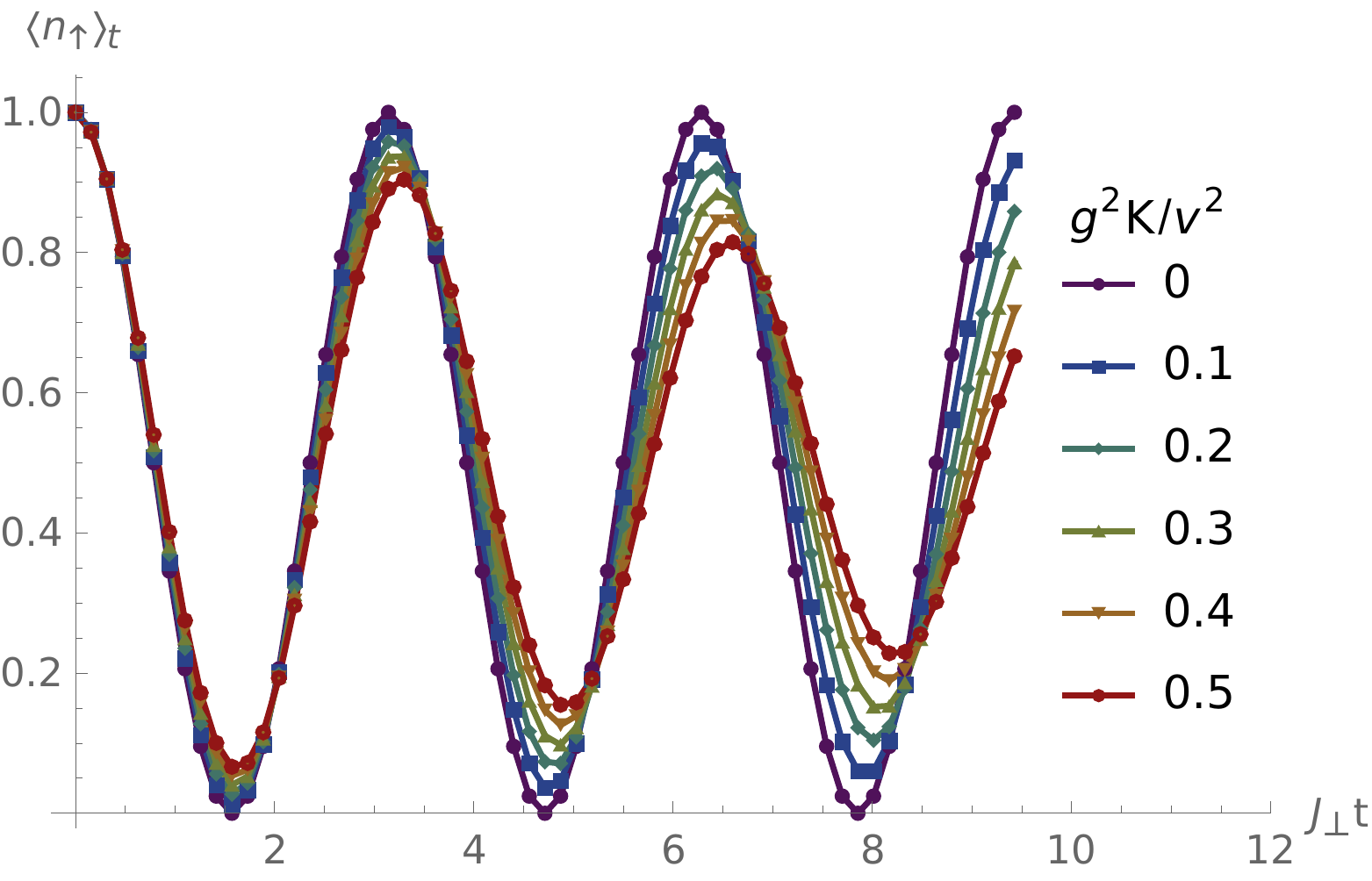}}
    \caption{Time evolution of $\expval{n_\uparrow}$ for the wave packet \eq~\eqref{eq: wave packet}. In plot (a) the coupling is kept fixed to $g^2K=0.5v^2$, while $\jp$ is varied, while in plot (b) we vary theg coupling while keeping a constant $\jp=0.05 Mv^2$. Notice that in (a) the time is expressed in units of $\jp^{-1}$, which is different for different plots.}
    \label{fig:impurity oscillations}
\end{figure}
In \figs~\ref{fig:impurity oscillations J} and \ref{fig:impurity oscillations g} we show the time evolution of $\expval{n_\uparrow}_t$, namely the probability of finding the impurity in bath $\uparrow$ at time $t$ (the probability for the other bath is simply $\expval{n_\downarrow}_t=1-\expval{n_\uparrow}_t$). We can observe that the interaction with the baths has two effects. First, the amplitude of the oscillations around the average value $1/2$ becomes a decreasing function of time. This decay is more pronounced for larger coupling (\fig~\ref{fig:impurity oscillations g}) and larger $\jp$ (\fig~\ref{fig:impurity oscillations J}, once the time is measured in $\jp$-independent units). Second, the frequency of the oscillations is decreased, by an amount that is larger for increasing coupling (\fig~\ref{fig:impurity oscillations g}) and decreasing $\jp$ (\fig~\ref{fig:impurity oscillations J}).
\par We can have an analytic insight on these observations by examining the expression \eqref{eq: nsigma} for $\expval{n_\uparrow}_t$. In particular, if we keep only the first term in the bath states overlap \eqref{eq: mu bar mu overlap}, which gives the leading contribution, and we use the asymptotic relations \eqs~\eqref{eq: asymptotics apmu} for $a_{p\mu}$, we find
\begin{equation}\label{eq: approx impurity oscillations}
    \expval{n_\sigma}_t\sim\tfrac{1}{2}+\sigma\Re\sum_{p}{\ee^{z_{pe}^\ast+z_{po}}c_{pe}^\ast c_{po}\ee^{-2\ii\tilde{J}_{\perp,p} t -2\gamma_p t}}~,
\end{equation}
where $\tilde{J}_{\perp,p}\equiv \jp +(\Delta\lambda_{po}-\Delta\lambda_{pe})/2$ is a renormalized inter-bath hopping amplitude (see \app~\ref{app: normalization factors}). From the above expression we can see that $\expval{n_\sigma}_t-1/2$ is  a superposition of damped oscillatory functions, one for each momentum in the wave packet. For comparison, in the non-interacting case we would have
\begin{equation}
    \expval{n_\sigma}_t^{(0)}=\tfrac{1}{2}+\sigma\Re\sum_{p}{c_{pe}^\ast c_{po}\ee^{-2\ii\jp t}}~,
\end{equation}
which describes the impurity periodically hopping from one bath to the other. In the weak coupling regime we are examining, the interaction with the baths decreases the hopping frequency ($\tilde{J}_{\perp,p}<\jp$), while the amplitude of the oscillations decreases exponentially with a decay time of $1/(2\gamma_p)$. While our solution ceases to be accurate beyond this decay time, it hints at $\expval{n_\sigma}_t\to1/2$ eventually. This limit is what we would expect intuitively as a result of dephasing and dissipation. 
\subsection{Impurity momentum}\label{subsec: impurity momentum}
\begin{figure}
    \centering
    \subfloat[\label{fig: impurity momentum g}]{\includegraphics[width=.9\linewidth]{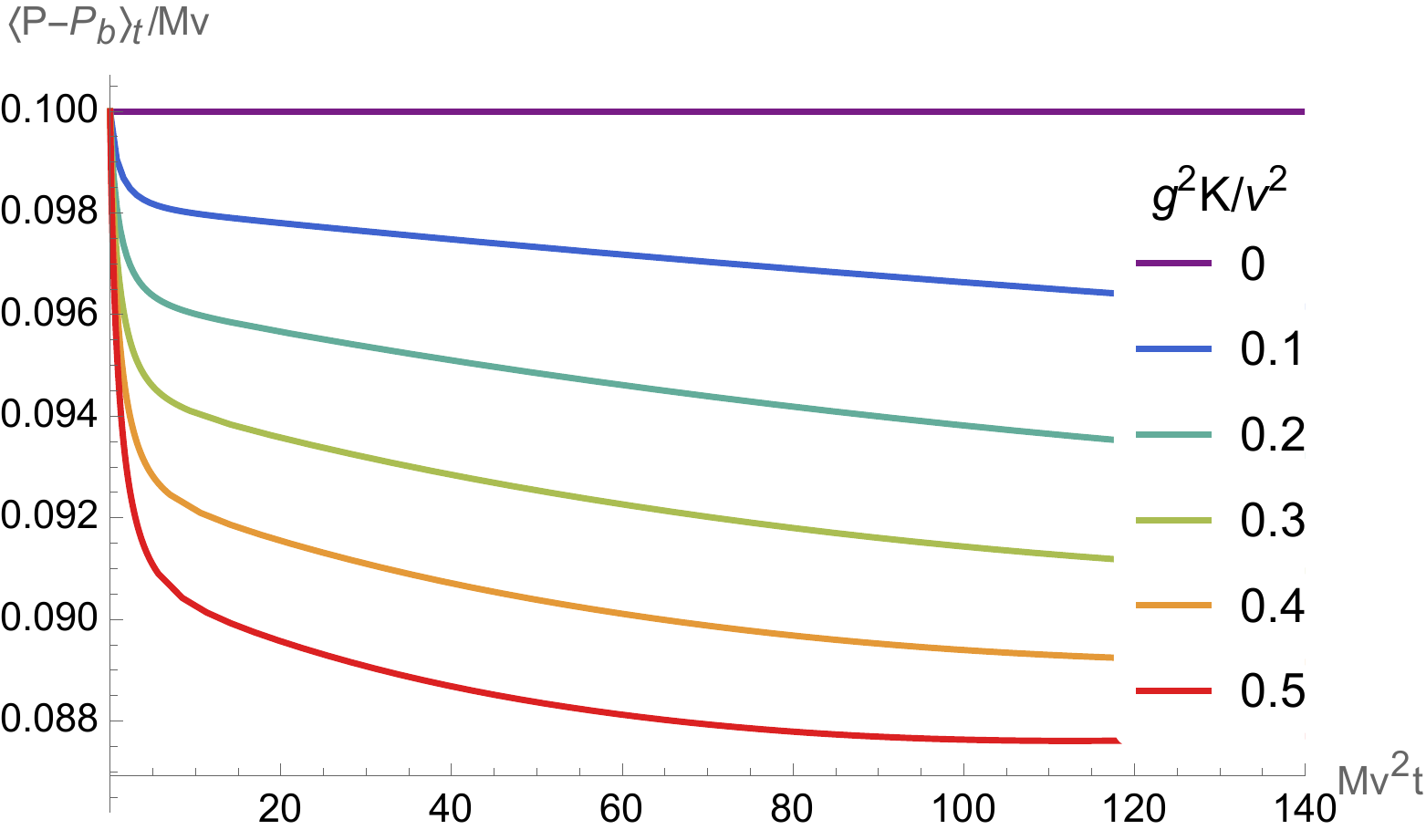}}\\
    \subfloat[\label{fig: impurity momentum p}]{\includegraphics[width=.9\linewidth]{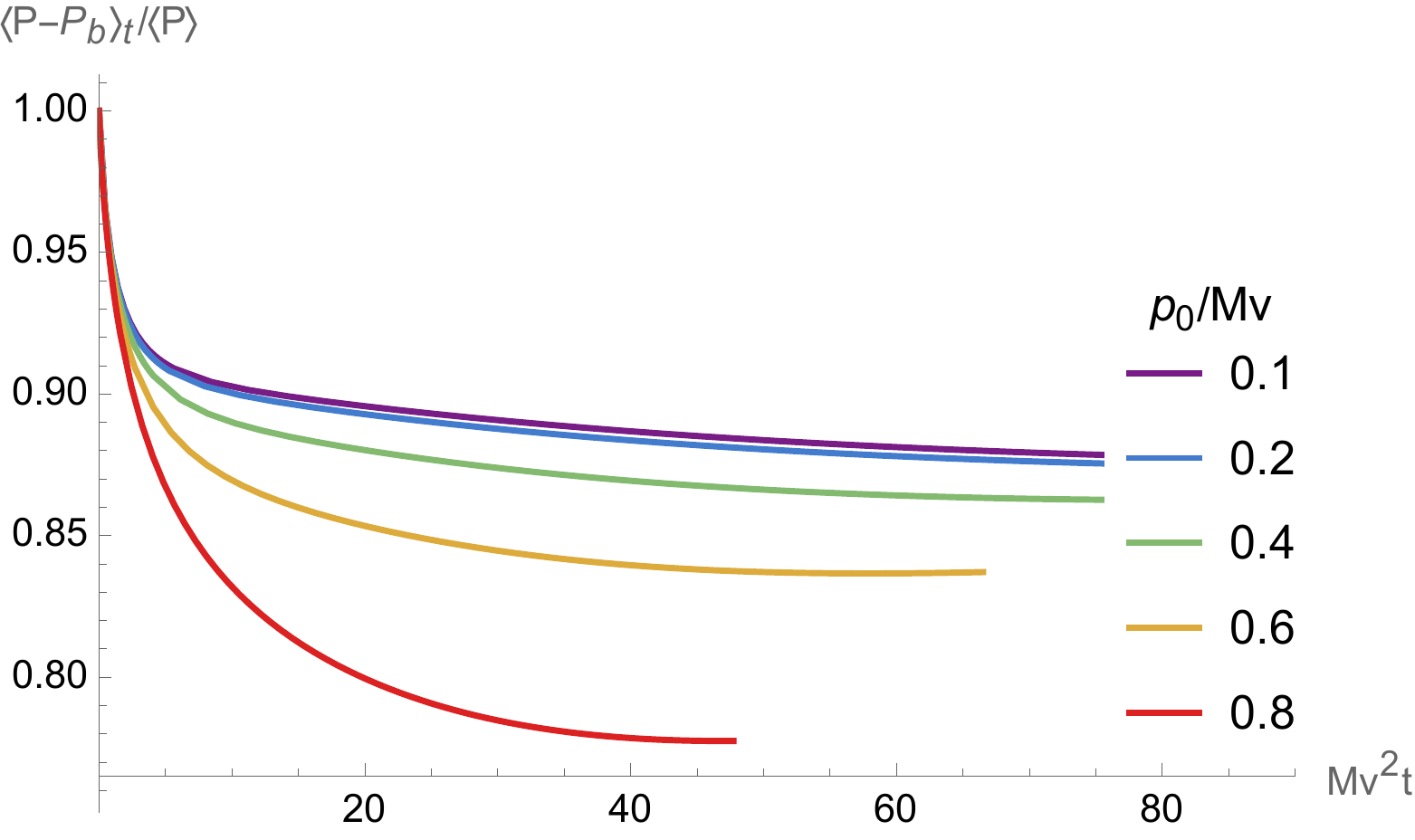}}\\
    \caption{Time evolution of the impurity momentum for the initial state $\ket{p_0\uparrow}_d$, and symmetric baths. Plot (a) has $p_0=0.1Mv$, $\jp=0.05Mv^2$ and shows the effect of increasing the coupling. Plot (b) is for $g^2K=0.5v^2$, $\jp=0.05Mv^2$ and shows the relative decrease of momentum $\expval{P_d}_t/p_0$ for increasing $p_0$. In the latter plot, the two curves at higher momentum are limited within the time domain in which where the loss of the state norm does not give rise to spurious effects.}
    \label{figs: impurity momentum 1}
\end{figure}
The time evolution of the impurity momentum is displayed in \fig~\ref{figs: impurity momentum 1}, as function of the various parameters of the model. In all plots, the impurity is initialized in the $\uparrow$ bath with a definite momentum\footnote{As we did with $\expval{n_\sigma}_t$, writing it as $1/2+\sigma\expval{\sigma_3}_t/2$, we take $\expval{P}_t=\expval{P}_{t=0}=p_0$, i.e. we ignore the effect of the loss of normalization on the conserved total momentum.} $p_0$. We observe that the momentum decay can be divided in two phases: an initial abrupt drop followed by a decrease at a milder rate. Figure \ref{fig: impurity momentum J} shows the effect of the inter-bath hopping on the momentum. We can see that the initial rapid decrease is basically unaffected by $\jp$, while the subsequent decay is faster for larger hopping. This difference suggests that the two phases of the decay originate from two different processes. 
We interpret the two phases of momentum decay in the following way: The first fast-decreasing region is caused by the baths relaxing to the injection of the impurity, while in the second phase the impurity momentum is carried away by the phonons generated from the deexcitation of the odd mode.
\par Indeed, this first phase occurs on a timescale that appears to be independent of $\jp$, and well before the impurity starts oscillating into the $\downarrow$ bath\footnote{This transient in the momentum evolution may be linked to a dimensional crossover, which is also shown by the Green's function \cite{PhysRevB.103.094310}.}. Initially, the first and last terms of \eq~\eqref{eq: bath momentum} contribute equally to the bath momentum. After the initial transient, the background contribution saturates to a constant value, while the first term of \eq~\eqref{eq: bath momentum} continues to grow, albeit at a slower rate (except for the unphysical decrease at late times). The timescale of this growth is shorter the larger is $\jp$, which is what we expect from the property that the odd mode decay constant $2\gamma_p$ is an increasing function of $\jp$. At the lowest inter-bath hopping we considered, $\jp=0.01Mv^2$, the timescale is so long that the contribution to $\expval{P_b}_t$ from the odd mode decay appears to converge to a value slightly below that of the background.
\par The dependence of the impurity momentum on the coupling is shown in \fig~\ref{fig: impurity momentum g}. As expected, the decrease is more marked for stronger coupling, while the two-phase structure is kept unaltered. A larger $gK^{1/2}/v$ increases the fraction of momentum that is lost in the initial transient, but not the timescale in  which it occurs. 
\par The following \fig~\ref{fig: impurity momentum p} shows the slowdown of the impurity momentum at increasing values of the initial momentum $p_0$, as a fraction of the latter. We can recognize the initial transient and the subsequent slower decay, with the former following a common shape for all momenta at small times. After the transient, we see that for increasing $p_0$ the relative amount of momentum that is transferred to the baths becomes larger, which also translates to a larger absolute decrease of the impurity momentum. This more pronounced decrease for faster impurities may be also traced back to the increase of the decay rate $2\gamma_{p_0}$ with $p_0$, which implies that the production of phonons caused by the deexcitation of the odd mode occurs earlier and more rapidly. At smaller momenta $\abs{p}\lesssim0.2Mv$ the ratio $\expval{P_d}_t/p_0$ tends to a common profile, independent of $p_0$. This property is explained by noticing that $R_1^\sigma(\ce;p_0)$ is linear in $p_0$ for small momentum $\abs{p_0}\ll Mv_\sigma$, hence for a single momentum component $\expval{P_b}(t)=\upsilon(t)p_0/(Mv)+\order{p_0^2/(Mv)^2}$, with $\upsilon(t)$ independent of $p_0$.
\subsection{Impurity density evolution}
\begin{figure*}
    \centering
    \subfloat[$g^2K=0$\label{fig: free impurity density}]{\includegraphics[width=0.4\linewidth]{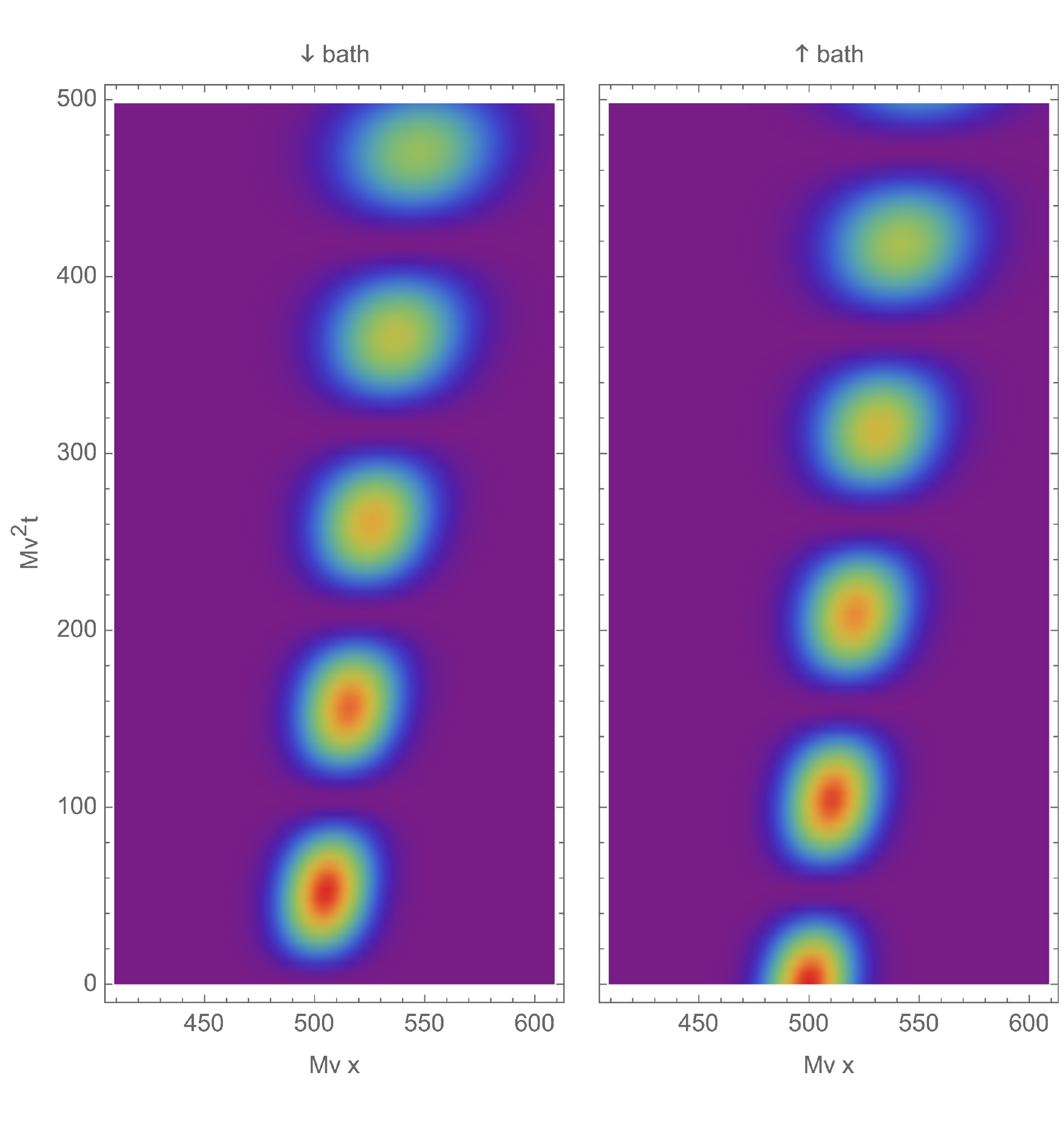}}
    \subfloat[$g^2K=0.5v^2$\label{fig: int impurity density}]{\includegraphics[width=0.4\linewidth]{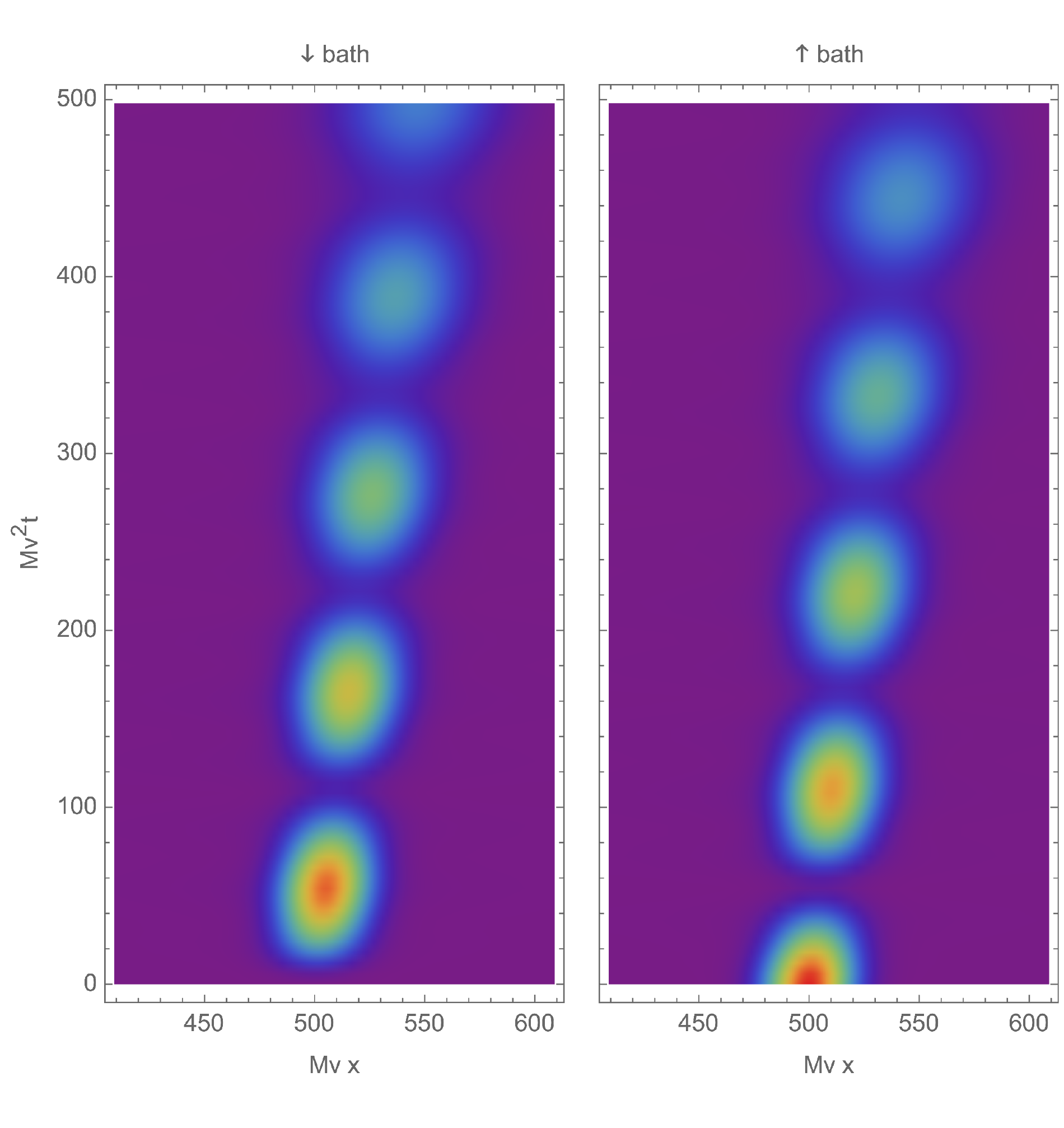}}
    \caption{Comparison of the time evolution of probability density for (a) the free and (b) the interacting impurity with symmetric baths, $g^2K=0.5v^2$ and $\jp=0.03Mv^2$. In both figures the wave packet is initialized with the  Gaussian distribution \eq~\eqref{eq: wave packet} of $N_p=64$ momenta around $p_0=0.1Mv$, with standard deviation $\delta p=0.04Mv$. The total length of the system is $L=1000\,(Mv)^{-1}$, but only its central part is depicted. The normalization of the color scale for the two figures is the same.}
    \label{figs: impurity density}
\end{figure*}

The typical time evolution of the probability density of the impurity is reported in \fig~\ref{fig: int impurity density}, compared with the free evolution \ref{fig: free impurity density}. Varying $\jp$ and $g_\sigma$ produces analogous results. The motion is qualitatively similar to the free one, namely the whole wave packet moves to the right at a speed slightly less than $p_0/M$ while oscillating with a frequency renormalized by the interactions as compared to the non-interacting value $\tilde{J}_{\perp p}$ [see \eq~\eqref{eq: renormalized Jp}]. In the perturbative regime the absolute effect of the baths on the impurity momentum and frequency of oscillation is typically small. 
\par A visible difference between the free and interacting time evolution of the wave packet is the larger spread of each peak in the ''time direction'', which means that the impurity never really leaves any of the baths for the other\footnote{This is in accord with the observation that in the interacting case $\expval{n_\sigma}_t$ never goes back to $0$ or $1$ for $t>0$.}. This phenomenon can be traced back to the inhomogeneous dephasing associated to the momentum dependence of the renormalized inter-bath hopping $\tilde{J}_{\perp p}$. The increased spread in the time direction can be interpreted as the initial evidence of the decay of the odd mode, which implies that eventually any oscillation should disappear.
\par There is also a second difference between the interacting and the free density evolution, namely the enhanced rate of decrease of the height of the wave packet. This decrease arises partly from the progressive loss of the norm of the state, and partly from the rapid flowing of probability towards the the tails of the distribution. While this could signal an actual tendency of the impurity to spread, we suspect that this behavior is an artifact induced by the perturbative method. On the other hand, the shape of the packet around its center remains Gaussian with good approximation.
\subsection{Bath density evolution}\label{subsec: density evolution}
\begin{figure}
    \centering
    \subfloat[][$\jp=0.1Mv^2$\label{fig:density J=0.1}]{\includegraphics[width=.6\linewidth]{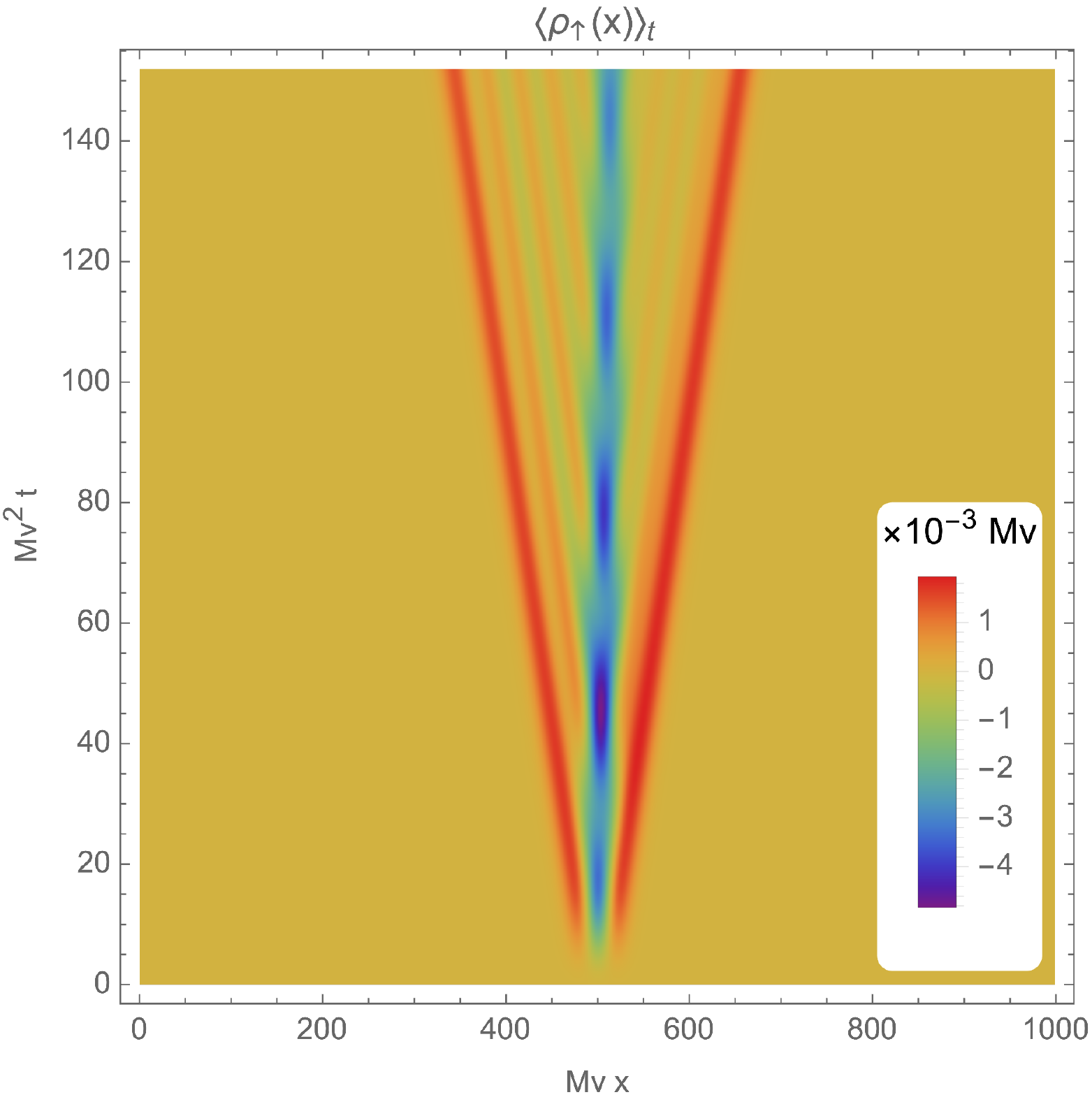}}\\
    \subfloat[][$\jp=0.03Mv^2$\label{fig:density J=0.03}]{\includegraphics[width=.6\linewidth]{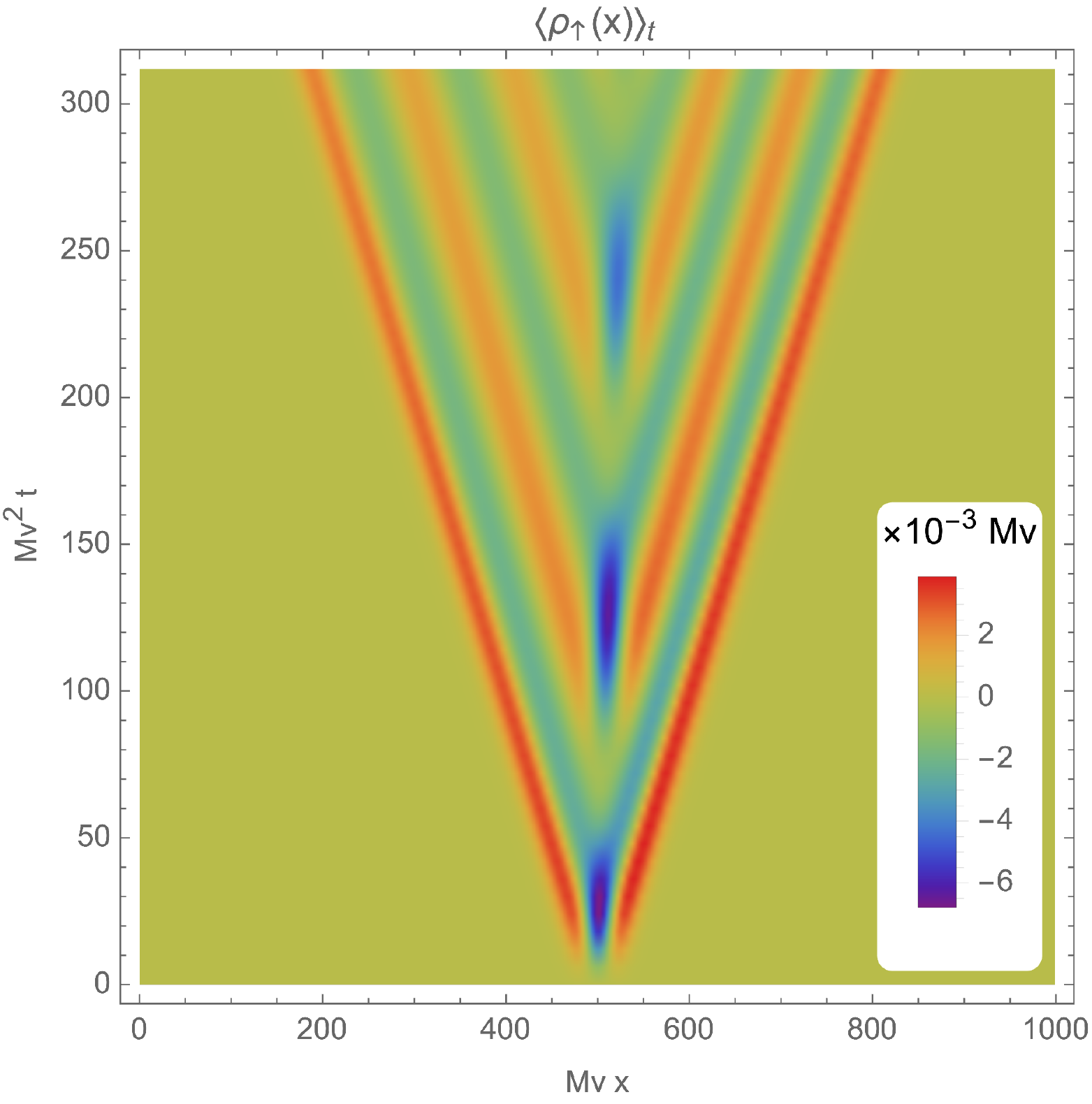}}\\
    \subfloat[][$\jp=0.01Mv^2$\label{fig:density J=0.01}]{\includegraphics[width=.6\linewidth]{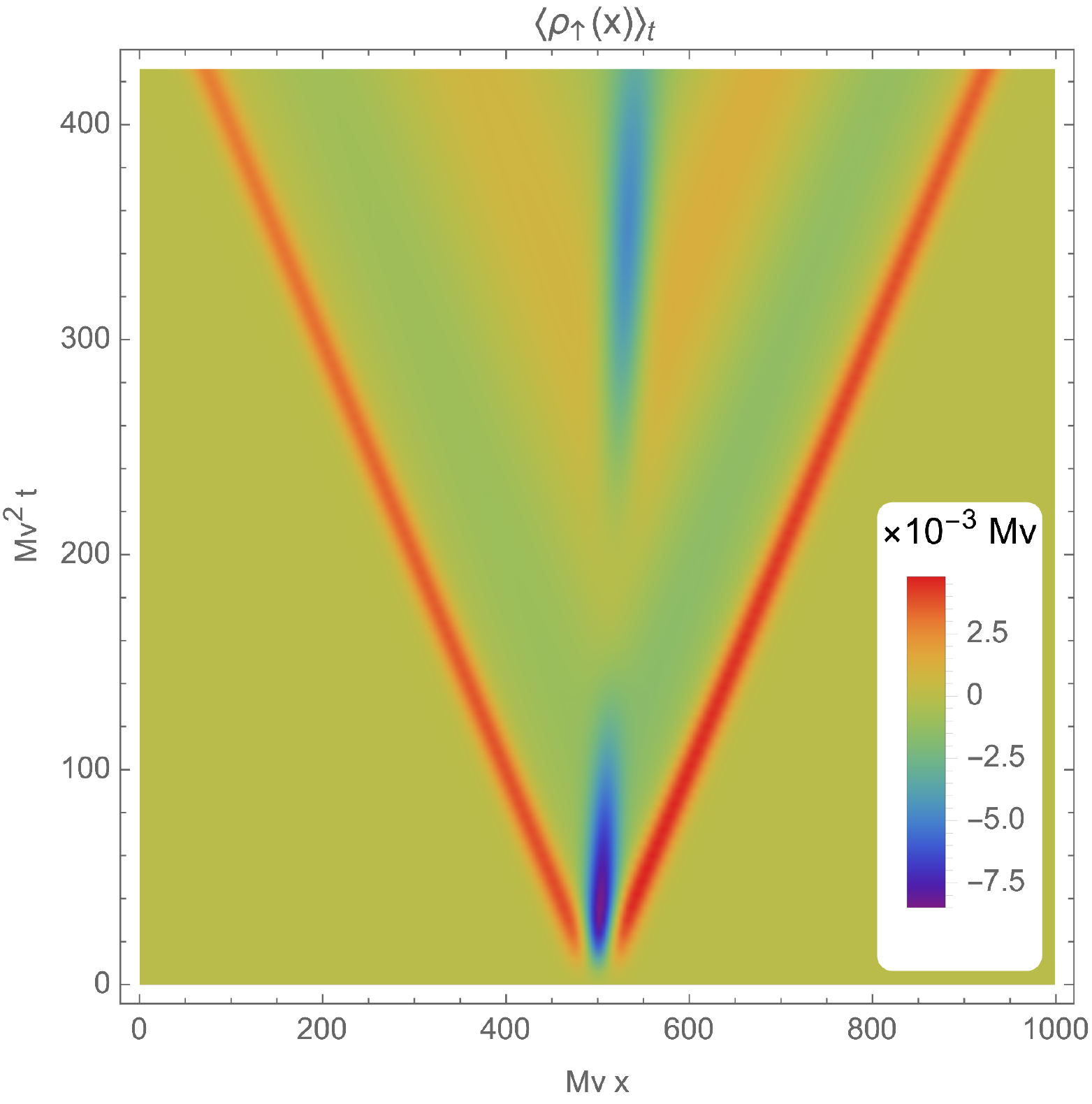}}
    \caption{Dynamics of the density perturbation of bath $\uparrow$ caused by the introduction of an impurity wave packet. The parameters are $g^2K=0.5v^2$, $K=2$ for both baths (we take $M=1$, $v=1$). The cutoffs are $Mv\alpha=0.5$ and $\Lambda=10Mv^2$. Plots (a) to (c) differ in the inter-bath hopping $\jp$.}
    \label{fig:density plots density}
\end{figure}
We have already presented an overview of the evolution of the density of the baths in \ref{sec:overview of the results}. In the following paragraphs, we will provide more details.
\par In \fig~\ref{fig:density plots density} we show the bath density evolution for increasing values of $\jp$, keeping all other parameters constant. The baths have identical properties. The initial impurity wave packet starts in the $\uparrow$ bath, with an average position $x_0=L/2$ and a momentum distribution of $N_p=64$ momenta that are centered around $p_0=0.1Mv$, with a width of $\delta p=0.04Mv$ (correspondingly, the spatial width of the wave packet is about\footnote{The relation $\delta x= 1/(2\delta p)$ is only approximately valid for our discrete Gaussian wave packet.} $\delta x\approx 1/(2\delta p)= 12.5 (M v)^{-1}$ ). These figures show that the density evolution has a common structure: a depletion that follows the impurity (notice how it is inclined to the right, owing to the nonzero average momentum $p_0$), flanked by two wave fronts that "radiate" in opposite directions at the speed of sound (which is $v=1$ in our units). As times goes on, the region between these features (the "light-cone" $\abs{x}\le vt$) becomes filled with density ripples, which can be interpreted as a manifestation of the emission of real phonons caused by the deexcitation from the odd to the even mode. This interpretation follows from the observation that these ripples ultimately come from the $b^\dag_\qs\ket{\omega_p(t)}$ terms (i.e. from the the last term of \eq~\eqref{eq: expv b}), which we identified as representing spontaneous emission. On the other hand, the depletion and the wave front are all contained in the coherent states $\ket{\omega_p(t)}$. More physically, the ripples have a wavelength that diminishes with increasing $\jp$, corresponding to the wave vectors $q_\pm(p_0)$ that solve $\Omega_{q_\pm \sigma}(p_0)=2\jp$. The negative and positive solutions of the latter equation apply to backward and forward emission, respectively, and are Doppler-shifted $q_+(p_0)>\abs{q_-(p_0)}$ because of the motion of the source (the impurity). These observations support our view of the perturbative solution as being decomposed into a "bath relaxation" (depletion and wave fronts) and spontaneous emission.
\par As $\jp$ becomes smaller, the ripples become higher and of longer wavelength, and the whole density profile becomes wider. At the same time, the depth of the minimum oscillates more and more evidently, as the impurity oscillation becomes slower. This behavior is consisted with the single-bath situation that is recovered for $\jp\to0$, in which the impurity remains in its initial bath.
\begin{figure}
    \centering
    \subfloat[\label{fig: bath comparison - density}]{\includegraphics[width=0.9\linewidth]{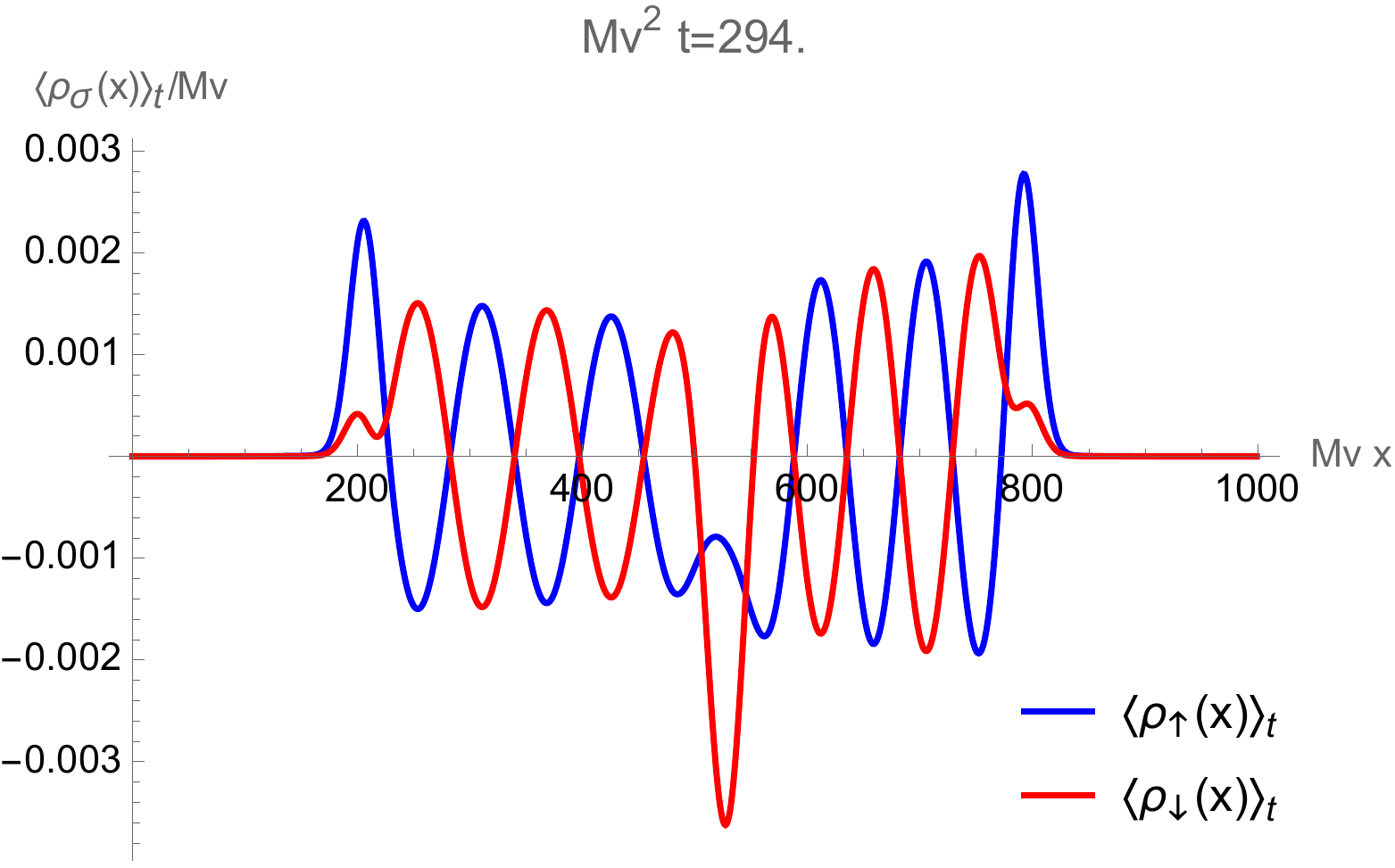}}\\
    \subfloat[\label{fig: bath density width comparison}]{\includegraphics[width=0.9\linewidth]{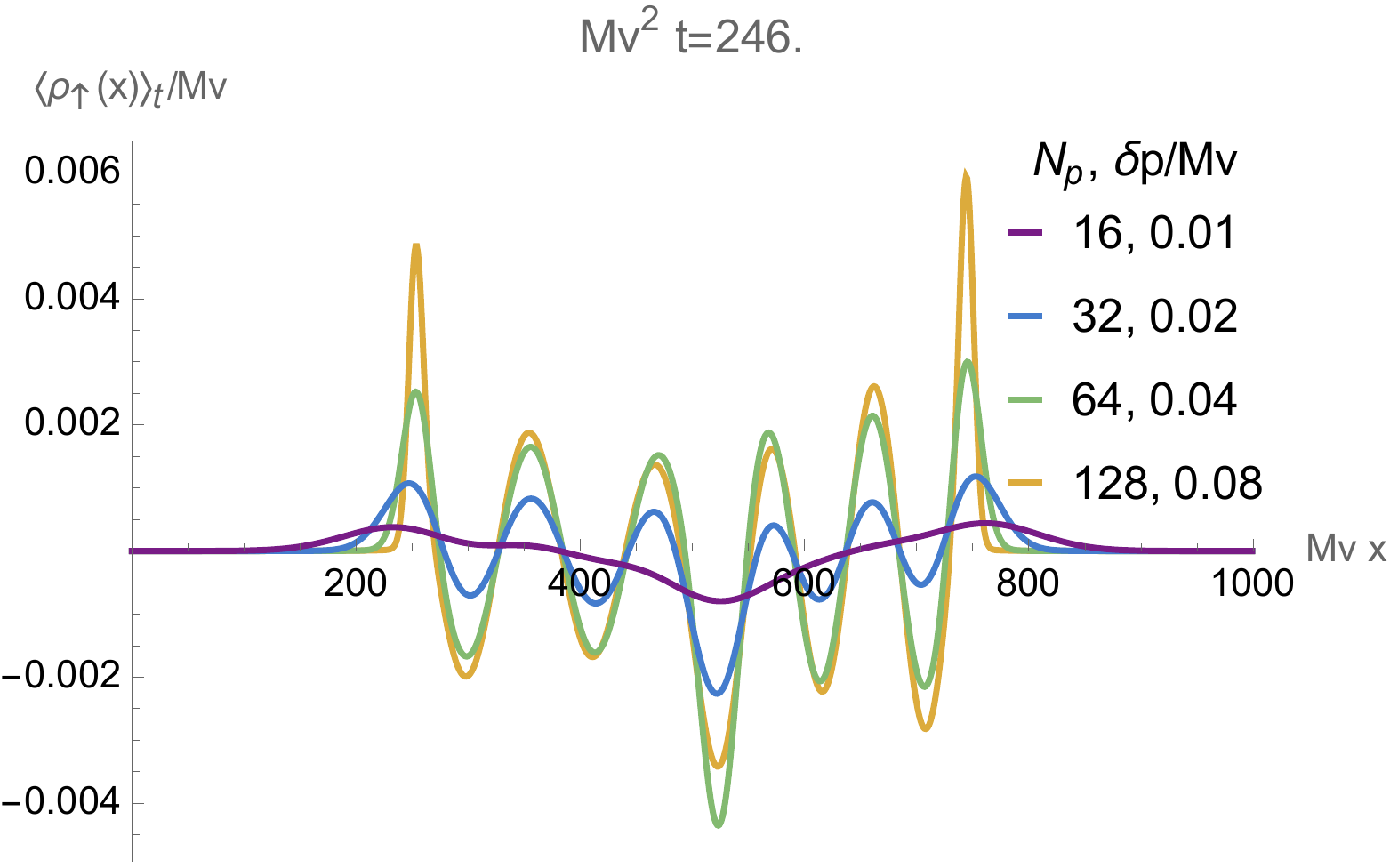}}
    \caption{(a) Density profile in both baths at a given instant of time. (b) Density profile in bath $\uparrow$ for various widths of the wave packet. The parameters are the same as \fig~\ref{fig:density plots density}.}
    \label{figs: bath comparisons}
\end{figure}
\par The bath are identical in their properties and initial state, the only source of asymmetry is the initial state of the impurity. The latter causes the $\downarrow$ bath to interact with the impurity a little later than the $\uparrow$ bath (roughly after a fraction of the bare oscillation period $2\pi/\jp$). Because of this, the density profile of the $\downarrow$ bath is qualitatively similar to the $\uparrow$ one, but it is "delayed" by the time it takes the impurity to change its initial bath. This characteristic is shown in \fig~\ref{fig: bath comparison - density}. At large $\jp$, the impurity is rapidly exchanged, and therefore the baths' density profiles are "synchronized", almost coinciding with each other, and with very little relative lag. As $\jp$ is decreased, the oscillations in the depletion depth become wider and wider and thus can be clearly seen to be out of phase, while the wave fronts are always in phase but show a visible lag. Moreover, the wave fronts in the $\downarrow$ bath decrease in height with respect to their $\uparrow$ counterparts when $\jp$ assumes smaller values. The ripples are rigorously out of phase. In fact, we have already remarked that the ripples come from the second term in the square brackets in \eq~\eqref{eq: expv b}, which is multiplied by $\sigma=\pm1$.
\par We point out that the depletion and wave fronts have a distinct shape from that of the ripples. This difference is also evident from the observation that the wavelength of the ripples depends on $\jp$, whereas the width of the other two does not. Indeed, it is not hard to guess that depletion and wave fronts are essentially ”images” of the Gaussian profile of the impurity wave packet, albeit slightly distorted. The shape of the ripples is instead more or less sinusoidal, and this suggests that their behavior is governed by the intrinsic dynamics of the baths, rather than by the details of the shape of the wave packet.
\par These observations are substantiated by investigating the effect of the initial wave packet width on the bath density profile. This analysis is shown in \fig~\ref{fig: bath density width comparison}, in which the density profile of the $\uparrow$ bath at a given time is compared for decreasing widths $\delta x \approx 1/(2\delta p)$, at fixed inter-bath hopping. To keep the wave packet shape unaltered while decreasing its standard deviation, we have kept the ratio $\delta p/N_p$ constant. The main effect of a smaller width is that the magnitude of the density fluctuations is increased. This effect is easily understood for the wave fronts and the central depletion if we trust the observation that they have the same shape as the wave packet, as a narrower normalized Gaussian is also taller. On the other hand, the influence on the height of the ripples can be partially understood by observing that wave packets with maximal width (equal to the length of the system) are made of only one momentum, and therefore [compare with \eqs~\eqref{eq: expv densities}] produce no density perturbation in the baths. Then, a continuity argument suggests that wider wave packets should give rise to smaller density fluctuations, including the ripples.
\par We also notice that the dependence of the density fluctuation amplitude is more prominent for the ripples than for the through and wave fronts, to the point that a wide enough wave packet is able to effectively suppress the ripples altogether. Indeed, from \figs~\ref{fig: bath density width comparison} we can see that at $\jp=0.03Mv^2$ the density perturbation for the wave packet $\delta p=0.01Mv$ lacks the ripples. We also verified that at $\jp=0.1Mv^2$, a standard deviation of $\delta p=0.02Mv$ is sufficient to cancel the ripples, whereas for $\jp=0.01Mv^2$ the ripples are present even for $\delta p=0.01Mv$. On the other hand, the wavelength of the ripples is not affected by the wave packet width, whereas the central dip and the wave fronts change their shape, becoming narrower and more peaked as $\delta p$ is increased. This is in accord with our observation that they should be shifted images of the initial wave packet.
\par In \app~\ref{app: density explanation} we show that the behavior of the ripples amplitude (i.e. their suppression for sufficiently large $\jp$ or small $\delta x$) can be explained as an interference effect, in which there is a cancellation between opposite-sign terms at different times in the past. This destructive interference occurs only if the initial wave packet is wide enough, according to the relation 
\begin{equation}\label{eq: critical delta x}
\begin{gathered}
\delta x \gtrsim \delta x_{c,\sigma}^{\pm}(p_{0},\jp),\\ \delta x_{c, \sigma}^{\pm}(p_{0},\jp)\equiv \left(v_{\sigma}\pm\tfrac{p_{0}}{M}\right)\frac{\pi}{2\tilde{J}_{\perp,p_{0}}}~.
\end{gathered}
\end{equation}
Equivalently, the ripples are suppressed if the momentum distribution is sufficiently narrow:
\begin{equation}\label{eq: critical delta p}
\begin{gathered}
\delta p \lesssim \delta p_{c,\sigma}^{\pm}(p_{0},\jp),\\
\delta p_{c, \sigma}^{\pm}(p_{0},\jp)\equiv \frac{\tilde{J}_{\perp,p_{0}}}{\pi \left(v_{\sigma}\pm\tfrac{p_{0}}{M}\right)}~.
\end{gathered}
\end{equation}
Ignoring for a moment the dependence on $p_{0}$, we see that $\delta p_{c, \sigma}^{\pm}(p_{0},\jp)$ is about $0.03Mv$ for $\jp=0.1Mv^2$ and $0.01Mv$ for $\jp=0.03Mv^2$. So, we see that the above inequalities are indeed satisfied in the above-mentioned cases in which there were no ripples. The positive or negative sign in \eqs~\eqref{eq: critical delta x} and \eqref{eq: critical delta p} refer to the ripples emitted backward or forward, respectively. This directional dependence allows us to justify the asymmetric height of the ripples, as the critical width for backward emission is larger than the one for the forward emission, which implies that the cancellation effect is less effective in the backward direction, resulting in the larger ripple amplitude that we observed.
\begin{figure}
    \centering
    \subfloat[][]{\includegraphics[width=.9\linewidth]{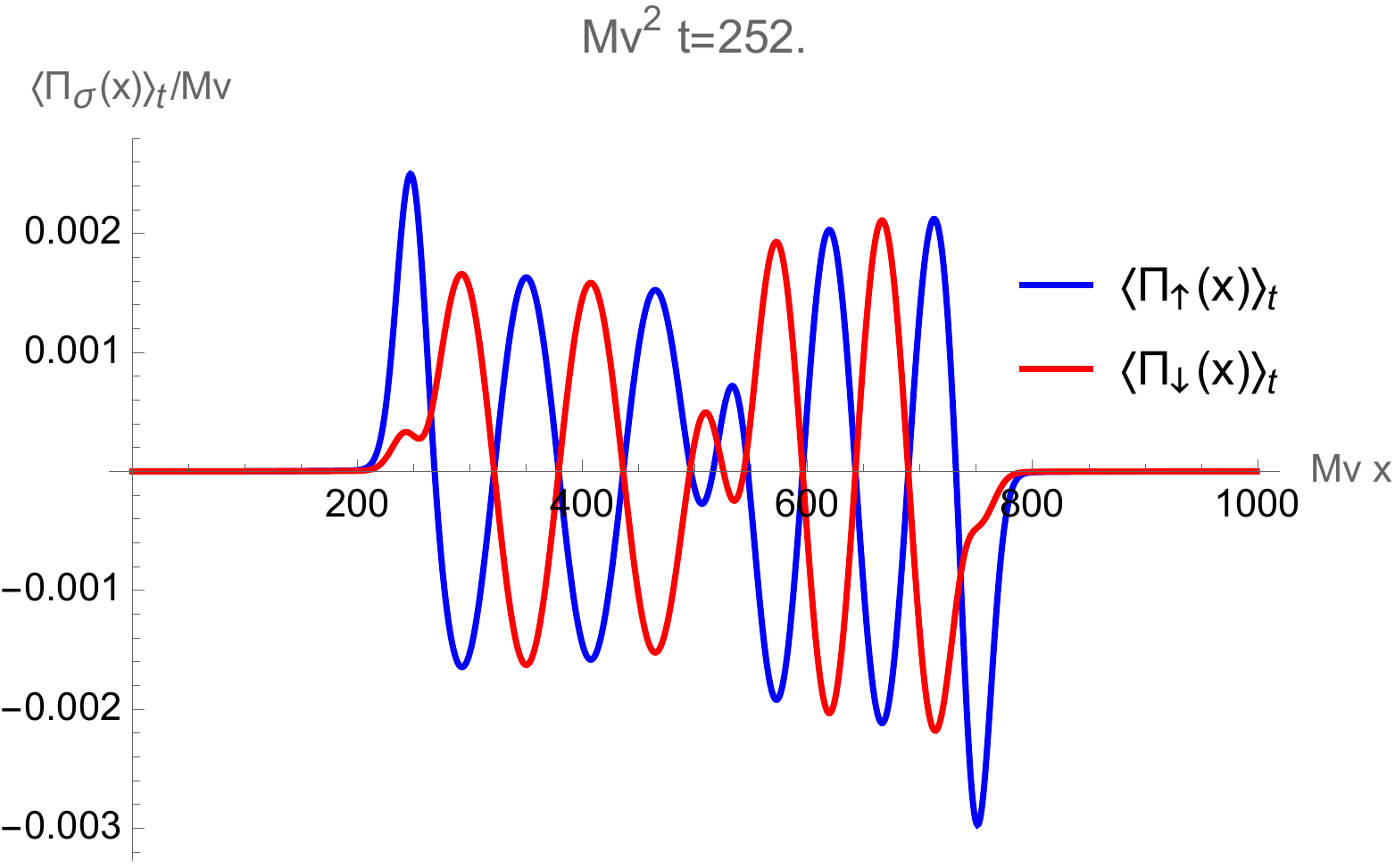}}\\
    \subfloat[][]{\includegraphics[width=.9\linewidth]{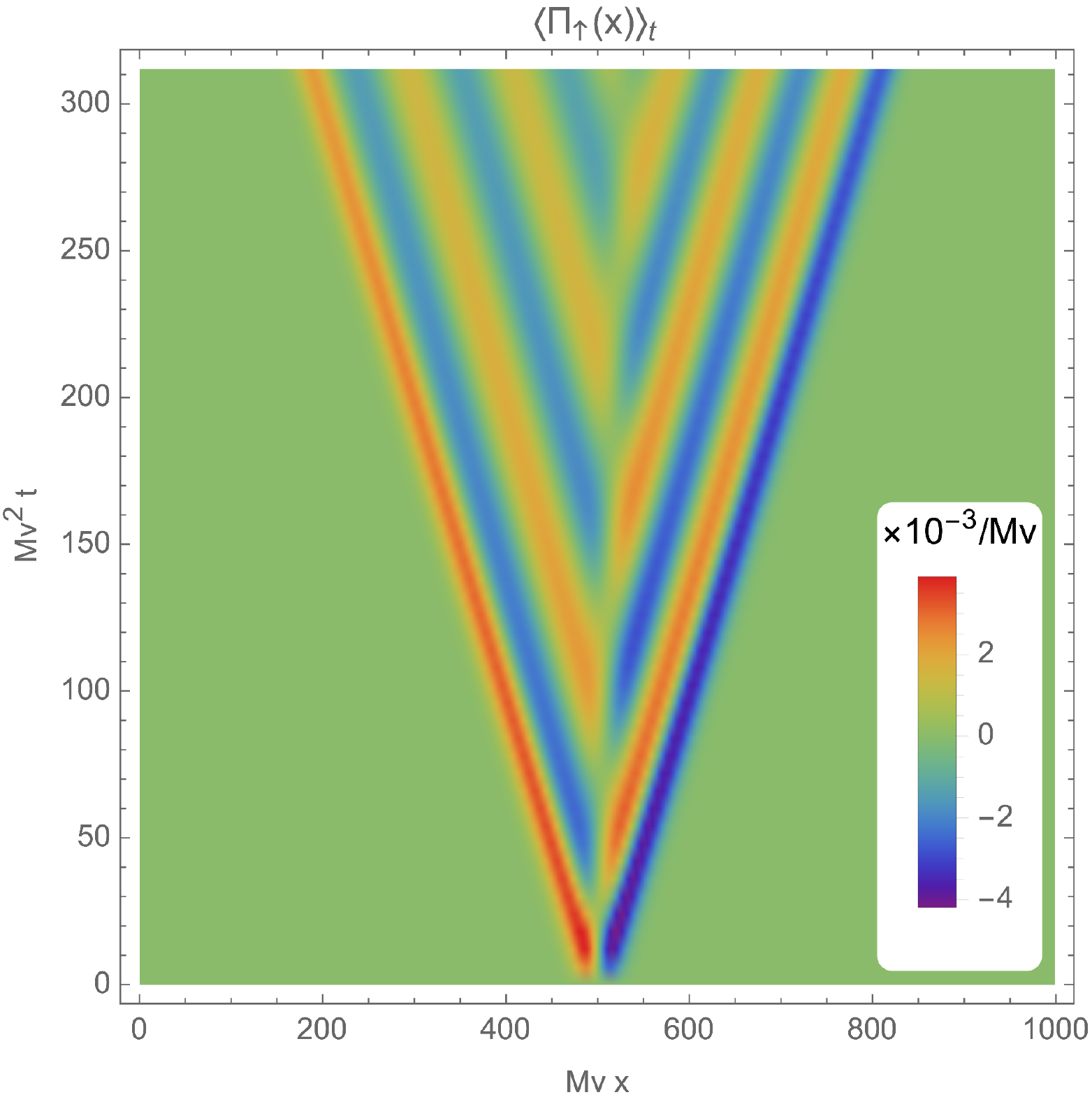}}
    \caption{Momentum density evolution. Figure (a) shows a snapshot of the momentum density profiles for both baths, while \fig~(b) shows the full evolution of the $\uparrow$ bath momentum density. The wave packet is Gaussian [\eq~\eqref{eq: wave packet}], with $N_p=64$ momenta distributed around $p_0=0.1Mv$ with a standard deviation $\delta p=0.04Mv$. The parameter settings are $g^2K=0.5v^2$, $K=2$ for both baths, and $\jp=0.03Mv^2$.}
    \label{fig:momentum density}
\end{figure}
\par We have also computed the momentum density $\expval{\Pi_\sigma(x)}_t$, and two examples are shown in \fig~\ref{fig:momentum density}. The time evolution of the momentum density shares many qualitative features with the density. There is a central part that follows the motion of the impurity, made of a relative minimum and a maximum that oscillate in time, cyclically exchanging their roles. From them, two trains of ripples expand in opposite directions, up to two wave fronts. Contrary to the density, these wave fronts are out of phase: the left-moving one is positive, while the right-moving one is negative. This property does not seem to be related to the sign of the momenta in the wave packet. As in the case of the density, the central part and the wave fronts are always present for any $\jp$, whereas the ripples increase their amplitude as $\jp$ becomes smaller. Moreover, all features except from the wave fronts are out of phase between the two baths.  
\par We want to briefly comment upon the scaling of the densities with Luttinger parameters $K_\sigma$. In the long-wavelength effective Hamiltonian \eqref{eq: H coupling}, $K_\sigma$ enters only through the combination $\gt_\sigma\equiv g_\sigma K^{1/2}_\sigma$, hence the expectation value \eq~\eqref{eq: expv b} of $\expval{\ee^{-\ii q X}b_\qs}$ only depends on $\gt_\sigma$. On the other hand, the densities \eqs~\eqref{eq: expv densities} explicitly contain $K_\sigma$, and we can express their scaling as
\begin{subequations}\label{eq:density scaling}
\begin{align}
    \expval{\rho_\sigma(x)}_t&=K_\sigma^{1/2}f_\rho(g_\uparrow^2 K_\uparrow,g_\downarrow^2 K_\downarrow)~,\\
    \expval{\Pi_\sigma(x)}_t&=K_\sigma^{-1/2}f_\Pi(g_\uparrow^2 K_\uparrow,g_\downarrow^2 K_\downarrow)~,
\end{align}
\end{subequations}
where $f_\rho$ and $f_\Pi$ are two appropriate functions that we do not need to specify here. From the equations above, we can deduce that the \emph{shape} of the density profiles is controlled only by the effective coupling $\gt_\sigma$, while if one varies $K_\sigma$ while keeping $\gt_\sigma=g_\sigma K^{1/2}_\sigma$ fixed the density or momentum profile only gets rescaled. Thus, each of the figures shown above can be taken to represent a family of density profiles.
\subsection{Correlation functions}\label{subsec:correlation functions}
\begin{figure}
    \centering
    \subfloat[][\label{fig:densityCorr J=0.1 Rr}$\jp=0.1Mv^2$]{\includegraphics[width=.9\linewidth]{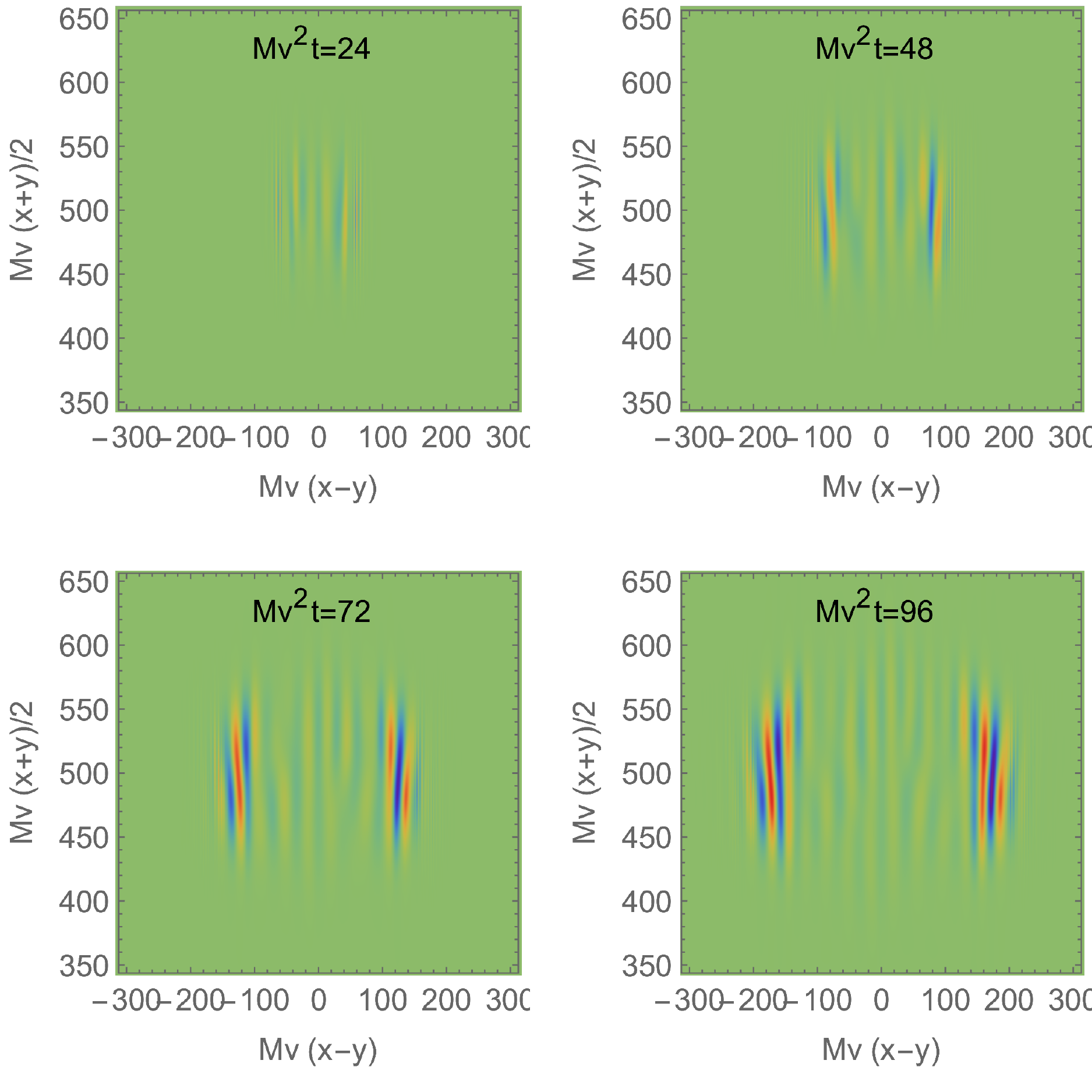}}\\
    \subfloat[][\label{fig:densityCorr J=0.01 Rr}$\jp=0.01Mv^2$]{\includegraphics[width=.9\linewidth]{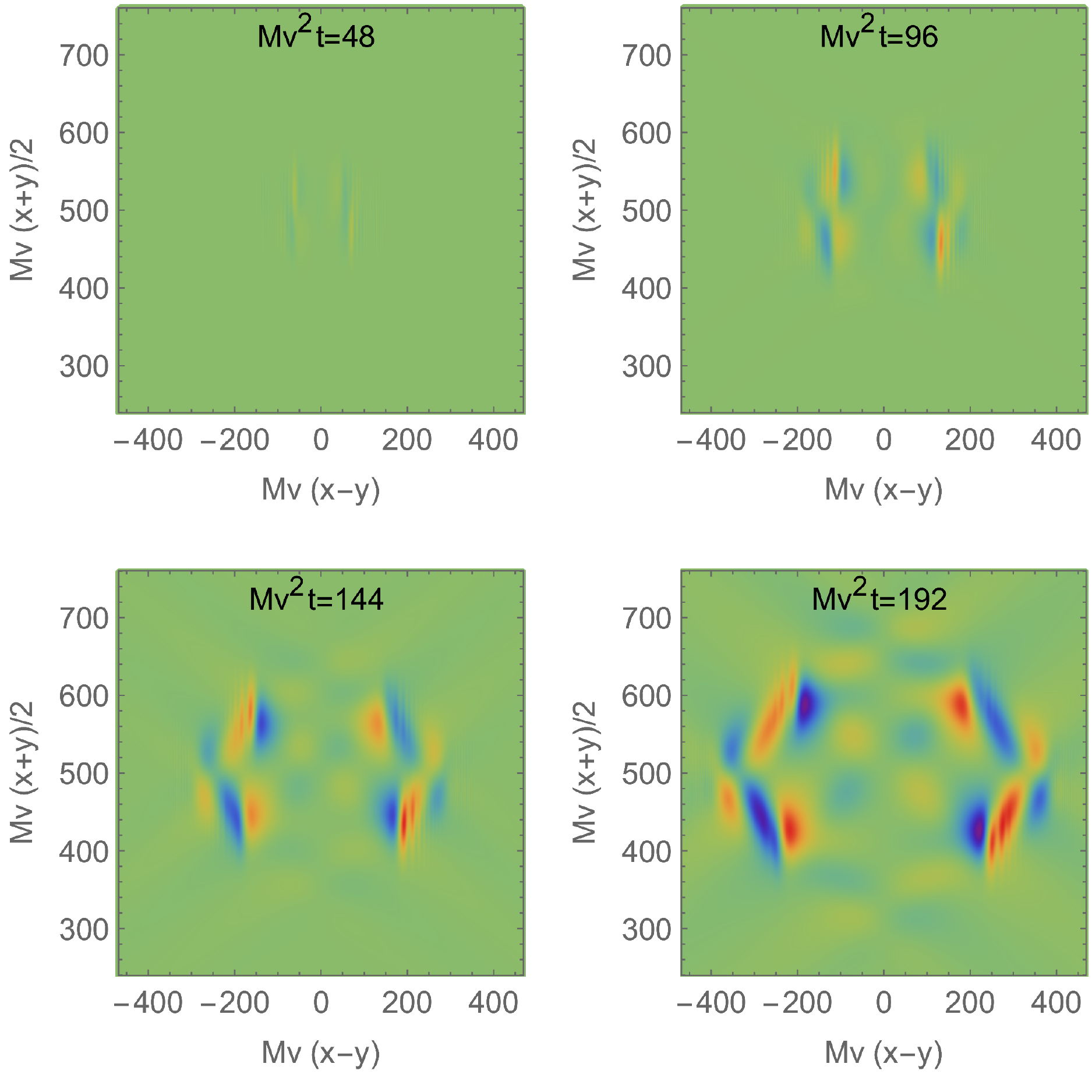}}
    \caption{Time evolution of the connected density-density correlation function. The baths are identical, with parameters $g^2K=0.5v^2$ and $K=2$. The wave packet is Gaussian [\eq~\eqref{eq: wave packet}], with $N_p=32$, $\delta p=0.02 Mv$ and $p_0=0.1Mv$. The color scale is normalized to the last ''snapshot'' of each set of plots.}
    \label{fig:full density corr}
\end{figure}

\begin{figure}
    \centering
    \subfloat[][\label{fig:densityCorr J=0.1 rt}]{\includegraphics[width=.9\linewidth]{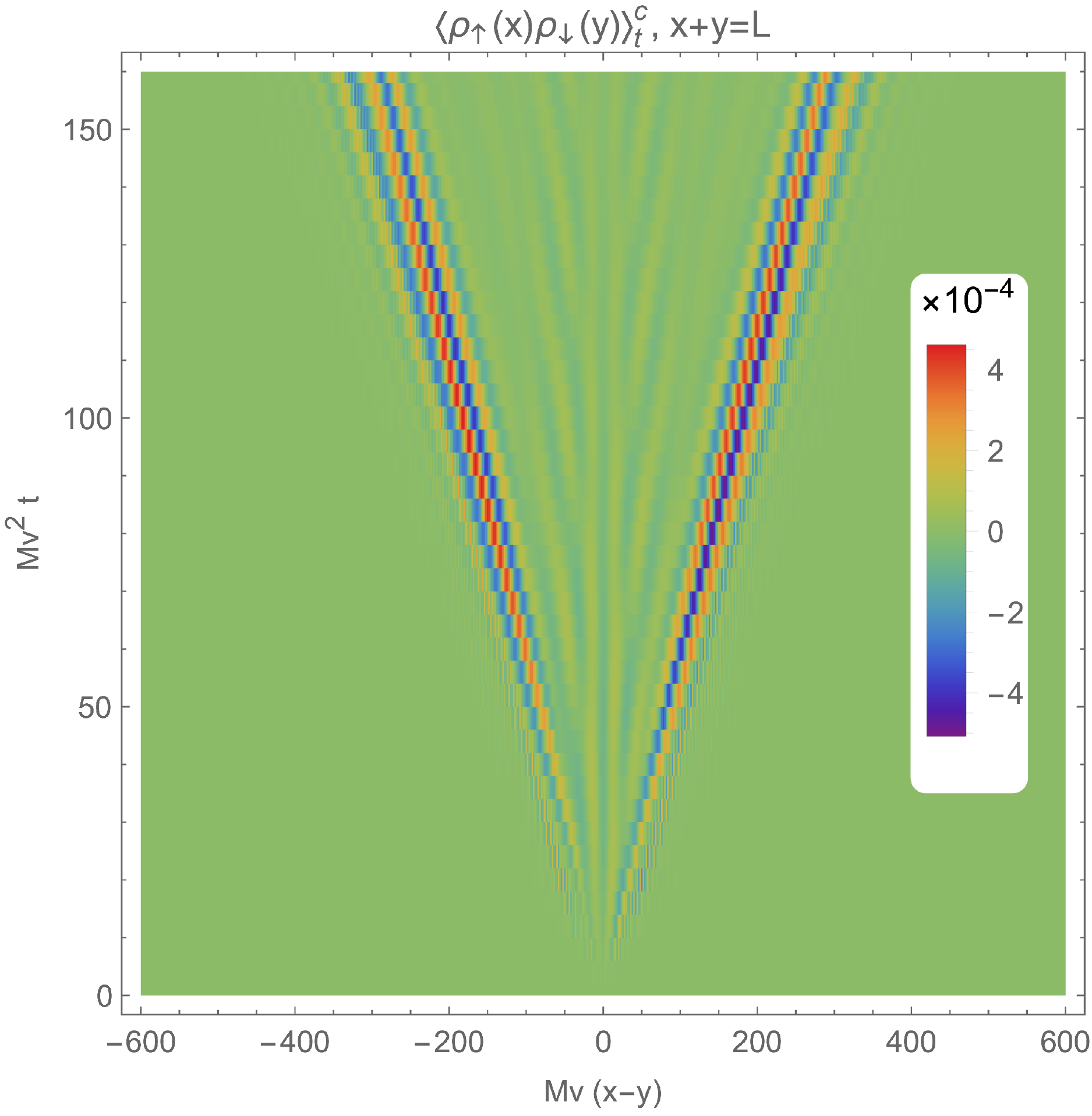}}\\
    \subfloat[][\label{fig:densityCorr J=0.1 Rt}]{\includegraphics[width=.9\linewidth]{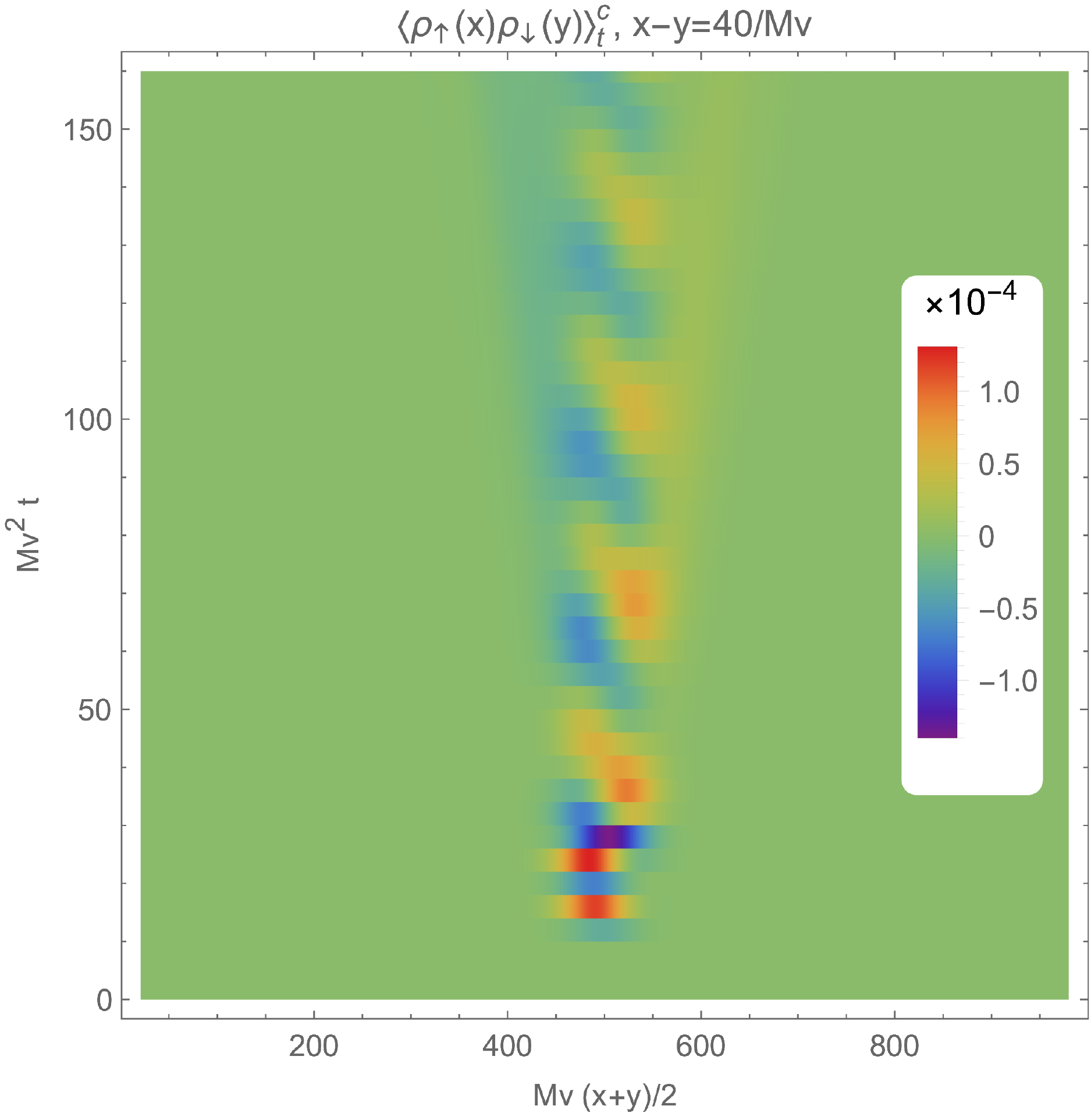}}
    \caption{Two ''slices'' of the density correlation function evolution at $\jp=0.1Mv^2$ (\fig~\ref{fig:densityCorr J=0.1 Rr}). The parameters are the same as \fig~\ref{fig:full density corr}. Figure (a) shows it as a function of the relative coordinate $r=x-y$,  while (b) uses the center of mass coordinate $R=(x+y)/2$.}
    \label{fig:densityCorr J=0.1}
\end{figure}
\begin{figure}
    \centering
    \subfloat[][\label{fig:densityCorr J=0.01 rt}]{\includegraphics[width=.9\linewidth]{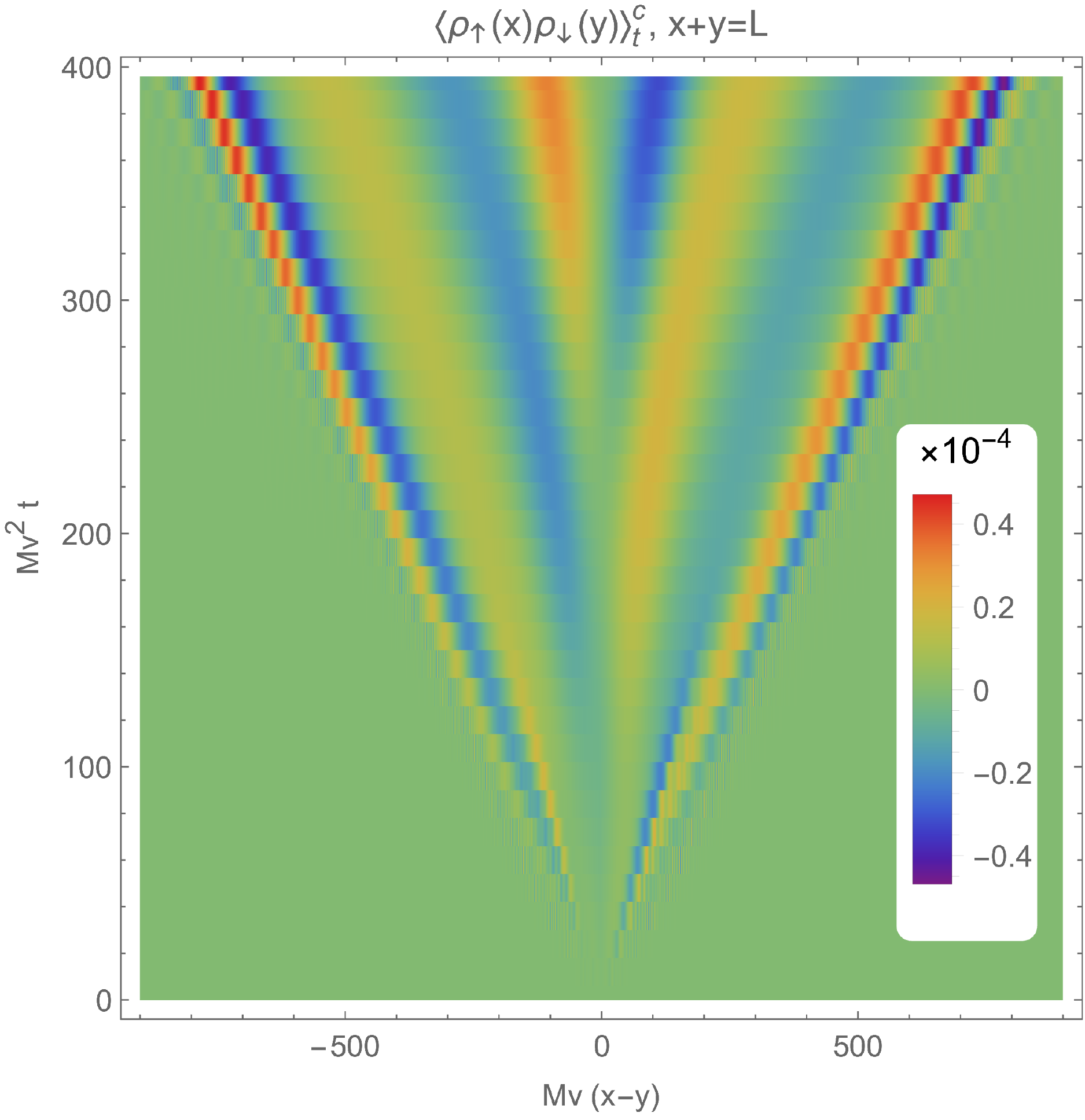}}\\
    \subfloat[][\label{fig:densityCorr J=0.01 Rt}]{\includegraphics[width=.9\linewidth]{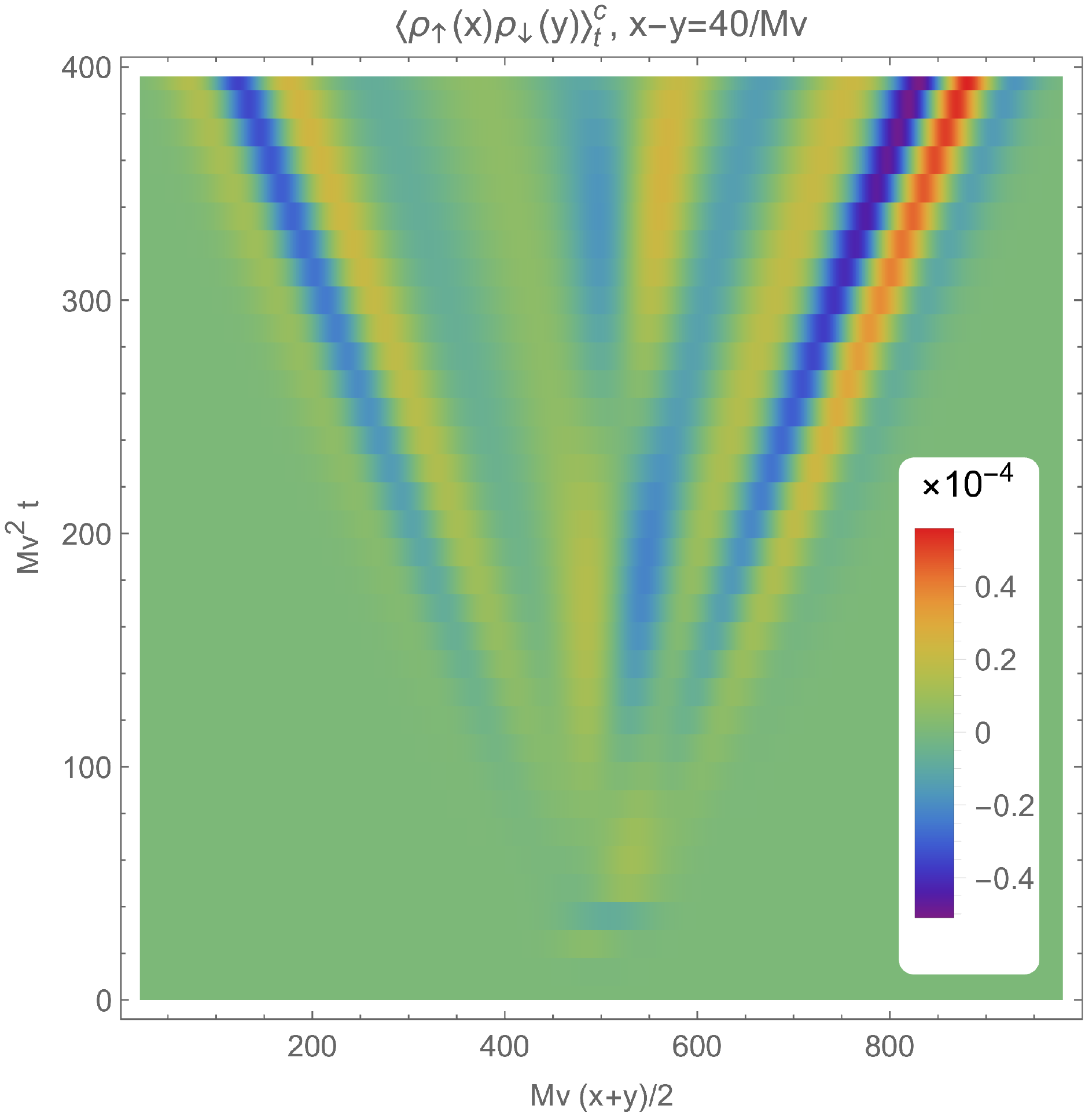}}
    \caption{Two ''slices'' of the density correlation function evolution at $\jp=0.01Mv^2$ (\fig~\ref{fig:densityCorr J=0.01 Rr}). The parameters are the same as \fig~\ref{fig:full density corr}. Figure (a) shows it as a function of the relative coordinate $r=x-y$,  while (b) uses the center of mass coordinate $R=(x+y)/2$}
    \label{fig:densityCorr J=0.01}
\end{figure}
The typical behavior of the equal-time correlation function is shown in \figs~\ref{fig:full density corr},~\ref{fig:densityCorr J=0.1} and \ref{fig:densityCorr J=0.01}, for the same initial conditions as in the discussion of the density evolution. All plots refer to the same $g_\sigma$, while we vary $\jp$.
\par Figures \ref{fig:densityCorr J=0.1 Rr} and \ref{fig:densityCorr J=0.01 Rr} show a sequence of ''snapshots'' of the full spatial behavior of the correlation function at various moments of time. These show that the correlations are concentrated within two ''lobes'', with a series of ripples between them. As time advances, the lobes move apart from each other, while both their amplitude and spatial width increase. The expansion is roughly ballistic, that is, all distances increase linearly in time, albeit with a larger speed in the relative $r=x-y$ direction than in the ''center-of-mass'' $R=(x+y)/2$ one. As this expansion takes place, in the region between the lobes a series of ripples form and increase their amplitude in time.
\par We remark that we have stopped all calculations before the light cone get too close to the edges of the system\footnote{For instance, notice that the ''snapshots'' in \fig~\ref{fig:full density corr} cover an area in $(r,\,R)$ space which is quite smaller than the whole allowed rectangle $[-L,L]\times[0,L]$.}, in order to avoid finite-size effects (besides the discretization of momenta). Therefore, the observed features should be a result of the intrinsic dynamics of the system under investigation, rather than an effect of interference through the boundaries.
\par These inter-bath density (or momentum) correlation functions are not symmetric under exchange of $x$ and $y$, even though $\rho_\uparrow(x)$ (or $\Pi_\uparrow(x)$) and $\rho_\downarrow(y)$ ($\Pi_\downarrow(y)$) commute and that the baths have identical properties. The asymmetry arises because the evolution is made asymmetric by the initial conditions, namely the impurity starting in bath $\uparrow$, with an average nonzero momentum. However, from \fig~\ref{fig:full density corr} it is easy to notice an approximate anti-symmetry with respect to the lines $r=0$ and $R=L/2$.
\par Analogously to the case of the density averages, the correlation functions obey a specific scaling with respect to the Luttinger parameters $K_\sigma$:
\begin{subequations}
\begin{align}
    \expval{\rho_\uparrow(x)\rho_\downarrow(y)}_t&=(K_\uparrow K_\downarrow)^{1/2}f_{\rho\rho}(g_\uparrow^2 K_\uparrow,g_\downarrow^2 K_\downarrow)~,\\
    \expval{\Pi_\uparrow(x)\Pi_\downarrow(y)}_t&=(K_\uparrow K_\downarrow)^{-1/2}f_{\Pi\Pi}(g_\uparrow^2 K_\uparrow,g_\downarrow^2 K_\downarrow)~.
\end{align}
\end{subequations}
The scaling of the densities, \eq~\eqref{eq:density scaling}, ensures that the same relation holds for the connected correlation functions. We have verified numerically that changing $g_\sigma K_\sigma^{1/2}$ only causes minor changes in the shape of the correlation functions, apart from an obvious change in the amplitude. The most relevant shape modifications are those induced by a change in $\jp$. At large $\jp$ (\fig~\ref{fig:densityCorr J=0.1 Rr}), the correlation function oscillates basically only in the relative $r$ direction, whereas the profile along $R$ shows less features. As $\jp$ is lowered (\fig~\ref{fig:densityCorr J=0.01 Rr}), the shape of the lobes becomes more complex, mainly because the correlation function oscillates also in the $R$ direction. Moreover, the ripples ''leak out'' of the inter-lobe region.
\par In the next figures we consider the behavior as a function of, respectively, the relative coordinate $r=x-y$ and center-of-mass coordinate $R=(x+y)/2$. 
\par In the time evolution of $\expval{\rho_\uparrow(x)\rho_\downarrow(y)}_t^c$ as a function of the coordinate difference, as shown in \figs~\ref{fig:densityCorr J=0.1 rt} and \ref{fig:densityCorr J=0.01 rt}, correlations appear only within a  "light-cone" $\abs{x-y}\leq v t$. The maximal amplitude occurs around the light-cone itself ($\abs{x-y}\approx v t$), while within the interior there are waves that appear to be radiated from $x=y$. A comparison with \fig~\ref{fig:full density corr} leads to identify the light-cone region with the lobes, while the waves in the interior are the ripples. The wavelength of the latter roughly corresponds to that of the phonons emitted during the deexcitation of the odd impurity mode. This identification, as in the case of the density, comes from the analytical expressions [\eqs~\eqref{eqs: expv bb}], and from the observation that the wavelength is essentially independent of $g_\sigma$ and $K_\sigma$, while it is inversely correlated with $\jp$, as can be appreciated by comparing \fig~\ref{fig:densityCorr J=0.1} and \fig~\ref{fig:densityCorr J=0.01}. As in the case of the average densities, the amplitude of these ''ripples'' increases for smaller $\jp$.
\par Summing up, the behavior of the correlation function along the relative coordinate basically reflects the ''relativistic'' nature of TLL bath dynamics, namely that inter-bath correlations are generated and propagated as linearly dispersing sound modes.
\par The situation looks different in the center-of-mass coordinate $R$, as \figs~\ref{fig:densityCorr J=0.1 Rt} and \ref{fig:densityCorr J=0.01 Rt} show. Here, we can distinguish a central area in which two trains of ripples oscillate out of phase, and an outer area formed of waves that radiate at the speed of sound from the central area. This distinction is sharp for higher $\jp$ (\fig~\ref{fig:densityCorr J=0.1 Rt}), as the amplitude of the emitted waves increase with decreasing $\jp$. The inner ripples occupy an area that spreads very slowly in space, and is centered along the trajectory $R=L/2+p_0 t/M$. Moreover, their oscillations in time occur with a period of about $\pi/\jp$, that is, half of the impurity oscillation period. These clues leads us to identify this ''section'' of the correlation function as the one more closely reflecting the motion of the impurity and the profile of its wave packet. In order to plot \figs~\ref{fig:densityCorr J=0.1 Rt} and \ref{fig:densityCorr J=0.01 Rt}, we chose a value for $r$. Changing it causes two main effects: first, the correlation function is zero up to a time that increases with $r$ (a light-cone effect). Second, as $r$ is decreased the oscillations in time get washed away by a featureless background contribution, until at $r=0$, i.e. $x=y$, there are no more visible oscillations. For all the parameters we checked, $\expval{\rho_\uparrow(x)\rho_\downarrow(x)}_t^c$ is always negative.
\begin{figure}
    \centering
    \includegraphics[width=.9\linewidth]{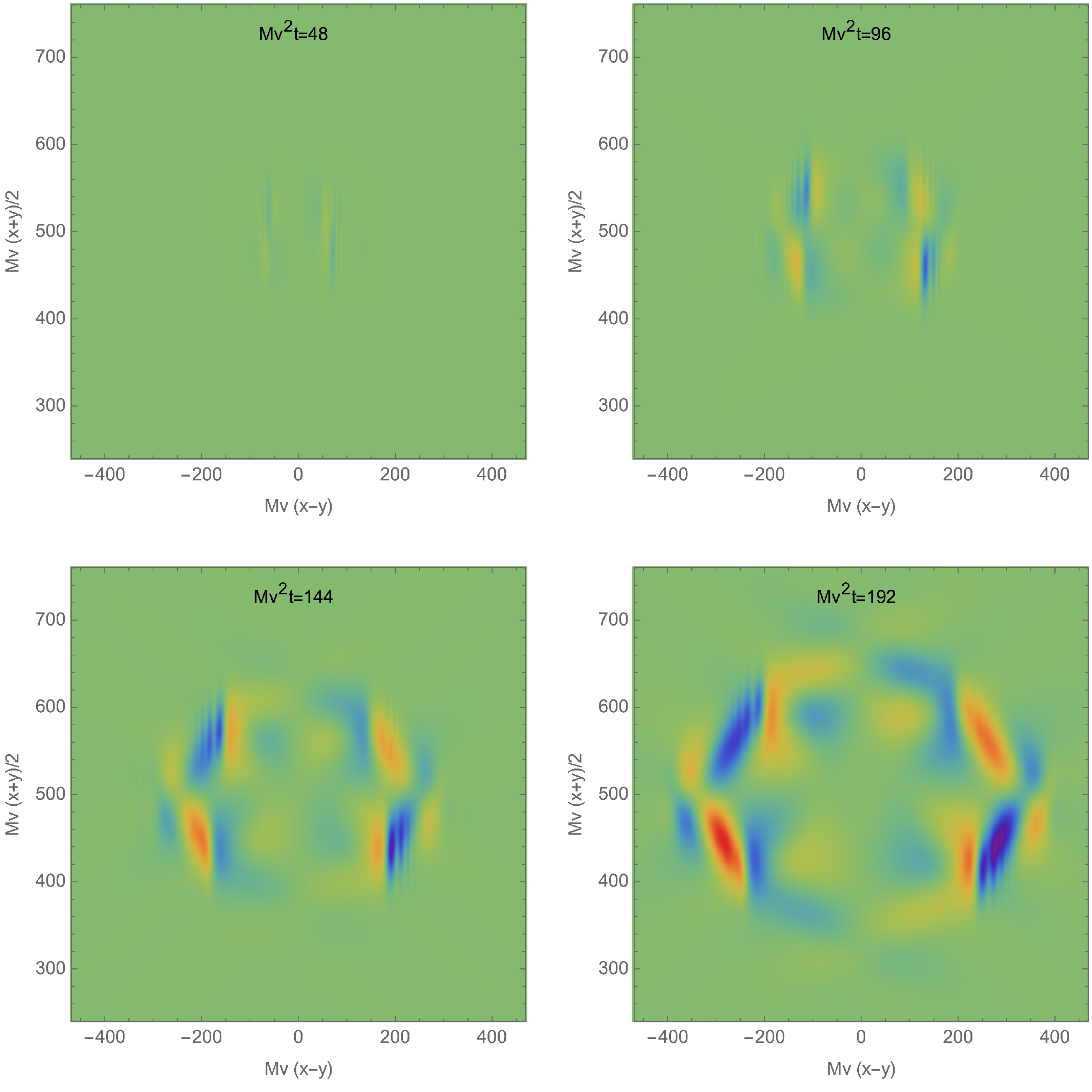}
    \caption{Time evolution of the connected momentum density correlation function. The baths are identical, with parameters $g^2K=0.5v^2$, $K=2$ and $\jp=0.01Mv^2$. The color scale is normalized to the last ''snapshot'' of each set of plots.}
    \label{fig:momDensityCorr J=0.01}
\end{figure}
\par The properties of the correlation functions can be rationalized using \eq~\eqref{eq: linear relation densities}. We already noticed that, since this equation is a relation between operators, it implies a whole hierarchy of relations that go beyond linear response theory. Indeed, we can compute \begin{multline}\label{eq: linear relation correlations}
\expval{\rho_\sigma(x)\rho_{\bar{\sigma}}(y)}^c_t=\\
=g_\sigma \int\dd{x_1}\dd{t_1}\chi_{\sigma}(x-x_1,t-t_1)\expval{N_\sigma(x_1,t_1) \rho_{\bar{\sigma}}(y,0)}+\\
+g_{\bar{\sigma}}\int\dd{x_2}\dd{t_2}\chi_{\bar{\sigma}}(x-x_2,t-t_2)\expval{\rho_\sigma(x,0) N_{\bar{\sigma}}(x_2,t_2)}+\\
+g_\uparrow g_\downarrow\int\dd{x_1}\dd{t_1}\dd{x_2}\dd{t_2}\chi_\sigma(x-x_1,t-t_1)\chi_{\bar{\sigma}}(x-x_2,t-t_2)\times\\
\times\expval{N_\sigma(x_1,t_1) N_{\bar{\sigma}}(x_2,t_2)}^c~,
\end{multline}
where we see that the bath density correlation function is related to the connected impurity density correlation function
\begin{multline}
    \expval{N_\sigma(x_1,t_1) N_{\bar{\sigma}}(x_2,t_2)}^c\equiv\\
    \equiv\expval{N_\sigma(x_1,t_1) N_{\bar{\sigma}}(x_2,t_2)}+\\
    -\expval{N_\sigma(x_1,t_1)}\expval{N_{\bar{\sigma}}(x_2,t_2)}~,
\end{multline}
which, unlike $\expval{\rho_\sigma(x)\rho_{\bar{\sigma}}(y)}^c_t$, correlates the impurity density at different times, and so it cannot be calculated from the knowledge of $\ket{\Psi(t)}$ only. Thus, in principle we could use \eq~\eqref{eq: linear relation correlations} to compute this impurity density correlation function. For now, we can just point out that thanks to this formula we have a hint on why $\expval{\rho_\sigma(x)\rho_{\bar{\sigma}}(y)}^c_t$ as a function of $(x+y)/2$ seems to mirror the time evolution of the impurity wave packet---indeed, we can now understand that it is keeping track of the impurity density (and impurity-bath density) correlation function.
\par We conclude this section by briefly discussing the connected momentum correlation function, $\expval{\Pi_\uparrow(x)\Pi_\downarrow(y)}_t^c$. An example is shown in \ref{fig:momDensityCorr J=0.01}. It shows the same qualitative features of the density correlation function, namely a pair of expanding lobes enclosing a region of smaller oscillations. The main differences with the density correlation are the more complex pattern of the ripples, and sign of the correlation for  $x=y$ which is positive for the momentum and negative for the density.
\section{Conclusions}\label{sec:conclusions}
 
 

In this paper, we have thoroughly studied an impurity hopping on a ladder whose two legs are described in terms of two Tomonaga-Luttinger Liquid baths.
\par We studied the problem from the perspective of the impurity and of the baths themselves, employing an improved perturbative technique. Albeit \textit{a priori} limited to small couplings, the method we used is rather simple, and has the advantage of providing analytical results for the whole system-bath state. Moreover, it is capable of treating the motion of wave packets with only a modest numerical effort. 
\par A comparison with a more conventional Green's function method, the Linked Cluster Expansion, showed that, at least in a symmetric setting where the two baths are identical, our perturbative technique yields the same result.
\par We analyzed the effect of the impurity motion on the bath density (and momentum density), finding that they mirror each other, with a behavior which reminds  the simple picture of a stone thrown in a pond. In each of the baths we observed two wave fronts generated by the insertion of the impurity and propagating away from it (i.e. the ''rings'' on the water surface in the pond analogy), a central density depletion following the impurity, and the emission of ripples as the impurity oscillates between the baths. The latter, in particular, are interpreted as a visualization of the phonon emission as the impurity loses its internal energy.
\par Then, we proceeded to examine the correlation between the two baths that is generated as they exchange the impurity. We did this by computing the inter-bath, connected density and momentum density correlation functions at equal times, unveiling a rich spatial structure. The correlation is non-vanishing only within an area in $(x,y)$ space, which expands ballistically. Two features can be distinguished: a pair of lobes and a series of ripples between them. Along the relative direction, $r=x-y$, the correlation function mainly shows the ''relativistic'' dynamics of the bath, with a clear light-cone as the phonons generated by the impurity spread the correlations. Along the center-of-mass direction $R=(x+y)/2$, instead, the light-cone of emitted phonons is superimposed with the density perturbation following the impurity wave packet in its motion.
\par There are various directions in which this work can be extended. An interesting direction is to explore the regime of stronger coupling between the impurity and the baths. A first step towards this regime would be to include the backscattering term in the interaction, and to see its effect within the perturbative formulation. A different approach would be to promote the perturbative expression of the state, \eq~\eqref{eq: sol wave packet2}, to a variational Ansatz whose coefficients would have to be found numerically. A second intriguing extension of this work would be to explore the effect of increasing the number of baths where the impurity can move.
\section{Acknowledgments}
We acknowledge useful discussions with A. Recati and F. Scazza.
We acknowledge financial support from MIUR through the PRIN 2017
(Prot. 20172H2SC4 005) program.

\appendix
\section{A different perspective on symmetric baths}\label{app: symmetric baths}
The physics of the system under investigation may be clearer in the case of symmetric baths, i.e. for $v_\sigma=v$, $K_\sigma=K$ and $g_\sigma=g$ (hence, $W_\qs\equiv W_q$). In this regime, it is useful to introduce even and odd bath modes,
\begin{equation}
    b_{q\mu=e/o}\equiv\tfrac{1}{2^{1/2}}(b_{q\uparrow}\pm b_{q\downarrow})~,
\end{equation}
and recast Hamiltonian \eqref{eq: LLP Hamiltonian} as
\begin{multline}\label{eq: HLLP in parity bath modes}
    \mathcal{H}_{LLP}(p)=\tfrac{(p-P_e-P_o)^2}{2M}-\jp\sigma_1+\\
    +\sum_{q}{v\abs{q}b_{qe}^\dag b_{qe}}+\sum_q\tfrac{W_q}{(2L)^{1/2}}(b_{qe}+b_{qe}^\dag)+\\
    +\sum_{q}{v\abs{q}b_{qo}^\dag b_{qo}}+\sigma_3\sum_q\tfrac{W_q}{(2L)^{1/2}}(b_{qo}+b_{qo}^\dag)~,
\end{multline}
where
\begin{equation}
    P_\mu\equiv\sum_q{q\, b_{q\mu}^\dag b_{q\mu}}~.
\end{equation}
It can be observed that the even and odd bath modes become partially independent of each other. In particular, only the odd bath modes are coupled to the impurity, and the deexcitation of the odd impurity band will generate odd bath modes. The even modes ''see'' the impurity only through the momentum-momentum coupling with the odd modes in the first term (that is, the kinetic energy of the impurity in the laboratory frame). The decoupling would be complete for a static impurity, namely for $M\to+\infty$.
\par With this separation in mind, it is natural to attempt a first approximation in which the bath parity modes evolve independently since odd bath modes are generated only by transitions between the even and odd impurity bands, involving a ''gap'' $2\jp$ in energy. This leads to our choice of the perturbative scheme, \eq~\eqref{eq:separation of HLLP}, that in this language reads
\begin{subequations}
\begin{align}
    \mathcal{H}_{LLP}(p)&=\mathcal{H}_0(p)+\Delta\mathcal{H}(p)~,\\
    \mathcal{H}_0(p)&\equiv E(p)-\jp\sigma_1+\notag\\
    &+\sum_{q}{\Omega_q b_{qe}^\dag b_{qe}}+\sum_q\tfrac{W_q}{(2L)^{1/2}}(b_{qe}+b_{qe}^\dag)+\notag\\
    &+\sum_{q}{\Omega_q b_{qo}^\dag b_{qo}}~,\\
    \Delta\mathcal{H}(p)&\equiv \sigma_3\sum_q\tfrac{W_q}{(2L)^{1/2}}(b_{qo}+b_{qo}^\dag)+\notag\\
    &+\tfrac{P_e P_o}{M}+\sum_\mu\tfrac{:P_\mu^2:}{2M}~.
\end{align}
\end{subequations}
According to this perspective, the perturbative expression \eq~\eqref{eq: sol pmu2} has a simple interpretation. The background coherent state $\ket{\omega_p(t)}$ contains only even bath modes, that are increasingly populated in time. On the other hand, the odd bath modes may contain only up to two phonons each, as described by the ''first'' and ''second-order'' corrections, proportional to $b_\qs^\dag$ and $b_\qs^\dag b_\qsprime^\dag$, respectively. The latter also introduces the correlation between even and odd bath modes as dictated by the $P_e P_o/M$ term in the Hamiltonian.  
\section{Normalization factors}\label{app: normalization factors}
In this \app, we shortly discuss the properties and  the computation of the normalization factors, \eq~\eqref{eq: apmu}. In particular, we often have to compute expressions in the form $\tfrac{1}{L}\sum_\qns{W_\qs^2f(\Omega_\qs)}$,
for a given function $f(\Omega_\qs)$. These can be conveniently computed by introducing an appropriate density of states $R^S(\ce)$:
\begin{equation}
    \tfrac{1}{L}\sum_\qns{W_\qs^2f(\Omega_\qs)}=\int_0^{+\infty}{\dd{\varepsilon}f(\varepsilon)R^S(\varepsilon)}~,
\end{equation}
where $R^S(\varepsilon)\equiv\sum_\qns{\tfrac{W_\qs^2}{L}\delta(\varepsilon-\Omega_\qs)}$. In the continuum limit $L\to+\infty$, $R^S(\ce)$ can be calculated exactly, and for subsonic momenta $\abs{p}<Mv_\sigma$ it reads:
\begin{multline}
    R^S(\varepsilon)=\frac{M}{(2\pi)^2}\theta(\ce)\ee^{-\ce/\Lambda}\times\\
    \times\sum_{\sigma,s=\pm1}{g_\sigma^2 K_\sigma\Big[1-\big(1+\tfrac{\ce}{k_{s\sigma}(p)}\big)^{-1/2}\Big]}~.
\end{multline}
In the above equation, $\theta(\ce)$ is the Heaviside theta function, $k_{s\sigma}(p)$ is defined as
\begin{equation}
    k_{s\sigma}(p)\equiv \frac{(Mv_\sigma+sp)^2}{2M},\quad s=\pm1
\end{equation}
and we introduced an energy cutoff $\Lambda$ instead of a momentum one, for simplicity.
\par This density of states is analogous to the ones that characterize the bath in the spin-boson or Caldeira-Leggett models \cite{RevModPhys.59.1}. At small energies $\varepsilon\ll Mv_\sigma^2$ they are linear, so that the Tomonaga-Luttinger liquid baths are classified as ohmic:
\begin{equation}
    R^S(\varepsilon)=\theta(\varepsilon)\varepsilon\sum_\sigma{\beta_\sigma^{\textup{sb}}(p)}+\order{\tfrac{\varepsilon^2}{M^2v_\sigma^4}}~,
\end{equation}
where
\begin{equation}
    \beta_\sigma^{\textup{sb}}=\frac{g_\sigma^2K_\sigma}{2\pi^2v_\sigma^2}\frac{1+(p/Mv_\sigma)^2}{(1-(p/Mv_\sigma)^2)^2}
\end{equation}
is the single-bath orthogonality exponent \cite{KSG, PhysRevB.103.094310}.
\par We have worked out a numerically friendly way of computing the exponents in \eq~\eqref{eq: apmu} in a previous paper \cite{PhysRevB.103.094310}. Here we quote only the final results:
\begin{align}
    -&\sum_{\qns}{\tfrac{W_{\qs}^{2}}{4L}\tfrac{1-\ii(\Omega_\qs\pm2\jp) t-\ee^{-\ii(\Omega_{\qs}\pm2\jp)t}}{(\Omega_{\qs}\pm2\jp)^2}}=\notag\\
    =&-\sum_{\sigma}{}\tfrac{M\gt_\sigma^2}{(4\pi)^2}\big\{f(\pm\jp,t)+\notag\\
    &+\sum_{s=\pm1}{}\big[\tfrac{1}{k_{s\sigma}(p)}\phi_1(\tfrac{\pm2\jp}{k_{s\sigma}(p)})-t \phi_2(\tfrac{\pm2\jp}{k_{s\sigma}(p)})+\notag\\ &+\ii\ee^{\pm2\ii\jp t}\int_0^{+\infty}\dd{u}\tfrac{\ee^{-u t}}{(\ii u\pm2\jp)^2}\tfrac{1}{(1-\ii u/k_{s\sigma}(p))^{1/2}}\big]\big\}
\end{align}
where
\begin{align}
    f(\pm\jp,t)&=\pi t\pm2t \mathrm{Si}(2\jp t)\mp\frac{1-\cos{2\jp t}}{\jp}+\notag\\
    &+2\ii t\bigg[\ln{\tfrac{2\jp}{\Lambda \ee^{-\gamma}}}+\Re E_1(2\ii\jp t)+\frac{\sin{2\jp t}}{2\jp t}\bigg]~,
\end{align}
$ \mathrm{Si}(z)$ and $E_1(z)$ are the sine and exponential integral functions \cite{DLMF}, respectively, and we introduced the functions
\begin{equation}
    \phi_1(x)=
    \begin{cases}
    \frac{1}{2x(1+x)^{3/2}}\Big[\pi x+\nonumber\\
    \;-2\ii\qty(\sqrt{x+1}+x\, \ash(\tfrac{1}{\sqrt{x}}))\Big],\quad x>0\\
    \frac{\ii}{\abs{x}(1-\abs{x})}-\frac{\ii}{(1-\abs{x})^{3/2}}\ach(\tfrac{1}{\sqrt{\abs{x}}}),\:\: x<0
    \end{cases}~,
\end{equation}
\begin{equation}
    \phi_2(x)=
    \begin{cases}
    \frac{1}{\sqrt{1+x}}\Big[\pi-2\ii\,\ash(\tfrac{1}{\sqrt{x}})\Big],\quad x>0\\
    -\frac{2\ii}{\sqrt{1-\abs{x}}}\ach(\tfrac{1}{\sqrt{\abs{x}}}),\quad x<0
    \end{cases}~,
\end{equation}
in which it is understood that
\begin{equation}\label{eq: cosh convention}
    \ach(x)=\ii\arccos{x}\qq{for}\abs{x}<1~.
\end{equation}
Analogously,
\begin{align}
    -&\sum_{\qns}{\tfrac{W_{\qs}^{2}}{4L}\tfrac{1-\ii\Omega_\qs t-\ee^{-\ii\Omega_{\qs}t}}{(\Omega_{\qs})^2}}=\notag\\
    &=-\sum_{\sigma}{}\tfrac{M\gt_\sigma^2}{(4\pi)^2}\big\{f(0,t)+\notag\\
    &-\sum_{s=\pm1}{\ii\int_0^{+\infty}\dd{u}\tfrac{1-ut-\ee^{-u t}}{u^2}\tfrac{1}{(1-\ii u/k_{s\sigma}(p))^{1/2}}}\big\}~,
\end{align}
where
\begin{equation}
    f(0,t)=\pi t-2\ii t\ln{\tfrac{\Lambda t}{\ee}}~.
\end{equation}
The manipulations done so far allow us to easily derive the asymptotic behavior of the normalization factors:
\begin{subequations}
\begin{align}
    \ln a_{pe}&=-\tfrac{M}{(4\pi)^2}\sum_\sigma\big[\tfrac{1}{\jp}+\ii\sum_{s=\pm}\tfrac{\phi_1(-2\jp/k_{s\sigma})}{k_{s\sigma}}\big]+\notag\\
    &-\ii\Delta\lambda_{pe}t+\order{1/t}\\
   \ln a_{po}&=-\tfrac{M}{(4\pi)^2}\sum_\sigma\big[-\tfrac{1}{\jp}+\ii\sum_{s=\pm}\tfrac{\phi_1(2\jp/k_{s\sigma})}{k_{s\sigma}}\big] +\notag\\
   &-\ii\Delta\lambda_{po}t -2\gamma_p t+\order{1/t}~,
\end{align}
\end{subequations}
where
\begin{subequations}
\begin{align}
    \Delta\lambda_{pe}&=\tfrac{M}{8\pi^2}\sum_\sigma g_\sigma^2K_\sigma\big[\ln\tfrac{2\jp}{\Lambda\ee^{-\eta}}+\notag\\
    &+\sum_{s=\pm1} \tfrac{1}{(1-2\jp/k_{s\sigma}(p))^{1/2}}\ach{\big(\tfrac{k_{s\sigma}(p)}{2\jp}\big)^{1/2}}\big]\\
    \Delta\lambda_{po}=&\tfrac{M}{8\pi^2}\sum_\sigma g_\sigma^2K_\sigma\big[\ln\tfrac{2\jp}{\Lambda\ee^{-\eta}}+\notag\\
    &+\sum_{s=\pm1} \tfrac{1}{(1+2\jp/k_{s\sigma}(p))^{1/2}}\ash{\big(\tfrac{k_{s\sigma}(p)}{2\jp}\big)^{1/2}}\big]\\
    2\gamma_p&=\tfrac{M}{16\pi}\sum_{s\sigma}g_\sigma^2K_\sigma\bigg[1-\tfrac{1}{(1+2\jp/k_{s\sigma}(p))^{1/2}}\bigg]~.
\end{align}
\end{subequations}
In the equations above, $\eta$ is the Euler-Mascheroni constant, and the convention \eq~\eqref{eq: cosh convention} is assumed once again. 
Physically, remembering that $a_{p\mu}$ is always multiplied by $\ee^{-\ii\lambda_{p\mu}t}$ [see \eqs~\eqref{eq: sol pmu1} and~\eqref{eq: sol pmu2}], we see that $\Delta\lambda_{p\mu}$ renormalize the two bands $\lambda_{p\mu}$, lowering them by a cutoff-dependent quantity and slightly altering their curvature. Moreover, the energy difference between the renormalized bands is reduced with respect to the noninteracting value $2\jp$. This energy difference can be seen as a renormalized inter-bath hopping
\begin{multline}\label{eq: renormalized Jp}
    2\tilde{J}_{\perp p}\equiv2\jp+\Delta\lambda_{po}-\Delta\lambda_{pe}=2\jp+\\
    -\tfrac{M}{8\pi^2}\sum_{\sigma,s=\pm1}g_\sigma^2 K_\sigma \bigg[\tfrac{1}{(1-2\jp/k_{s\sigma}(p))^{1/2}}\ach{\big(\tfrac{k_{s\sigma}(p)}{2\jp}\big)^{1/2}}+\\
    -\tfrac{1}{(1+2\jp/k_{s\sigma}(p))^{1/2}}\ash{\big(\tfrac{k_{s\sigma}(p)}{2\jp}\big)^{1/2}}\bigg]~.
\end{multline}
The correction to $2\jp$ is always negative (that is, the inter-bath hopping is suppressed, as it usually happens for polarons \cite{Mahan} and spin-boson models \cite{RevModPhys.59.1}), and it is quite small in the low-momentum, weak coupling regime that we considered. However, notice that the expressions for these renormalized quantities are non-analytic for $\jp=0$.
\par The quantity $2\gamma_p$ is the width of the odd mode, and is proportional to $\jp$ at small inter-bath hopping while saturating at a maximum value for large $\jp$. However, the latter behavior has to be taken with caution, because in the large-$\jp$ regime bosonization is not applicable. See the end of \sect~\ref{sec:the perturbative technique}. 
\section{Second-order terms}\label{app: second order terms}
In this \app~we give details on the determination of the second-order correction to the state evolution, \eq~\eqref{eq: second order term}. In order to implement the perturbative procedure, we have to find the various second-order contributions among all terms in the expansion, which we will indicate with $(\dots)^{(2)}$:
\begin{subequations}
\begin{align}
    &(\Delta\mathcal{H}(t)\ket{p\mu,\omega})^{(2)}=\ee^{2\ii\mu\jp t}\expval{V(t)}\ket{p\bar{\mu},\omega}+\notag\\
    &+\sum_{\qs,\qsprime}{}\tfrac{q q^\prime}{2M}\tfrac{W_\qs W_\qsprime}{4L}\chi_t^\ast(\Omega_\qs)\chi_t^\ast(\Omega_\qsprime)b_\qs^\dag b_\qsprime^\dag \ket{p\mu,\omega}\label{eq: 2 from 0th}~,\\
    &\bigg(\Delta\mathcal{H}(t)\ket{\phi_{p\mu}^{(1)}}(t)\bigg)^{(2)}=\Delta E_{p\mu}^{(2)}\ket{p\mu,\omega}+\notag\\
    &+\sum_{\qs,\qsprime}{}\sigma\sigma^\prime \tfrac{W_\qs W_\qsprime}{4L}\chi_t^\ast(\Omega_\qs^{+\mu}(p))\ee^{\ii\Omega_\qsprime^{-\mu}t}b^\dag_\qs b^\dag_\qsprime\ket{p\mu,\omega}+\notag\\
    &+\sum_{\qs,\qsprime}{}\sigma\tfrac{q q^\prime}{M}\tfrac{W_\qs W_\qsprime}{4L}\chi_t^\ast(\Omega_\qs^{+\mu}(p))\chi_t^\ast(\Omega_\qsprime(p))\times\notag\\
    &\quad\times b^\dag_\qs b^\dag_\qsprime\ket{p\bar{\mu},\omega}~,\label{eq: 2 from 1st}\\
    &\bigg(\Delta\mathcal{H}(t)\ket{\delta_2\phi_{p\mu}^{(2)}}(t)\bigg)^{(2)}=\tfrac{:P_b^2:}{2M}\ket{\phi_{p\mu}^{(2)}(t)}=\notag\\
    &=\sum_\nu{\sum_{\qs,\qsprime}\tfrac{q q^\prime}{M}A^{\mu\nu}_{\qs,\qsprime}(p,t)b^\dag_\qs b^\dag_\qsprime\ket{p\nu,\omega}}~.\label{eq: 2 from 2nd}
\end{align}
\end{subequations}
The first term on the right-hand side (rhs) of \eq~\eqref{eq: 2 from 0th} yields what we called $\ket{\delta_1\phi_{p\mu}^{(2)}(t)}$ [see \eq~\eqref{eq: delta1 phi2}]. The first term on the rhs of \eq~\eqref{eq: 2 from 1st} would give rise to secular behavior, and it is canceled by the perturbative procedure [see \eq~\eqref{eq: equation set}]. All the remaining terms contain two phonon creation operators, and thus contribute to what we called $\ket{\delta_2\phi_{p\mu}^{(2)}(t)}$. The latter is determined by an equation of the form
\begin{multline}
    \ii\dv{}{t}\ket{\delta_2\phi_{p\mu}^{(2)}(t)}=\\
    =\sum_\nu{\sum_{\qs,\qsprime}\ii\dv{}{t}A^{\mu\nu}_{\qs,\qsprime}(p,t)b^\dag_\qs b^\dag_\qsprime\ket{p\nu,\omega}}=\\
    =\sum_\nu{}\sum_{\qs,\qsprime}\big(\tfrac{q q^\prime}{M}A^{\mu\nu}_{\qs,\qsprime}(p,t)+\\
    +S^{\mu\nu}_{\qs,\qsprime}(p,t)\big)b^\dag_\qs b^\dag_\qsprime\ket{p\nu,\omega}~,
\end{multline}
where we put all terms in \eqs~\eqref{eq: 2 from 0th} and \eqref{eq: 2 from 1st} involving two creation operators into the matrix $S^{\mu\nu}_{\qs,\qsprime}(p,t)$. Multiplying the above equation by $\bra{p\nu,\omega}b_\qsprime b_\qs$ to the left, and taking into account that $A^{\mu\nu}_{\qs,\qsprime}(p,t)$ is symmetric under exchange of $\qs$ and $\qsprime$, while $S^{\mu\nu}_{\qs,\qsprime}(p,t)$ is not, we find
\begin{multline}
    \ii\dv{}{t}A^{\mu\nu}_{\qs,\qsprime}(p,t)=\tfrac{q q^\prime}{M}A^{\mu\nu}_{\qs,\qsprime}(p,t)+\\
    +\tfrac{1}{2}\big(
    S^{\mu\nu}_{\qs,\qsprime}(p,t)+S^{\mu\nu}_{\qsprime,\qs}(p,t)\big)~,
\end{multline}
which is readily integrated:
\begin{multline}
    A^{\mu\nu}_{\qs,\qsprime}(p,t)=-\ii\ee^{-\ii\tfrac{q q^\prime}{M}t}\times\\
    \times\int_0^t \dd{t^\prime}\ee^{\ii\tfrac{q q^\prime}{M}t^\prime}\tfrac{1}{2}\big(
    S^{\mu\nu}_{\qs,\qsprime}(p,t^\prime)+S^{\mu\nu}_{\qsprime,\qs}(p,t^\prime)\big)~.
\end{multline}
Performing the integral, we arrive at the results quoted in the main text, \eqs~\eqref{eq: Amumu} and \eqref{eq: Amubarmu}.
\section{Comparison with the Linked-Cluster Expansion}\label{app: comparison with LCE}
In a previous paper \cite{PhysRevB.103.094310}, we calculated the Green's function (or fidelity, if $\sigma^\prime=\sigma$)
\begin{equation}
    G_{\sigma^\prime\sigma}(p,t)=-\ii\ev{d_{p\sigma^\prime}\ee^{-\ii\mathcal{H}t}d^\dag_{p\sigma}}{0,\omega}
\end{equation}
using the Linked Cluster Expansion (LCE) perturbative technique. In the symmetric-bath case, it was calculated to be
\begin{multline}\label{eq: G LCE}
    G^{\textup{LCE}}_{\sigma^\prime\sigma}(p,t)=-\tfrac{\ii}{2}\ee^{F(0,t)}\times\\
    \times\big(\ee^{-\ii\lambda_{pe}t+F(-\jp,t)}+\sigma\sigma^\prime \ee^{-\ii\lambda_{pe}t+F(\jp,t)}\big)~,
\end{multline}
where
\begin{multline}
    F(J,t)\equiv\int_0^{+\infty}\dd{\ce}\tfrac{1-(\ce-2J)t-\ee^{-\ii(\ce-2J)t}}{(\ce-2J)^2}R^S(\ce)~.
\end{multline}
It is easy to see that the perturbative approach that we developed in the present article is capable of reproducing the same result. In general
\begin{subequations}
\begin{align}
    G_{\sigma\sigma}(p,t)&=-\ii\theta(t)\mel{p\sigma}{\ee^{-\ii\mathcal{H}t}}{p\sigma}=\notag\\
    &=\tfrac{1}{2}\sum_\mu{\big[G_{\mu\mu}(p,t)+\sigma G_{\bar{\mu}\mu}(p,t)\big]}~,\\
    G_{\bar{\sigma}\sigma}(p,t)&=-\ii\theta(t)\mel{p\bar{\sigma}}{\ee^{-\ii\mathcal{H}t}}{p\sigma}=\notag\\
    &=-\tfrac{1}{2}\sum_\mu{\big[G_{\mu\mu}(p,t)-\sigma G_{\bar{\mu}\mu}(p,t)\big]}~,\end{align}
\end{subequations}
where
\begin{equation}
    G_{\mu\nu}(p,t)\equiv-\ii\theta(t)\mel{p\mu,\omega}{\ee^{-\ii\mathcal{H}t}}{p\nu,\omega}~.
\end{equation}
In our perturbative approximation $\ee^{-\ii\mathcal{H}t}\ket{p\mu,\omega}\approx\ket{\psi_{p\mu}}$ [\eq~\eqref{eq: sol pmu1} or~\eqref{eq: sol pmu2}], hence
\begin{subequations}
\begin{align}
G_{\mu\mu}(p,t)&=-\ii a_{p\mu}(t)\ee^{-\ii\lambda_{p\mu}t}\braket{\omega}{\omega_p(t)}~,\\
G_{\bar{\mu}\mu}(p,t)&=-\ii a_{p\mu}(t)\ee^{-\ii\lambda_{p\bar{\mu}}t}\braket{\omega}{\omega_p(t)}\tilde{B}_{p\mu}(t)~.
\end{align}
\end{subequations}
One can readily identify $a_{p\mu}(t)\equiv\ee^{F(-\mu\jp,t)}$, and calculate that $\braket{\omega}{\omega_p(t)}=\ee^{F(0,t)}$. Therefore, in the symmetric case when $\tilde{B}_{p\mu}(t)=0$, we recover exactly \eq~\eqref{eq: G LCE}.
\par The comparison with the LCE Green's function in the asymmetric case is less clear, because both in the LCE and in the perturbative technique of this paper the corresponding function depends explicitly on the cutoff $\Lambda$.
\section{Bath density evolution from linear response}\label{app: density explanation}
\begin{figure}
\centering
\subfloat[$\jp=0.1Mv^{2}$\label{fig: integrand annotated 0.1}]{\includegraphics[width=0.9\linewidth]{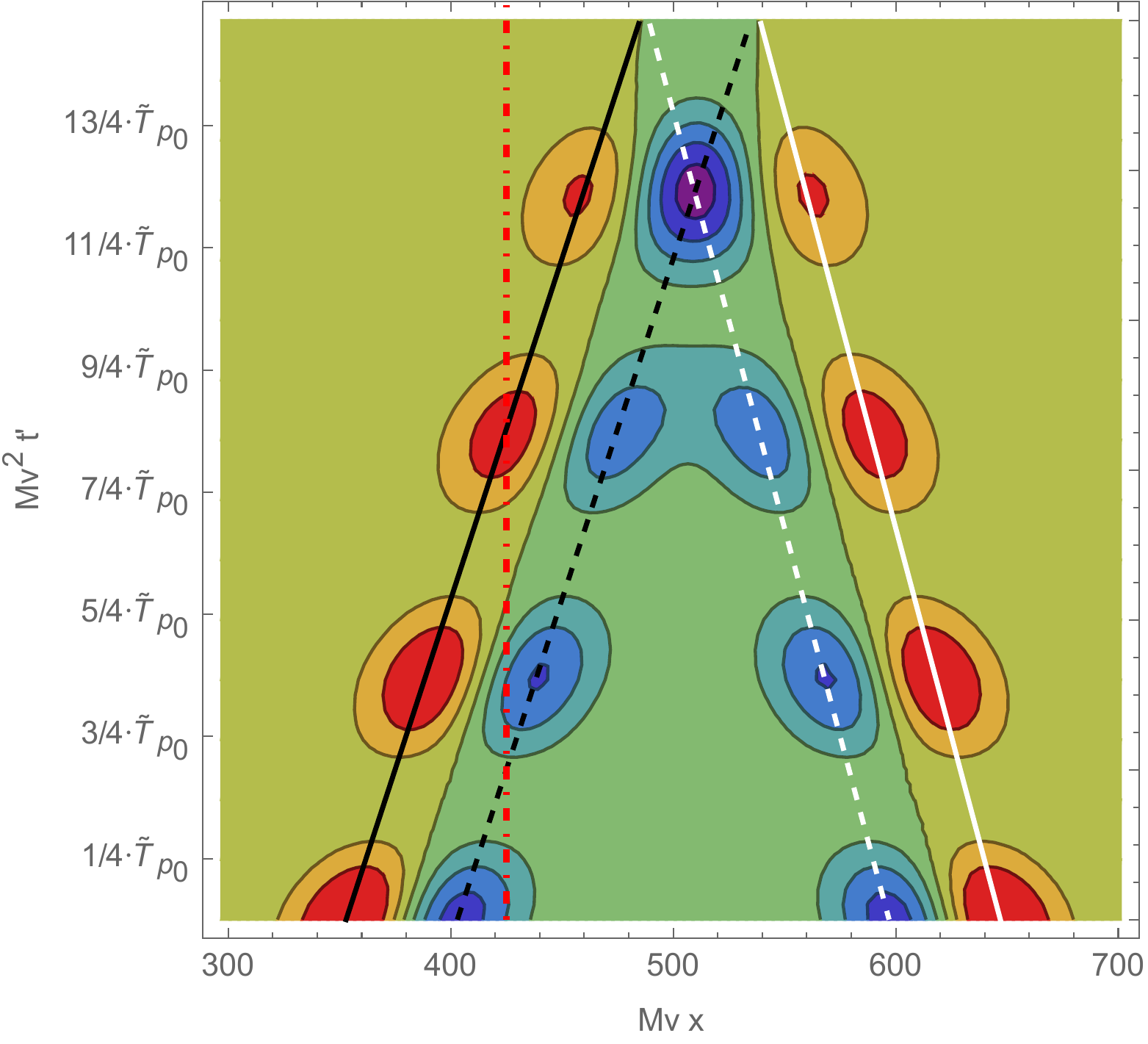}}\\
\subfloat[$\jp=0.03Mv^{2}$\label{fig: integrand annotated 0.03}]{\includegraphics[width=0.9\linewidth]{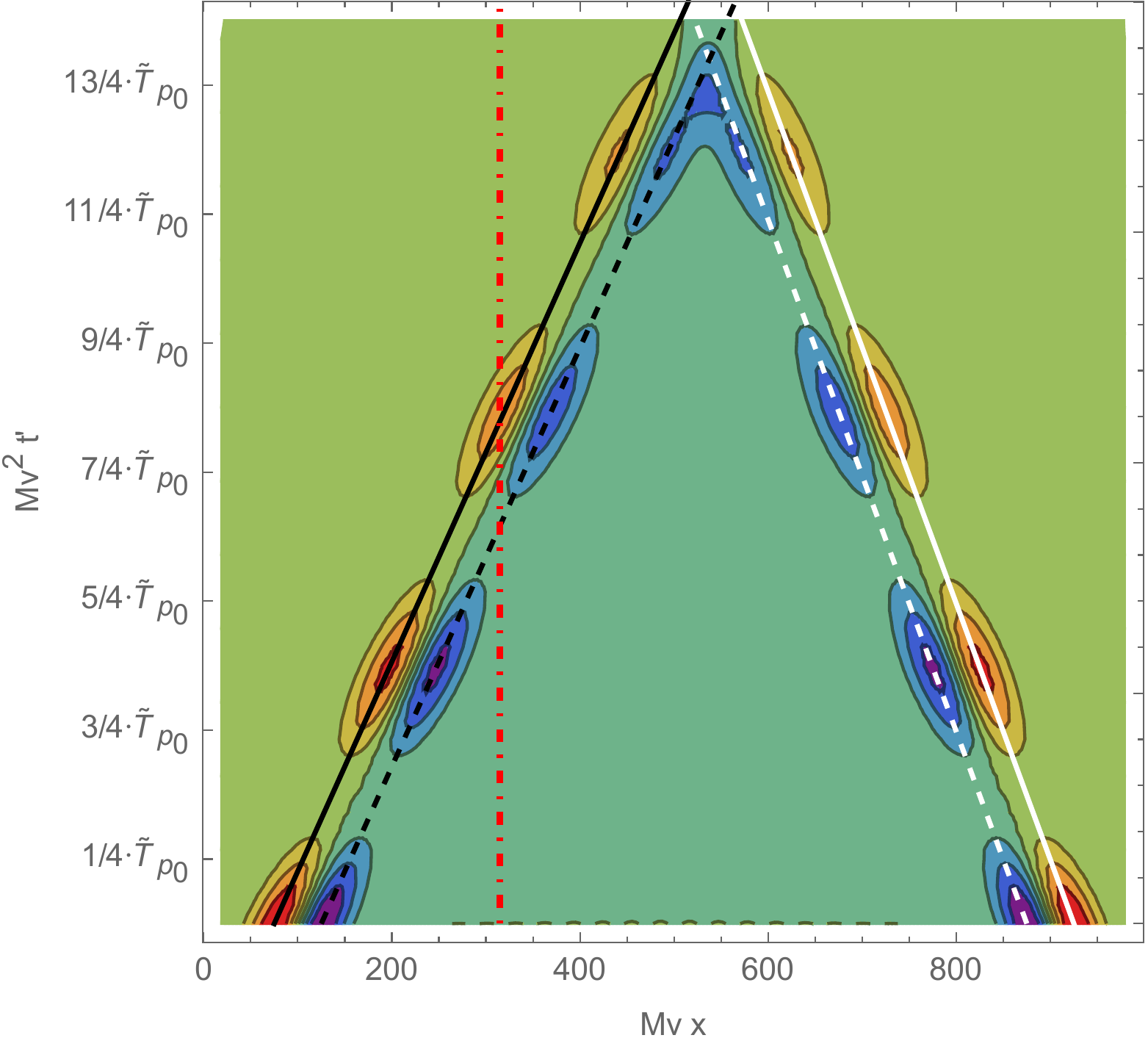}}
\caption{Contour plot of the integrand of \eq~\eqref{eq: continuum density linear response}, computed with the impurity density obtained numerically. Warm colors correspond to positive values, while cold colors indicate negative values. The coupling is $g^{2}K=0.5v^{2}$ (symmetric baths), with a Gaussian wave packet composed of $N_p=32$ momenta with standard deviation $\delta p=0.02Mv$. Time is measured in units of the interacting density oscillation period $\Tilde{T}_{p_0}=\pi/\tilde{J}_{\perp,p_0}$ (the absolute timescales of the two plots are therefore different). For the meaning of the various lines, see the text.}
\label{figs: integrand annotated}
\end{figure}
In this \app, we use \eq~\eqref{eq: continuum density linear response} to obtain a qualitative and quantitative understanding of the observed behavior of the time evolution of the bath density profiles. 
\par We repeat the equation here, for clarity:
\begin{multline*}
    \expval{\rho_\sigma(x)}_t=\tfrac{g_\sigma K_\sigma}{2\pi}\int_0^t\dd{t^\prime}\bigg[\partial_{x^\prime}\expval{N_\sigma(x^\prime)}_{t^\prime}\eval_{x^\prime=x+v_\sigma(t-t^\prime)}+\\-\partial_{x^\prime}\expval{N_\sigma(x^\prime)}_{t^\prime}\eval_{x^\prime=x-v_\sigma(t-t^\prime)}\bigg]~.
\end{multline*}
In \figs~\ref{figs: integrand annotated} we show the integrand of \eq~\eqref{eq: continuum density linear response}, using a numerical lattice derivative of the interacting impurity density we computed in \ref{subsec:impurity observables} for $g^{2}K=0.5v^{2}$,in the $\uparrow$ bath and for two different values of the inter-bath hopping. The wave packets are composed of $N_p=32$ momenta, and are initially Gaussian with standard deviation $\delta p=0.02Mv$ in momentum, which translates to a spatial width of $\delta x\approx1/(2\delta p)=25 (Mv)^{-1}$. In the notation of \eq~\eqref{eq: continuum density linear response}, the time $t$ at which we want to calculate the bath density is fixed in each plot (and coincides with the maximum time shown), while the horizontal and vertical axes of the figures run along the desired position $x$ and the integration time $t^\prime$, respectively. Therefore, the bath density at a given position is obtained by integration along a vertical line, as the red dotted-dashed lines shown as examples. To highlight the periodicity of the oscillations, we measure time in the renormalized period of density oscillations, $\Tilde{T}_{p_0}\equiv\pi/\tilde{J}_{\perp,p_0}$. As the impurity density is essentially a Gaussian, its derivative has both a positive and a negative part, depicted in warm and cold colours, respectively, and this two-lobe structure is repeated along the lines $t^\prime=t+(x-x_0-p_0 t/M)/v_{-\sigma}$ and $t^\prime=t+(x-x_0-p_0 t/M)/v_{+\sigma}$, as dictated by the causality structure of \eq~\eqref{eq: continuum density linear response} (we are ignoring the small slowing down of the impurity momentum) and by the periodic oscillations from one bath to the other. Recall that $v_{\pm\sigma}\equiv v_\sigma\pm p_0/M$. The four tilted lines show the approximate loci of the maxima and minima: $t^\prime=t+(x-x_0-p_0 t/M\pm \delta x)/v_{-\sigma}$ for the black lines and $t^\prime=t-(x-x_0-p_0 t/M\pm \delta x)/v_{-\sigma}$ for the white ones. Notice that we are neglecting the increase in width of the wave packet during its dynamics. It can be seen in the \figs~that it does not seem to play a relevant role, so we take $\delta x$ to be the initial standard deviation.
\par If we compare the bath densities in \figs~\ref{fig: bath density width comparison}, we see that \fig~\ref{fig: integrand annotated 0.1} corresponds to a situation in which there are no ripples, while \fig~\ref{fig: integrand annotated 0.03} gives rise to ripples. Now it is easy to understand how this situation emerges from \eq~\eqref{eq: continuum density linear response}. Let us take the position corresponding to the red line in \fig~\ref{fig: integrand annotated 0.1}. We see that during the time integration we encounter a positive contribution and part of two negative lobes belonging to the previous two impurity oscillations. The results will thus be close to zero. On the other hand, the integration path in \fig~\ref{fig: integrand annotated 0.03} only encounters a positive lobe, and therefore it will give rise to the positive part of a ripple. If we change position, the same situation occurs: for $\jp=0.1Mv^2$, any vertical line will always cross regions of both signs, with the result that it will always close to zero, while for $\jp=0.03Mv^2$ it will alternatively cross positive and negative regions, resulting in oscillations of the density, i.e. the ripples. Thus, we see that the ripples emerge from an interference effect between subsequent oscillations of the impurity. The extent of this interference is regulated the interplay between the periodic impurity oscillations, the sound speed and the width of the wave packet. We can make a quantitative estimate of the parameters needed for a destructive interference: it happens whenever the oscillation period is such that the position of a positive lobe overlaps with the position of the negative lobe of the previous oscillation. With the help of \figs~\ref{figs: integrand annotated}, this translates to
\begin{equation}
    v_{\pm\sigma} \tilde{T}_{p_0}\lesssim 2\delta x~,
\end{equation}
which directly leads to \eqs~\eqref{eq: critical delta x} and \eqref{eq: critical delta p} in the main text.
\par The only regions that are exempted from this interference mechanism are the farthest positions reachable by causality, $\abs{x-x_0}\approx vt$, and the ones around the center $x\approx x_0+p_0 t/M$, which are easily identified with the wave fronts and the central depletion of the bath density, respectively. Indeed, from \figs~\ref{figs: integrand annotated} we can see that for $\abs{x-x_0}\approx vt$ the time integral intersects only positive lobes, while around $x= x_0+p_0 t/M$ there is a region with only negative contributions. Therefore, we obtain the features we observed in \ref{subsec: density evolution}, namely that the wave fronts are always positive, while the depletion is always negative. These arguments also show that the wave fronts are images of the impurity density at the initial time, whereas the depletion is sensitive only to the density in the near past. In the main text, we claimed that the wave fronts and the central dip were images of the impurity density. This claim would be exactly true if the density evolution were given simply by a translation: $\expval{d^\dag_\sigma(x^\prime)d_\sigma(x^\prime)}_{t^\prime}=N_0(x-x_0-p_0 t/M)$, where $N_0(x-x_0)$ is the initial profile shape:
\begin{multline*}
    \expval{\rho_\sigma(x,t)}_t=\tfrac{g_\sigma K_\sigma}{2\pi}\Big[\tfrac{1}{v_{+\sigma}}N_0(x-x_0+v t)+\\
    -\left(\tfrac{1}{v_{+\sigma}}-\tfrac{1}{v_{-\sigma}}\right)N_0(x-x_0+p_0 t/M)+\\
    +\tfrac{1}{v_{-\sigma}}N_0(x-x_0-v t)\Big]~.
\end{multline*}
We can easily recognize the first and the last terms as the two counter-propagating wave fronts, which are translated images of the wave packet, and a negative depletion that follows the impurity. We also see that the heights of the wave fronts are different from each other, with the backward being shorter than the forward one, the difference being larger the fastest is the impurity. The \eq~above is valid only in a highly idealized situation, in which there is only one bath ($\jp=0$) and the wave packet does not spread. In our situation, both hypotheses are false, but we can guess that the most relevant phenomena are caused by the retardation effects given by the density oscillations\footnote{Indeed, it is possible to obtain an analytic expression of the bath density if we take $\expval{d^\dag_\sigma(x^\prime)d_\sigma(x^\prime)}_{t^\prime}=N_0(x-x_0-p_0 t/M)(\cos\jp t^\prime)^2$, that is, if we again discard the wave packet spreading. In the solution, we can still recognize the presence of the wave fronts and the depletion.}. For instance, the ratio of the wave fronts heights of the bath density we computed numerically tends to $v_{+\sigma}/v_{-\sigma}$ at long times.
\begin{figure}
    \centering
    \includegraphics[width=0.9\linewidth]{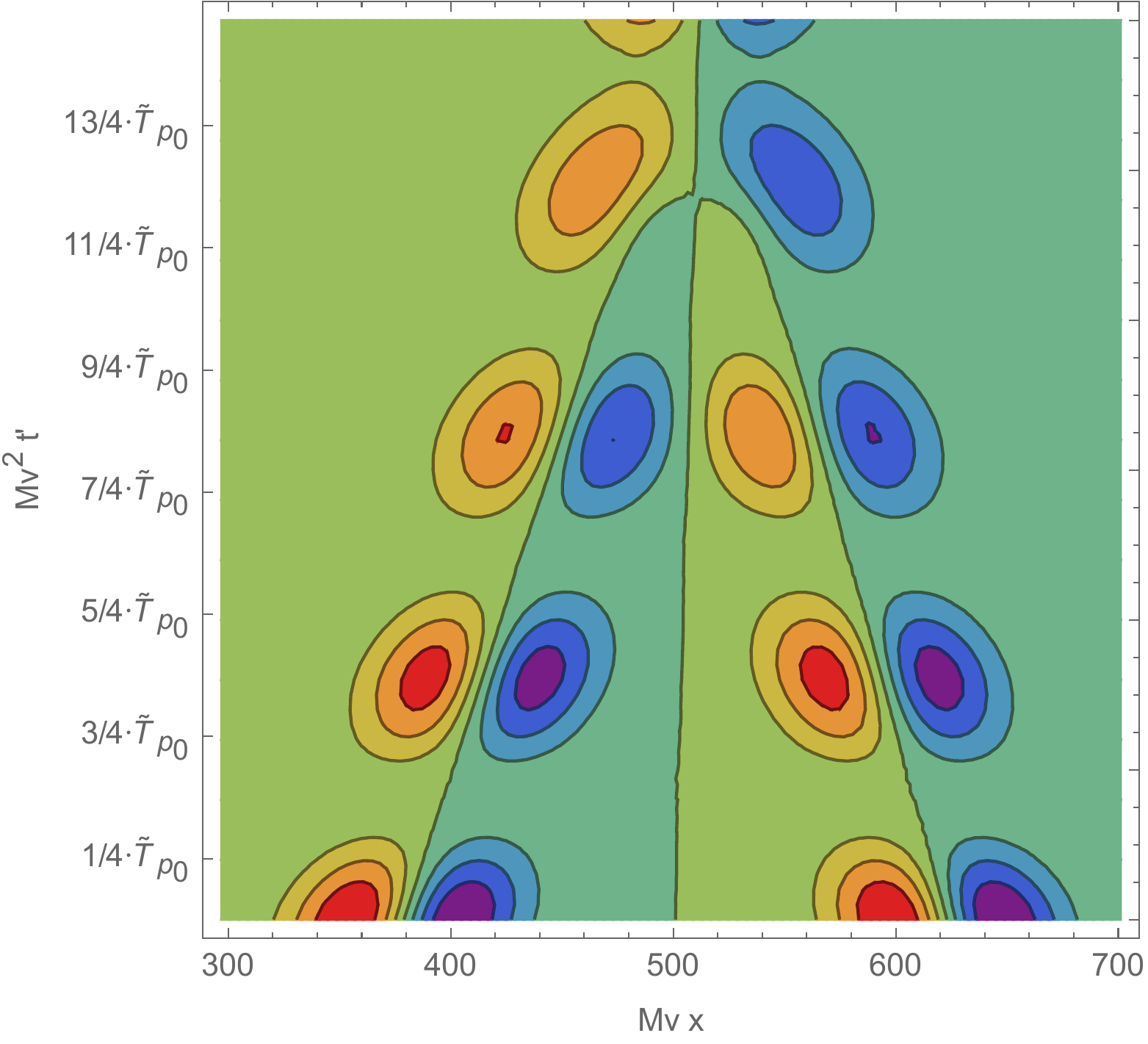}
    \caption{Same as \fig~\ref{fig: integrand annotated 0.1} (i.e. for $\jp=0.1Mv^2$), but for the bath momentum density.}
    \label{fig: integrand bath momentum}
\end{figure}
\par As a final remark, we point out that the same arguments given above can be repeated for the bath momentum density, which is connected to the impurity density through an equation analogous to \eqref{eq: linear relation densities}, with the response function
\begin{equation}
    \chi^{\Pi\rho}_\sigma(x,t)=-\theta(t)\tfrac{1}{2\pi}\left[\delta_\alpha^\prime(x+v_\sigma t)+\delta_\alpha^\prime(x-v_\sigma t)\right]~.
\end{equation}
We can see that we would obtain the analogous of \eq~\eqref{eq: continuum density linear response}, but with the two translated density gradients added to each other instead of being subtracted. We would then obtain an integrand depicted in \fig~\ref{fig: integrand bath momentum}, which clearly shows the characteristic feature of the bath momentum density that we observed in the main text, namely that the fluctuations emitted forward and backward are approximately inverted images of each other, instead of being approximately mirror images as in the case of the density.
\bibliography{biblioPaper2.bib}
\end{document}